\documentstyle[preprint,aps,version2]{revtex}
\newfont{\titlefont}{cmbx10 scaled 1270}      %font for maintext
\newfont{\titlefonta}{cmbx10 scaled 1100}      %font for maintext
\catcode`\@=11
\long\def\@makefntext#1{
\protect\noindent \hbox to 3.2pt {\hskip-.9pt
$^{{\ninerm\@thefnmark}}$\hfil}#1\hfill}                %CAN BE USED
\def\@makefnmark{\hbox to 0pt{$^{\@thefnmark}$\hss}}  %ORIGINAL
\def\ps@myheadings{\let\@mkboth\@gobbletwo
\def\@oddhead{\hbox{}
\rightmark\hfil\ninerm\thepage}
\def\@oddfoot{}\def\@evenhead{\ninerm\thepage\hfil
\leftmark\hbox{}}\def\@evenfoot{}
\def\sectionmark##1{}\def\subsectionmark##1{}}
\renewcommand{\thefootnote}{\fnsymbol{footnote}}
\newcounter{sectionc}\newcounter{subsectionc}\newcounter{subsubsectionc}
\renewcommand{\section}[1] {\vspace*{0.6cm}\addtocounter{sectionc}{1}
\setcounter{subsectionc}{0}\setcounter{subsubsectionc}{0}\noindent
        {\normalsize\bf\thesectionc. #1}\par\vspace*{0.4cm}} 
\renewcommand{\subsection}[1] {\vspace*{0.6cm}\addtocounter{subsectionc}{1}
        \setcounter{subsubsectionc}{0}\noindent
    {\normalsize\it\thesectionc.\thesubsectionc. #1}\par\vspace*{0.4cm}}
\renewcommand{\subsubsection}[1]
{\vspace*{0.6cm}\addtocounter{subsubsectionc}{1}
\noindent {\normalsize\rm\thesectionc.\thesubsectionc.\thesubsubsectionc.
    #1}\par\vspace*{0.4cm}}

\newcounter{appendixc}
\newcounter{subappendixc}[appendixc]
\newcounter{subsubappendixc}[subappendixc]

\renewcommand{\appendix}[1] {\vspace*{0.6cm}
        \refstepcounter{appendixc}
        \setcounter{figure}{0}
        \setcounter{table}{0}
        \setcounter{equation}{0}
        \renewcommand{\thefigure}{\Alph{appendixc}.\arabic{figure}}
        \renewcommand{\thetable}{\Alph{appendixc}.\arabic{table}}
        \renewcommand{\theappendixc}{\Alph{appendixc}}
        \renewcommand{\theequation}{\Alph{appendixc}.\arabic{equation}}
        \noindent{\bf Appendix \theappendixc #1}\par\vspace*{0.4cm}}

\renewenvironment{thebibliography}[1]
    {\begin{list}{\arabic{enumi}.}
    {\usecounter{enumi}\setlength{\parsep}{0pt}
\setlength{\leftmargin 1.25cm}{\rightmargin 0pt}
     \setlength{\itemsep}{0pt} \settowidth
    {\labelwidth}{#1.}\sloppy}}{\end{list}}
\newcounter{itemlistc}
\newcounter{romanlistc}
\newcounter{alphlistc}
\newcounter{arabiclistc}

\newcommand{\fcaption}[1]{
        \refstepcounter{figure}
        \setbox\@tempboxa = \hbox{\footnotesize Fig.~\thefigure. #1}
        \ifdim \wd\@tempboxa > 6in
           {\begin{center}
        \parbox{6in}{\footnotesize\baselineskip=15pt Fig.~\thefigure. #1}
            \end{center}}
        \else
             {\begin{center}
             {\footnotesize Fig.~\thefigure. #1}
              \end{center}}
        \fi}
\newcommand{\tcaption}[1]{
        \refstepcounter{table}
        \setbox\@tempboxa = \hbox{\footnotesize Table~\thetable. #1}
        \ifdim \wd\@tempboxa > 6in
           {\begin{center}
        \parbox{6in}{\footnotesize\baselineskip=15pt Table~\thetable. #1}
            \end{center}}
        \else
             {\begin{center}
             {\footnotesize Table~\thetable. #1}
              \end{center}}
        \fi}
 1 %2
 1  %2
 1  %2
 1  %half
 1  %half
 1  %half
 1  %half

\font\ninerm=cmr9

\evensidemargin=\oddsidemargin  
\def\today{\number\day
           \space\ifcase\month\or
             January\or February\or March\or April\or May\or June\or
             July\or August\or September\or October\or November\or December\fi
           \space\number\year}
\evensidemargin=\oddsidemargin  
\def\doublespaced{\baselineskip=\normalbaselineskip\multiply
    \baselineskip by 150\divide\baselineskip by 100}
\doublespaced

\begin{document}

%%%%%%%%%%%%%%%%%%%%%%   HHHHHHHHHHHHHHHHHHHHHHHHH   %%%%%%%%%%%%%%%%%%%%%%%%%
\thispagestyle{empty}
\renewcommand{\thefootnote}{\fnsymbol{footnote}}
\begin{flushright}
%{\small ISSN 0418-9833} \hfill   {\small July, 1996}\\[-0.25cm]
{\small August, 1996} \hfill {hep-ph/9611316}\\[-0.25cm]
{\small DESY-96-148}\\[-0.26cm]
{\small TUIMP-TH-96/78}\\[-0.26cm]
{\small MSUHEP-60615}
\end{flushright}
\vspace{0.3cm}

%%%%%%%%%%%%%%%%%%%%%%%%%%%%%%%%%%%%%%%%%%%%%%%%%%%%%%%%%%%%%%%%%%%%%%%%%%%%%%%
\centerline{\titlefont
Estimating the Sensitivity of the LHC to Electroweak Symmetry Breaking:}
\baselineskip=17pt
\centerline{\titlefonta             
Longitudinal-Goldstone Boson Equivalence as a Criterion}
%%%%%%%%%%%%%%%%%%%%%%%%%%%%%%%%%%%%%%%%%%%%%%%%%%%%%%%%%%%%%%%%%%%%%%%%%%%%%%%%

\vspace*{1.3cm}
\centerline{\small  
{\bf Hong-Jian He}~$^{(a)}$,~~~~
{\bf Yu-Ping Kuang}~$^{(b)}$,~~~~
{\bf C.--P. Yuan}~$^{(c)}$ }
\vspace*{0.4cm}

\baselineskip=17pt
\centerline{\small\it
$^{(a)}$ { Theory Division, Deutsches Elektronen-Synchrotron DESY, 
Hamburg, Germany
\footnote{Mailing address.}}}
\baselineskip=14pt
\centerline{\small\it 
{ Department of Physics and Institute of High Energy Physics}}
\baselineskip=14pt
\centerline{\small\it
Virginia Polytechnic Institute and State University, VA 24061, USA}
\vspace*{0.28cm}
\centerline{\small\it
$^{(b)}$ {
CCAST ( World Laboratory ), P.O.Box 8730, Beijing 100080, China }}
\baselineskip=14pt
\centerline{\small\it
Institute of Modern Physics, Tsinghua University, Beijing 100084, China
$^{\dagger}$}
\vspace*{0.28cm}
\centerline{\small
$^{(c)}$
{\it Department of Physics and Astronomy, Michigan State University } }
\baselineskip=14pt
\centerline{\small\it East Lansing, Michigan 48824, USA }

\vspace{-1.2cm}
\begin{abstract}
\baselineskip=17pt

\begin{center}
{\bf Abstract}
\end{center}
\noindent
Based upon our recent study, 
we reveal the profound physical content of 
the longitudinal-Goldstone boson  equivalence theorem  as 
being able to discriminate physical processes 
which are sensitive/insensitive to probing the 
electroweak symmetry breaking (EWSB) sector.  
We then develop a precise electroweak power counting rule 
(a generalization from Weinberg's counting method)
to {\it separately} count the power dependences on the energy $E$ 
and all relevant mass scales.
With these, we analyze the complete set of the bosonic 
operators in the electroweak chiral Lagrangian 
and systematically estimate and classify the sensitivities 
for testing all these effective operators at the CERN LHC 
via the weak-boson fusions and quark-anti-quark annihilations.  
These two kinds of processes are shown to be {\it complementary} 
in probing the EWSB mechanism.

%\vspace{0.3cm}
\noindent
{\small PACS number(s): 11.30.Qc, 11.15.Ex, 12.15.Ji, 14.70.--e}
\hfill ({\it Phys. Rev. D}, in press) 
\begin{flushright}
\pacs{11.30.Qc, 11.15.Ex, 12.15.Ji, 14.70}
\end{flushright}

% ( Version to be Published in {\it Physical Review D} )
%\end{center}
\end{abstract}

\normalsize
\baselineskip=19.5pt
\setcounter{footnote}{00}
\renewcommand{\thefootnote}{\alph{footnote}}
\renewcommand{\baselinestretch}{1.3}

%%%%%%%%%%%%%%%%%%%%%%%%%%%%%%%%%%%%%%%%%%%%%%%%%%%%%%%%%%%%%%%%%%%%%%%%%%
% Contents:
% 1. Introduction
% 2. Formulating the ET as a criterion for probing the EWSB
% 3. Generalized precise power counting for the EW chiral Lagrangian
% 4. Classification of sensitivities at the level of S-matrix elements
% 5. Probing EWSB mechanisms at the CERN LHC via weak-boson scatterings
% 6. Acknowledgements
% 7. Appendix-A: Validity of the ET in some special kinematic regions
%    Appendix-B: Electroweak Power Counting Rule for Linearly Realized
%                Effective Lagrangians   
% 8. References
%%%%%%%%%%%%%%%%%%%%%%%%%%%%%%%%%%%%%%%%%%%%%%%%%%%%%%%%%%%%%%%%%%%%%%%%%%

\newpage
\setcounter{page}{1}
\noindent
{\bf 1. Introduction}
\vspace{0.3cm}

Despite the impressive success of the Standard Model (SM) over the years,
its scalar part, the electroweak symmetry breaking (EWSB) sector,
remains as the greatest mystery. 
Due to Veltman's screening theorem~\cite{screening}, 
the current low energy data, 
allowing the SM Higgs boson mass to range 
from $65.2$\,GeV to about $O(1)$\,TeV~\cite{LEP2},
tell us little about the EWSB mechanism.
Therefore, it is important to probe {\it all possible} EWSB mechanisms,
either weakly or strongly interacting
as long as the light Higgs particle remains undetected. 
Even if a light resonance is detected at future colliders, it is still 
crucial to further test whether it is associated with a strong dynamics,
because it is unknown {\it a priori} if such a resonance simply serves
as the SM Higgs boson or comes from a more complicated mechanism~\cite{tc}.
If the EWSB is driven by a strong dynamics with no new resonance well
below the TeV scale, the probe will be more difficult at the future
high energy colliders. It is this latter case that
we shall currently investigate for the CERN Large Hadron Collider (LHC).

While the transverse components $V_T^a$ of $~W^{\pm},~Z^0~$ 
are irrelevant to the EWSB mechanism, the longitudinal weak-bosons 
( $V_L^a=W^{\pm}_L,~Z^0_L~$), as the products of 
the spontaneously symmetry-breaking
mechanism, are expected to be sensitive to
probing the EWSB sector. However, even for the strongly
coupled case, studying the $V_L$-scatterings does not guarantee 
probing the EWSB sector in a sensitive and unambiguous way 
because the spin-$0$ Goldstone bosons (GB's) are 
invariant under the proper Lorentz 
transformations, while, on the contrary, both $V_L$ and $V_T$ 
are Lorentz non-invariant 
(LNI). After a Lorentz transformation, the $V_L$
 component can mix with  or even turn into a pure $V_T$.
Thus a conceptual ambiguity arises: How can 
the LNI $~V_L$-amplitudes be used to probe the 
EWSB sector of which the physical mechanism should evidently be 
independent of the choices of the Lorentz frames? 
This motivated our recent precise 
formulation of the electroweak longitudinal-Goldstone boson Equivalence 
Theorem (ET) in Ref.~\cite{et3}.
In the high energy region ($E\gg M_W$), 
the ET provides a general and quantitative
relation between any $V_L$-amplitude and its corresponding 
GB-amplitude to all loop orders~\cite{et3}-\cite{etIR};\footnote{
A simple pedagogical discussion on the ET was given 
at the tree level for amplitudes with one external $V_L$-line
in a new textbook by Peskin and Schroeder~\cite{peskin}.
For an earlier textbook introduction of the ET, 
see Ref.~\cite{donoghue}.}~ the former is physically measurable while
the latter carries information about the EWSB sector. 
Hence, the ET allows us to probe the EWSB sector
by relating it to the $~V_L$-scattering experiments.
As will be shown later, the {\it difference} 
between the $V_L$- and GB-amplitudes
is intrinsically related to 
the ambiguous LNI part of the $V_L$-scattering
which has the same origin as the $V_T$-amplitude
and is thus insensitive to probing the EWSB sector.
When the LNI contributions can be safely ignored and the 
Lorentz invariant (LI) scalar  GB-amplitude dominates
the experimentally measured $V_L$-amplitudes, 
the physical $V_L$-scatterings can then
{\it sensitively  and unambiguously} probe the EWSB mechanism.
Since the ratio of the LI GB-amplitude to the LNI contributions 
is {\it process-dependent}, it can thus provide a 
useful theoretical determination on the sensitivities of various 
scattering processes to probing the EWSB sector. 

At the scale below new heavy resonances, 
the EWSB sector can be parametrized by means of the
electroweak chiral Lagrangian (EWCL) in which the 
$SU(2)_L \otimes U(1)_Y$ gauge symmetry is nonlinearly realized.
Without experimental observation of any
new light resonance in the EWSB sector, this effective
field theory approach provides the most economic description of
the possible new physics effects 
and is thus {\it complementary} to those specific model buildings.
In the present analysis, taking this conservative and
general EWCL approach, we shall concentrate on 
studying the effective bosonic operators among which 
the leading order operators are universal 
and the next-to-leading-order (NLO) operators describe 
the model-dependent effects in the EWCL.
We show in this paper that, for a given process, the ratio 
of the scalar GB-amplitude to the LNI part of the $V_L$-amplitude,
which determines the validity of the ET,
varies for different effective operators.
The larger this ratio is, the more sensitive 
this process will be to an operator. 
{\it  Therefore, this ratio can be used to discriminate sensitivities 
to all the NLO effective operators as well as to the scattering processes
for probing the EWSB sector.}  
By formulating the ET as a physical criterion, we shall systematically 
classify the sensitivities to all these effective
operators at the LHC\footnote{
The actual sensitivity of the LHC to probing these operators will
also depend on the detection efficiency for suppressing 
the backgrounds to observe the specific decay mode of the final state
for each given process (as discussed in Ref.~\cite{wwlhc}). 
This is beyond the scope of our present theoretical
global analysis. In this paper we take the same spirit
as Ref.~\cite{BDV} and leave the detection issue to a
future detailed and precise numerical study with this work 
as a useful guideline.}.~  We show that the ET is 
not just a technical tool 
for explicitly computing $V_L$-amplitudes via GB-amplitudes; 
{\it as a criterion, it has an even more profound physical content for 
being able to theoretically discriminate sensitivities to
different effective operators via different processes for probing the
EWSB mechanism}~\cite{et3}.  

In performing such a global analysis, in contrast
to just studying a few operators, we 
need to estimate the contributions of all the 
NLO operators to various high energy scattering processes. For this
purpose, we construct   
a precise electroweak power counting rule
for the EWCL formalism through a natural generalization of Weinberg's
counting method for non-linear sigma model~\cite{wei}. 
This simple power counting rule is 
proven to be extremely convenient and useful for our global analysis.

This paper is organized as follows. 
We first formulate the ET
as a criterion for probing the EWSB mechanism in Sec.~2, 
and derive a precise electroweak power counting rule
for the EWCL formalism in Sec.~3. 
Then, based upon these 
we classify the sensitivities of all effective
operators at the level of the $S$-matrix elements in Sec.~4. 
Finally we further analyze, in Sec.~5, 
the probe of the EWSB sector at the LHC via weak-boson fusions 
and quark-anti-quark annihilations.
Conclusions are given in Sec.~6. Also, a detailed analysis on the 
the validity of the ET in some special kinematic regions 
and its implication in probing the EWSB sector is
presented in Appendix A.
Appendix B is devoted to derive a set of power counting 
rules for the linearly realized effective Lagrangian formalism
which includes the light Higgs SM at the lowest order.

\vspace{0.5cm}
\noindent
{\bf 2. Formulating the ET as a Criterion for Probing the EWSB}
\vspace{0.3cm}

The precise conditions for the longitudinal-Goldstone boson equivalence,
i.e. for the validity of the ET, have been derived in our recent 
study~\cite{et3}. In this section, after a further exploration on the
physical implications of these conditions we formulate 
the ET as a criterion for probing the EWSB mechanism at the level
of both the $S$-matrix elements and the total cross sections.

We start from  the following general 
identity for the renormalized $S$-matrix elements (cf. 
the 2nd paper in Ref.~\cite{et2} for a rigorous derivation): 
$$
T[V^{a_1}_L,\cdots ,V^{a_n}_L;\Phi_{\alpha}]
= C\cdot T[-i\pi^{a_1},\cdots ,-i\pi^{a_n};\Phi_{\alpha}]+ B  ~~,
\eqno(2.1)                                               %(2.1)
$$
$$
\begin{array}{ll}
C & \equiv C^{a_1}_{\rm mod}\cdots C^{a_n}_{\rm mod}
~=~1+O({\rm loop}) ~, \\[0.3cm]
B & \equiv\sum_{l=1}^n (~C^{a_{l+1}}_{\rm mod}\cdots C^{a_n}_{\rm mod}
T[v^{a_1},\cdots ,v^{a_l},-i\pi^{a_{l+1}},\cdots ,
-i\pi^{a_n};\Phi_{\alpha}]  + ~{\rm permutations}~)~~,  \\[0.3cm]
v^a & \equiv v^{\mu}V^a_{\mu} ~, 
~~~~v^{\mu}~\equiv \epsilon^{\mu}_L-k^\mu /M_V = O(M_V/E) ~,
~~~(M_V=M_W,M_Z)~,
\end{array}
\eqno(2.1a,b,c)                                            %(2.1a,b,c)
$$
where $~\pi^a$'s  are GB fields, $\Phi_{\alpha}$ 
denotes other possible physical in/out states. The finite 
constant modification factor $~C_{\rm mod}^a~$ has been systematically 
studied in Ref.~\cite{et2,et3}.\footnote{The loop factor $~C_{\rm mod}^a-1
=O\left(\frac{g^2}{16\pi^2}\right)~$ for the EWCL formalism if the 
wavefunction renormalization constant $Z_{\pi^a}$ is subtracted at a 
scale of $O(M_W)$; and $~C_{\rm mod}^a=1~$ to all orders in some 
convenient renormalization schemes for all $R_\xi$-gauges~\cite{et2}.}~
From the above identity, we see that the 
LNI $~V_L$-amplitude can be decomposed 
into two parts: the 1st part is 
$~C\cdot T[-i\pi;\Phi_{\alpha}]~$ which is LI; the 2nd
part is the $~v_\mu$-suppressed $B$-term which is  LNI 
because it contains the external {\hbox{spin-1}} $V_{\mu}$-field(s).  
(Without losing generality~\cite{et3}, 
here we have assumed that $~\Phi_\alpha~$ contains 
possible physical scalars, photons and light fermions.)
Such a decomposition  
shows the {\it essential difference} between the $V_L$- and 
the $V_T$-amplitudes: the former contains a LI GB-amplitude that can 
yield a large $V_L$-amplitude in the case of
strongly coupled EWSB sector, but the latter does not.
We note that only the LI part (the GB-amplitude) of the $V_L$-amplitude 
is {\it sensitive} to probing the EWSB sector, 
while its LNI part, containing a
significant {\it Lorentz-frame-dependent} $~B$-term (which is related to
the $V_L$-$V_T$ mixing effects under proper Lorentz transformations),
 is {\it insensitive} to the EWSB mechanism.
Thus, for a sensitive and unambiguous probe of the EWSB,
we must find conditions under which the LI GB-amplitude
dominates the $V_L$-amplitude and the LNI $B$-term is negligible.
It is obvious that one can technically improve the prediction for the 
$V_L$-amplitude from the right-hand side (RHS) of (2.1) 
by including the complicated $B$-term 
( or part of $B$ )~\cite{gk} or even directly calculate its
left-hand side (LHS) of (2.1) despite the complexity.
However, this is {\it not} an improvement of the 
longitudinal-Goldstone boson {\it equivalence} and thus the sensitivity 
of probing the EWSB mechanism via  $V_L$-scattering experiments.
 {\it The physical content of the ET is essentially
independent of how to numerically compute the $V_L$-amplitude.}

In Ref.~\cite{et3}, we estimated the $B$-term from a detailed analysis on the 
LNI $V_L$-amplitude and concluded
$$
B \approx O\left(\frac{M_W^2}{E_j^2}\right)
  T[ -i\pi^{a_1},\cdots , -i\pi^{a_n}; \Phi_{\alpha}] +   
  O\left(\frac{M_W}{E_j}\right)T[ V_{T_j} ^{a_{r_1}}, -i\pi^{a_{r_2}},
                      \cdots , -i\pi^{a_{r_n}}; \Phi_{\alpha}]~~. 
\eqno(2.2)                                                   %(2.2)
$$
We see that the condition $~~ E_j \sim k_j  \gg  M_W , ~
(j=1,2,\cdots ,n) ~~$  for each external longitudinal weak-boson is
necessary for making the GB-amplitude much larger than the $B$-term 
(and its Lorentz variation). We thus deduced the general 
and precise formulation of the ET as~\cite{et3}:\footnote{
Note that the $B$-term on the RHS of (2.3),
as specified, is only $~O(M_W/E_j)$-suppressed {\it relative} to the
leading contributions in the GB-amplitude and is therefore {\it not
necessarily} of the $O(M_W/E_j)$ in magnitude. (2.2) explicitly shows that 
the magnitude of the $B$-term depends on the size of the amplitudes 
$T[-i\pi^{a_1},\cdots ]$ and $T[V_{T_j}^{a_{r_1}},-i\pi^{a_{r_2}},\cdots ]$ 
so that it can be {\it either larger  or smaller than $O(M_W/E_j)$} 
[cf. Eq.~(2.5)].  }
$$                                                                             
T[V^{a_1}_L,\cdots ,V^{a_n}_L;\Phi_{\alpha}]                                   
= C\cdot T[-i\pi^{a_1},\cdots ,-i\pi^{a_n};\Phi_{\alpha}]+ 
O(M_W/E_j){\rm -suppressed} ~~,
\eqno(2.3)                                               %(2.3)
$$                                                                             
$$                                                                             
E_j \sim k_j  \gg  M_W , ~~~~~(~ j=1,2,\cdots ,n ~)~~,           
\eqno(2.3a)                               %(2.3a)
$$
$$
C\cdot T[-i\pi^{a_1},\cdots ,-i\pi^{a_n};\Phi_{\alpha}]\gg B ~~. 
\eqno(2.3b)                               %(2.3b)                        
$$                                                                             
Here, (2.3a,b) are the precise conditions for 
the validity of the {\it equivalence} in (2.3).
We emphasize that, in principle, the complete set of diagrams 
(including those with internal gauge boson lines) 
has to be considered when calculating 
$~T[-i\pi^{a_1},\cdots ,-i\pi^{a_n};\Phi_{\alpha}]~$, 
as already implied in (2.3).
If not, this equivalence might not manifest for 
scattering processes involving $t$- or $u$- channel diagram
in either forward or backward direction.
A detailed discussion on this point is given in Appendix A. 
Furthermore, the ET (2.3) and its high energy condition (2.3a) 
indicate the absence of infrared (IR)
power divergences (like $~\left(\frac{E_j}{M_W}\right)^r,~r>0~$) 
in the $~M_W\rightarrow 0~$ limit for fixed
energy $~E_j\sim k_j~$~\cite{etIR}. 
This can be understood by noting
that the limit $~M_W\rightarrow 0~$ implies $~g\rightarrow 0~$
after fixing the physical vacuum expectation value (VEV) at
$~f_\pi =246$~GeV. Taking $~g\rightarrow 0~$ limit 
leads to the well-defined un-gauged linear or non-linear sigma-model
which suggests turning off the gauge coupling to be a smooth procedure.
The smoothness of the $~g\rightarrow 0~$ limit indicates the 
absence of IR power divergences for $~M_W\rightarrow 0~$. 
In our present formalism, we shall fix
the gauge boson mass $M_W(M_Z)$ at its experimental value and take
the validity of the ET as a physical criterion for sensitively
probing the EWSB sector in the specified high energy regime
of a given collider.

The amplitude $T$, to a finite order, can be written as 
$~~T= \sum_{\ell=0}^N T_\ell ~$ in the perturbative calculation.
Let $~~ T_0 > T_1,\cdots, T_N \geq T_{\min}~$, where
$~T_{\min}= \{ T_0, \cdots , T_N \}_{\min}~$, then condition (2.3b)
and eq.~(2.2) imply
$$
\begin{array}{l}
~~~~~~~~~~
T_{\min}[-i\pi^{a_1},\cdots ,-i\pi^{a_n};\Phi_{\alpha}] ~\gg ~\\[0.35cm]
O\left(\frac{M_W^2}{E_j^2}\right) \,
T_0[ -i\pi^{a_1},\cdots , -i\pi^{a_n}; 
   \Phi_{\alpha}] + 
  O\left(\frac{M_W}{E_j}\right) \,T_0[ V_{T_j}^{a_{r_1}}, -i\pi^{a_{r_2}},
                      \cdots , -i\pi^{a_{r_n}}; \Phi_{\alpha}] ~~.
\end{array}                                                                    
\eqno(2.4)                                                 %(2.4)             
$$                                                                             
Note that the above formulation of the ET discriminates processes
which are insensitive to probing the EWSB sector when either
(2.3a) or (2.3b) fails.
Furthermore, {\it as a necessary criterion, condition (2.4) determines 
whether or not the corresponding $~V_L$-scattering process 
in (2.3) is sensitive to probing 
the EWSB sector to the desired precision in perturbative calculations.}

From (2.2) or the RHS of (2.4) and the precise electroweak power counting
rule (cf. Sec.~3.2), we can directly estimate the 
{\it largest} model-independent $B$-term to be
$~~~B_{\max} = O(g^2)f_{\pi}^{4-n}~~$ in the EWCL 
formalism, which comes from the $n$-particle pure $V_L$-amplitude.
(This conclusion also holds for the heavy Higgs SM.)
It is crucial to note that $B_{\max}$ is of the same order 
of magnitude as the leading $V_T$-amplitude:
$$
B_{\max} \approx T_0[V^{a_1}_T,\cdots ,V^{a_n}_T] 
= O(g^2)f_{\pi}^{4-n}~~.
\eqno(2.5)                                 %(2.5)
$$
Since both the largest $B$-term and the leading $V_T$-amplitude are of 
$O(g^2)$~, they are therefore irrelevant to the EWSB mechanism 
as pointed out in the above discussion. Thus, {\it  (2.4) provides a 
useful criterion for discriminating physical processes 
which are sensitive, marginally sensitive, or insensitive to 
the EWSB sector.}

In conclusion, our formulation of the ET 
provides a necessary criterion for probing the EWSB sector as follows. 
If the ET is valid for a scattering process to the order of
$~T_{\min}~$ [i.e. $~T_{\min}\gg B_0~$, cf.~(2.4),
where $B_0$ is the tree-level leading contribution to $B$.], 
this process is classified to be {\it sensitive} to probing $~T_{\min}~$.
Otherwise, we classify this process to be either 
{\it marginally sensitive} (for $~T_{\min}> B_0$ but 
$T_{\min}\not\gg B_0$~) or {\it insensitive} (for $~T_{\min}\leq B_0~$)
to testing $~T_{\min}~$. 
This classification is given at the level of the 
$S$-matrix elements. Up to the next-to-leading order, 
$~T_{\min}=T_1~$. For this case, 
by simply squaring both side of the condition (2.4)
and integrating over the phase space, 
we can easily derive the corresponding condition
(criterion) at the level of the constituent cross sections for 
$~\hat{\sigma}_{1}\simeq \int_{\rm phase} 2T_0T_{1}~$ and 
$~\hat{\sigma}_{B}\simeq \int_{\rm phase} 2T_0B_0~$, 
where $~\int_{\rm phase}~$ denotes the phase space integration. 
The constituent cross sections ($\hat{\sigma}$)
are functions of the invariant mass 
( $\sqrt{\hat{s}}$ ) of the final state weak bosons.
Defining the differential parton luminosity (for either 
the incoming light fermion or the weak boson) as
$~\frac{{\rm d} \,{\cal L}_{\rm partons}}{{\rm d} \,{\hat{s}}} ~$,
the total cross section is thus given by
$$  
\sigma  ~=~ \int {\rm d} \, {\hat{s}}
\frac{{\rm d} \, {\cal L}_{\rm partons}}{{\rm d} \, {\hat{s}}} 
\hat{\sigma}
({\hat{s}}) ~~.
\eqno(2.6)
$$ 
Using (2.6) we can further derive the corresponding 
conditions for total event rates $R_1$ 
(calculated from $\hat{\sigma}_1$) and
$R_B$ (calculated from $\hat{\sigma}_B$), 
and then define the corresponding criterion for 
testing the sensitivities of various 
operators and processes at the level of event rates.
They are:
(i). Sensitive, if $~R_1 \gg R_B~$; (ii). Marginally sensitive,
  if $~R_1 > R_B~$, but $~R_1\not\gg R_B~$; (iii). Insensitive,
  if $~R_1 \leq R_B~$. 
A specific application to the LHC physics is given  in Sec.~5. 
We note that, at the event rate level, the above criterion is 
{\it necessary} but not sufficient since the leading $B$-term, of the
same order as the LNI $V_L$-$V_T$ mixing effects [cf. (2.2)] and as
an intrinsic background to any strong $V_L$-$V_L$ scattering process,
denotes a universal part of the full backgrounds~\cite{mike-EWA}.
The sufficiency will of course require detailed numerical analyses
on the detection efficiency for suppressing the full backgrounds to 
observe the specific decay mode of the final state (as discussed in
Ref.~\cite{wwlhc}). This is beyond our present first step theoretical 
global study (see also footnote-$b$).

Before concluding this section, we note that
in the following power counting analysis
(cf. Secs.~3-5), 
{\it both the GB-amplitude and the $B$-term are explicitly estimated}
(cf. Tables 1-4 in Sec.4).
The issue of numerically
including/ignoring $B$ in an explicit calculation
is {\it essentially  irrelevant} here. If $~T_1 \leq B~$, this
means that the sensitivity is poor so that the probe of $T_1$ is
experimentally harder and requires a higher experimental 
precision of at least the order of $B$ to test $T_1$.

\vspace{0.5cm}
\noindent
{\bf 3.  Electroweak Chiral Lagrangian  and 
         A Generalized Precise Power Counting Rule }

\vspace{0.1cm}

In this section, we first define the EWCL which we investigate
in this paper and analyze the current constraints on the
next-to-leading order EWCL parameters (whose values are model-dependent
and reflect the underlying  dynamics). Then,
we generalize Weinberg's counting method~\cite{wei} and
develop a precise counting rule for the EWCL  
in the energy region $~~M_W, m_t\ll E < \Lambda~$,
where the effective cutoff $~\Lambda ~$  is the 
upper limit of $E$ at which the EWCL formalism 
ceases to be applicable. 
[The generalization of Weinberg's counting method to the linearly
realized effective Lagrangian (including the light Higgs SM)
is given in Appendix B.] 
As will be shown in Sec.~4 and 5, such a precise
electroweak power rule is particularly convenient and useful for correctly
estimating all high energy scattering amplitudes in performing a
global analysis. 

\vspace{0.2cm}
\noindent
{\bf 3.1. Electroweak Chiral Lagrangian and the Current Constraints }
       
The electroweak chiral Lagrangian (EWCL)  
gives the most economical description of the EWSB sector below the
scale of new heavy resonance and can be 
constructed as follows~\cite{app,peccei}:
$$
\begin{array}{ll}
{\cal L}_{\rm eff}& 
= \displaystyle\sum_n 
\ell_n\displaystyle\frac{f_\pi~^{r_n}}{\Lambda^{a_n}}
{\cal O}_n(W_{\mu\nu},B_{\mu\nu},D_\mu U,U,f,\bar{f})
= {\cal L}_{\rm G} + {\cal L}_{\rm S} + {\cal L}_{\rm F}
\end{array}
\eqno(3.1.1)                          %(3.1.1)
$$
where 
$$
\begin{array}{l}
D_{\mu}U =\partial_{\mu}U + ig{\bf W}_{\mu}U -ig^{\prime}U{\bf B}_{\mu}~,
\\[0.25cm]
U=\exp [i\tau^a\pi^a/f_\pi ]~,~~~
{\bf W}_{\mu}\equiv W^a_{\mu}\displaystyle\frac{\tau^a}{2}~,~~~
{\bf B}_{\mu}\equiv B_{\mu}\displaystyle\frac{\tau^3}{2}~.\\[0.20cm]
\end{array}
$$
$f$($\bar{f}$) is the SM fermion with mass $~m_{f}\leq 
O(m_t)\simeq O(M_W)~$.
$~{\cal L}_{\rm G}~$, $~{\cal L}_{\rm S}~$ and $~{\cal L}_{\rm F}~$ denote
 gauge boson kinetic terms, scalar boson interaction terms
(containing GB self-interactions and gauge-boson-GB interactions), 
and fermion interaction terms, respectively.
Here we concentrate on probing 
new physics from all possible bosonic operators so that we shall 
not include the next-to-leading order fermionic operators 
in $~{\cal L}_F~$. 
For clearness, we have factorized out the dimensionful parameters 
$~f_\pi~$ and $~\Lambda~$ in the coefficients so that the dimensionless 
factor $~~\ell_n\sim O(1)~$. This makes our definitions of the
$~\ell_n$'s different from the $~\alpha_i$'s in Ref.~\cite{app}
by a factor of $~(f_{\pi}/\Lambda )^2~$.
We note that  $~f_\pi~$  and $~\Lambda~$ 
are the two essential scales in any effective Lagrangian
that describes the spontaneously broken symmetry.  The former determines the 
symmetry breaking scale while the latter determines the scale at which
new resonance(s) besides the light fields 
(such as the SM
weak bosons, would-be Goldstone bosons and fermions) may appear. 
In the non-decoupling scenario, 
the effective cutoff scale $\Lambda$ cannot be arbitrarily large:
$~~\Lambda = \min (M_{SB},\Lambda_0 )\leq\Lambda_0~~$ \cite{georgi}, 
where $~~\Lambda_0\equiv 4\pi f_\pi\simeq 3.1$~TeV and
$~M_{SB}~$ is the mass of the 
lightest new resonance in the EWSB sector.
In (3.1.1), $~~ r_n=4+a_n-D_{{\cal O}_n} ~$, where $~D_{{\cal O}_n}
={\rm dim}({\cal O}_n)~$.  
The power factor $~\Lambda^{a_n}~$ associated with each operator
$~{\cal O}_n~$ can be counted by the naive dimensional analysis 
(NDA)~\cite{georgi}\footnote{In this paper, 
the NDA is only used to count the 
$~\Lambda$-powers  ($~\Lambda^{a_n}~$) associated with the 
operators $~{\cal O}_n~$'s in the chiral Lagrangian (3.1.1). 
This is irrelevant to the derivation of the
power counting rule for $~D_E~$ in the following (3.2.4).}.
For the bosonic part of EWCL, we have~\cite{app}:
$$
\begin{array}{ll}
{\cal L}_{\rm G} &
 =~ -\frac{1}{2}{\rm Tr}({\bf W}_{\mu\nu}{\bf W}^{\mu\nu})
              -\frac{1}{4}B_{\mu\nu}B^{\mu\nu}  ~~,\\[0.25cm]
{\cal L}_{\rm S} & 
 = {\cal L}^{(2)}+{\cal L}^{(2)\prime}+
             \displaystyle\sum_{n=1}^{14} {\cal L}_n ~~,\\[0.25cm]
{\cal L}^{(2)} & =
   \frac{f_\pi^2}{4}{\rm Tr}[(D_{\mu}U)^\dagger(D^{\mu}U)]   ~~,\\[0.20cm]
%\end{array}
%$$
%$$
%\begin{array}{ll}
{\cal L}^{(2)\prime} & =\ell_0 (\frac{f_\pi}{\Lambda})^2~\frac{f_\pi^2}{4}
               [ {\rm Tr}({\cal T}{\cal V}_{\mu})]^2 ~~,\\[0.2cm]
{\cal L}_1 & = \ell_1 (\frac{f_\pi}{\Lambda})^2~ \frac{gg^\prime}{2}
B_{\mu\nu} {\rm Tr}({\cal T}{\bf W^{\mu\nu}}) ~~,\\[0.2cm]
{\cal L}_2 & = \ell_2 (\frac{f_\pi}{\Lambda})^2 ~\frac{ig^{\prime}}{2}
B_{\mu\nu} {\rm Tr}({\cal T}[{\cal V}^\mu,{\cal V}^\nu ]) ~~,\\[0.2cm]
{\cal L}_3 & = \ell_3 (\frac{f_\pi}{\Lambda})^2 ~ig
{\rm Tr}({\bf W}_{\mu\nu}[{\cal V}^\mu,{\cal V}^{\nu} ]) ~~,\\[0.2cm]
{\cal L}_4 & = \ell_4 (\frac{f_\pi}{\Lambda})^2 
                     [{\rm Tr}({\cal V}_{\mu}{\cal V}_\nu )]^2 ~~,\\ [0.2cm] 
{\cal L}_5 & = \ell_5 (\frac{f_\pi}{\Lambda})^2 
                     [{\rm Tr}({\cal V}_{\mu}{\cal V}^\mu )]^2 ~~,\\  [0.2cm]
{\cal L}_6 & = \ell_6 (\frac{f_\pi}{\Lambda})^2 
[{\rm Tr}({\cal V}_{\mu}{\cal V}_\nu )]
{\rm Tr}({\cal T}{\cal V}^\mu){\rm Tr}({\cal T}{\cal V}^\nu) ~~,\\[0.2cm]
{\cal L}_7 & = \ell_7 (\frac{f_\pi}{\Lambda})^2 
[{\rm Tr}({\cal V}_\mu{\cal V}^\mu )]
{\rm Tr}({\cal T}{\cal V}_\nu){\rm Tr}({\cal T}{\cal V}^\nu) ~~,\\[0.2cm]
{\cal L}_8 & = \ell_8 (\frac{f_\pi}{\Lambda})^2~\frac{g^2}{4} 
[{\rm Tr}({\cal T}{\bf W}_{\mu\nu} )]^2  ~~,\\[0.2cm]
{\cal L}_9 & = \ell_9 (\frac{f_\pi}{\Lambda})^2 ~\frac{ig}{2}
{\rm Tr}({\cal T}{\bf W}_{\mu\nu}){\rm Tr}
        ({\cal T}[{\cal V}^\mu,{\cal V}^\nu ]) ~~,\\[0.2cm]
{\cal L}_{10} & = \ell_{10} (\frac{f_\pi}{\Lambda})^2\frac{1}{2}
[{\rm Tr}({\cal T}{\cal V}^\mu){\rm Tr}({\cal T}{\cal V}^{\nu})]^2 ~~,\\[0.2cm]
{\cal L}_{11} & = \ell_{11} (\frac{f_\pi}{\Lambda})^2 
~g\epsilon^{\mu\nu\rho\lambda}
{\rm Tr}({\cal T}{\cal V}_{\mu}){\rm Tr}
({\cal V}_\nu {\bf W}_{\rho\lambda}) ~~,\\[0.2cm]
{\cal L}_{12} & = \ell_{12}(\frac{f_\pi}{\Lambda})^2 ~2g
                    {\rm Tr}({\cal T}{\cal V}_{\mu}){\rm Tr}
                  ({\cal V}_\nu {\bf W}^{\mu\nu}) ~~,\\[0.2cm]
{\cal L}_{13} & = \ell_{13}(\frac{f_\pi}{\Lambda})^2~ 
      \frac{gg^\prime}{4}\epsilon^{\mu\nu\rho\lambda}
      B_{\mu\nu} {\rm Tr}({\cal T}{\bf W}_{\rho\lambda}) ~~,\\[0.2cm]     
{\cal L}_{14} & = \ell_{14} (\frac{f_\pi}{\Lambda})^2~\frac{g^2}{8} 
\epsilon^{\mu\nu\rho\lambda}{\rm Tr}({\cal T}{\bf W}_{\mu\nu})
{\rm Tr}({\cal T}{\bf W}_{\rho\lambda})   ~~,\\[0.2cm]
\end{array}
\eqno(3.1.2)                                         %(3.1.2)
$$

\noindent
where 
$~{\bf W}_{\mu\nu} =\partial_{\mu}{\bf W}_{\nu}-\partial_{\nu}{\bf W}_{\mu}
                      +ig [{\bf W}_\mu , {\bf W}_\nu ] ~$,
$~{\cal V}_{\mu}\equiv (D_{\mu}U)U^\dagger~$,  
and $~{\cal T}\equiv U\tau_3 U^{\dagger} ~$.
There is certain arbitrariness in choosing the complete set of operators 
which can be related to another set after applying the equation of motion,
but this will not affect the physical results~\cite{eom}.
Eq.~(3.1.2) contains fifteen bosonic 
NLO effective operators among which
there are twelve $CP$-conserving operators 
($~{\cal L}^{(2)\prime},{\cal L}_{1\sim 11}~$) and three $CP$-violating
operators (~${\cal L}_{12\sim 14}~$).
Furthermore, the operators $~{\cal L}_{6,7,10}~$ 
violate custodial $SU(2)_C$ symmetry 
(~even after $g^{\prime}$ being turned off~) contrary to
 $~{\cal L}_{4,5}~$ which contain $SU(2)_C$-invariant 
 pure GB interactions.
The coefficients ( $\ell_n$'s ) of all the above operators are 
model-dependent and carry 
information about possible new physics beyond the SM.
The dimension-$2$ custodial $SU(2)_C$-violating operator 
$~{\cal L}^{(2)\prime} ~$  has a coefficient at most of 
$~O(f_\pi^2/\Lambda^2 )~$ since it is  proportional to
$~\delta\rho \approx O(m_t^2/(16\pi^2 f_\pi^2))
\approx O(f_\pi^2 /\Lambda^2 )~$ for the top Yukawa coupling being 
of $~O(1)~$. 

In the non-decoupling scenario~\cite{georgi,review},
the coefficients for all 
NLO dimension-$4$ operators are suppressed 
by a factor $~(f_\pi/\Lambda )^2\approx 1/(16\pi^2)~$ 
relative to that of the universal dimension-$2$ operator ${\cal L}^{(2)}$,
because of the derivative expansion in terms of $~(D_\mu/\Lambda )^2~$.
If we ignore the small $CP$-violating effects from the 
Cabibbo-Kobayashi-Maskawa mixings in the lowest order 
fermionic Lagrangian $~{\cal L}_{\rm F}~$,
all the one-loop level new divergences generated from 
$~{\cal L}_{\rm G}+{\cal L}_{\rm F}+{\cal L}^{(2)}~$ are thus $CP$-invariant. 
Therefore, the $CP$-violating operators $~{\cal L}_{12\sim 14}~$ are 
actually {\it decoupled} at this level, and their
coefficients can have values significantly larger or smaller than that from
the naive dimensional analysis~\cite{georgi}.
Since the true mechanism for $CP$-violation remains un-revealed,  we
shall consider in this paper the coefficients $~\ell_{12\sim 14}~$
to be around of $~O(1)~$.

Before proceeding further, let us discuss the current
constraints on these EWCL parameters which will be useful for our 
study in Sec.~5. First, the coefficients $~\ell_{0,1,8}~$ are 
related to the low energy $S,T,U$ parameters~\cite{P-T} through the
oblique corrections:
$$
\begin{array}{l}
\ell_0~ =\displaystyle\left(\frac{\Lambda}
          {\Lambda_0}\right)^2 8\pi^2\alpha T 
        =\left(\frac{\Lambda}{\Lambda_0}\right)^2 8\pi^2 \delta\rho 
        =\displaystyle\left(\frac{\Lambda}{\Lambda_0}\right)^2 
         \frac{32\pi^3\alpha}{c_{\rm w}^2s_{\rm w}^2M_Z^2}
         [\Pi_{11}^{\rm new}(0)-\Pi_{33}^{\rm new}(0)] ~~,   \\[0.35cm]
\ell_1~ = -\displaystyle\left(\frac{\Lambda}{\Lambda_0}\right)^2 \pi S 
        = \displaystyle\left(\frac{\Lambda}{\Lambda_0}\right)^2 
          8\pi^2 \Pi_{3Y}^{{\rm new}~\prime}(0) ~~,\\[0.3cm]
\ell_8~ = -\displaystyle\left(\frac{\Lambda}{\Lambda_0}\right)^2 \pi U 
        = \displaystyle\left(\frac{\Lambda}{\Lambda_0}\right)^2 
          16\pi^2[\Pi_{33}^{{\rm new}~\prime}(0)
          -\Pi_{11}^{{\rm new}~\prime}(0)] ~~,
\end{array}
\eqno(3.1.3)                        %(3.1.3)
$$
where $~\Lambda_0\equiv 4\pi f_\pi\simeq 3.1$~TeV, and 
$~s_{\rm w}=\sin\theta_{\rm W},~c_{\rm w}=\cos\theta_{\rm W},
~\alpha =\frac{e^2}{4\pi}~$ are measured at $Z$-pole 
in the $\overline{\rm MS}$
scheme. The factor $~\left(\frac{\Lambda}{\Lambda_0}\right)^2~$ in (3.1.3)
reduces to one for the case  $~\Lambda =\Lambda_0~$.
The updated global fit to the low energy data gives \cite{PDG}:
$$
\begin{array}{lll}
S & = & -0.36 \pm 0.19  ~~,\\
T & = & -0.03\pm  0.26   ~~,\\
U & = & -0.31\pm 0.54    ~~,
\end{array}                         %(3.1.4)
\eqno(3.1.4)
$$ 
for $~s^2_{\rm w}=0.2311\pm 0.0003$ and $m_t=181\pm 12~$GeV. 
For the present analysis, we have specified the reference value of 
the SM Higgs mass as $~m_H=1$~TeV. 
Results for other values of $m_H$ can be found in Ref.~\cite{PDG}.
Since the experimental errors in (3.1.4) are still quite large, the
parameters ~$T$~ and ~$U$~ ($~\ell_0$ and $~\ell_8~$) can 
be either positive or negative within $1\sigma$ error and 
the parameter $~S~$ ($~\ell_1$~) 
is only about $~-1.89\sigma~$ below the SM value.\footnote{
This is at the same deviation level as the present $R_b$ anomaly 
which is $~+1.75\sigma~$ above the SM value~\cite{warsaw}.}

From (3.1.3) and (3.1.4), for $\Lambda\simeq 3.1$~TeV,
we get the following $1\sigma$ level constraints on $~\ell_{0,1,8}~$
at the scale $~\mu =M_Z$:~\footnote{The $2\sigma$ bounds at
$\mu =M_Z$~ are:  $~~-0.34 \leq \ell_0 \leq 0.30~$,~
$~-0.063 \leq \ell_1 \leq 2.33~$,~ $~-2.42 \leq \ell_8 \leq 4.37~$,
which allow ~$\ell_{0,1,8}$~ to be either positive or negative at 
$~O(1)~$ or larger.}
$$
\begin{array}{rllll}
-0.18 & \leq & \ell_0  & \leq & 0.14  ~~,\\
 0.53 & \leq & \ell_1  & \leq & 1.73  ~~, \\
-0.72 & \leq & \ell_8  & \leq & 2.67  ~~.
\end{array}
\eqno(3.1.5)                        %(3.1.5)
$$
 We can further
deduce the bounds at the TeV scale ({\it e.g.}, $\mu =1$~TeV)
by incorporating the running effects from the 
renormalization $\log$-terms. By the one-loop calculation~\cite{long}, 
we find, for instance, the coefficients
$~\ell_{0,1,8}~$ are renormalized as:
$$
\begin{array}{l}
\ell_0 (\mu )=\ell_0^{b} - 
              \displaystyle\left(\frac{\Lambda}{\Lambda_0}\right)^2 
              \frac{3}{4}(g^{\prime})^2
              \left(\frac{1}{\hat\epsilon} + c_0 \right) ~~,\\[0.3cm] 
\ell_1 (\mu )=\ell_1^{b} - 
              \displaystyle\left(\frac{\Lambda}{\Lambda_0}\right)^2 
              \frac{1}{6}
              \left(\frac{1}{\hat\epsilon} + c_1\right) ~~,\\[0.3cm] 
\ell_8 (\mu )=\ell_8^{b} - 
             \displaystyle\left(\frac{\Lambda}{\Lambda_0}\right)^2 c_8 ~~,
\end{array}
\eqno(3.1.6)                            %(3.1.6)
$$
where 
$~~\displaystyle
\frac{1}{\hat\epsilon}\equiv\frac{1}{4-n}- \ln\mu
~~$  and the superscript ``$~^{b}~$''  denotes the bare quantity. 
In (3.1.6), the $c_i~$'s are finite constants
whcih depend on the subtraction scheme and are irrelevant to the running
of $\ell_n(\mu )~$'s. From (3.1.5) and (3.1.6) we can deduce the
$1\sigma$ bounds at other scales, {\it e.g.}, at $\mu =1$~TeV 
for $\Lambda=\Lambda_0=3.1~$TeV,
$$
\begin{array}{rllll}
0.045 & \leq & \ell_0(1~{\rm TeV}) & \leq & 0.37  ~~,\\
 0.93 & \leq & \ell_1(1~{\rm TeV}) & \leq & 2.13  ~~, \\
-0.72 & \leq & \ell_8(1~{\rm TeV}) & \leq & 2.67  ~~,
\end{array}
\eqno(3.1.7)                        %(3.1.7)
$$
where the ranges for ~$\ell_{0,1}$~ are slightly moved toward positive
direction due to the running effects. 
(3.1.7) shows that $~\ell_{0,1,8}~$  are 
allowed to be around of $O(1)$ except that the
parameter space for $~\ell_0$~ is
about a factor of ~$5\sim 10$~ smaller than the others.
All those NLO coefficients $\ell_n$'s in (3.1.2) 
varies for different underlying theories and thus must be
{\it independently tested} since the real underlying theory is unknown
and these operators are inequivalent by the equation of motion. 
For example, the $SU(2)_C$-violating operator
${\cal L}^{(2)\prime}$ (containing two ${\cal T}=U\tau_3U^\dagger$~'s)
is constrained to be significantly below $O(1)$ within $1\sigma$ bound
[cf. (3.1.7)], while the updated data still allow 
the coefficients of other $SU(2)_C$-violating
operators such as ${\cal L}_1$ (containing one ${\cal T}$ operator) and  
${\cal L}_8$ (containing two ${\cal T}$ operators) 
to be around of $O(1)$ or even larger [cf. (3.1.5) and (3.1.7)].

Besides the bounds from the oblique corrections,
the tests on triple gauge boson couplings (TGCs) 
at LEP and Tevatron~\cite{TGC} impose further constraints on more operators 
at the tree level.  
In the conventional notation~\cite{TGC,TGC1}, the TGCs are parameterized as
$$
\begin{array}{ll}
\displaystyle\frac{{\cal L}_{WWV}}{g_{WWV}}= &
ig^V_1 [W^+_{\mu\nu}W^{-\mu}V^\nu - W^{-}_{\mu\nu}W^+_{\mu}V^\nu ]
+i\kappa_V W^+_\mu W^-_\nu V^{\mu\nu} + \frac{i\lambda_V}{\Lambda^2}
W^+_{\mu\nu}W^{-\nu}_{~~\rho}V^{\rho\mu}\\
& -g_4^V W^+_\mu W^-_\nu
  [\partial^\mu V^\nu + \partial^\nu V^\mu ]
 + g_5^V\epsilon^{\mu\nu\rho\lambda}[W^+_\mu\partial_\rho W^-_\nu
                              -W^-_\nu\partial_\rho W^+_\mu ]V_\lambda\\
& + i\tilde{\kappa}_V W^+_\mu W^-_\nu \tilde{V}^{\mu\nu}
+ i\frac{\tilde{\lambda}_V}{\Lambda^2}W^+_{\mu\nu}W^{-\nu}_{~~\rho}
  \tilde{V}^{\rho\mu} ~~,\\[0.4cm]
& V= Z, ~{\rm or }~ \gamma~,~~
  W^{\pm}_{\mu\nu}=\partial_{\mu}W^\pm_\nu-\partial_\nu W^\pm_\mu~,~~
 \tilde{V}_{\mu\nu} =\frac{1}{2}\epsilon_{\mu\nu\rho\lambda}V^{\rho\lambda}~.
\end{array}
\eqno(3.1.8)                  %(3.1.8)
$$
We summarize the tree level contributions from all $15$ NLO
order operators listed in (3.1.2) to the TGCs defined in (3.1.8) as follows:
$$
\begin{array}{l}
g_1^Z-1 \equiv \Delta g_1^Z = \displaystyle\frac{f^2_\pi}{\Lambda^2}
        \left[\frac{1}{c^2-s^2}\ell_0 +\frac{e^2}{c^2(c^2-s^2)}\ell_1
                                 +\frac{e^2}{s^2c^2}\ell_3 \right] ~~,~~~
g_1^\gamma -1 =0 ~~,\\[0.4cm]
g_4^Z=-\displaystyle\frac{f^2_\pi}{\Lambda^2}
\frac{e^2}{{\rm s}^2_{\rm w}{\rm c}^2_{\rm w}}\ell_{12}~~,~~~
g_4^\gamma =0 ~~,\\[0.4cm]
g_5^Z =\displaystyle\frac{f^2_\pi}{\Lambda^2}\frac{e^2}{s^2c^2}\ell_{11} ~~,
~~~g_5^\gamma =0 ~~,\\[0.4cm]
\kappa_Z -1\equiv\Delta\kappa_Z = \displaystyle\frac{f^2_\pi}{\Lambda^2}
\left[\frac{1}{c^2-s^2}\ell_0 +
 \frac{2e^2}{c^2-s^2}\ell_1 -\frac{e^2}{c^2}\ell_2
 +\frac{e^2}{s^2}(\ell_3 -\ell_8 +\ell_9 ) \right]~~,\\[0.5cm]
\kappa_\gamma -1 \equiv \Delta\kappa_\gamma
 =\displaystyle\frac{f^2_\pi}{\Lambda^2}\frac{e^2}{s^2}
    \left[ -\ell_1 +\ell_2 +\ell_3 -\ell_8 +\ell_9 \right]~~,\\[0.5cm]
\tilde{\kappa}_Z = \displaystyle\frac{f^2_\pi}{\Lambda^2}
\left[\frac{e^2}{{\rm c}_{\rm w}^2}\ell_{13}-
      \frac{e^2}{{\rm s}_{\rm w}^2}\ell_{14}\right]~~,~~~
\tilde{\kappa}_\gamma = -\displaystyle\frac{f^2_\pi}{\Lambda^2}
    \frac{e^2}{{\rm s}_{\rm w}^2}[\ell_{13}+\ell_{14}] ~~,
\end{array}
\eqno(3.1.9)                     %(3.1.9)
$$
which coincide with Ref.~\cite{app2} after taking into account the
difference in defining the coefficients.
(3.1.9) shows that, at the first non-trivial order 
[i.e., $O\left(\frac{f_\pi^2}{\Lambda^2}\right)$ ],  
only operators $~{\cal L}^{(2)\prime}~$ and
$~{\cal L}_{1,2,3,8,9,11\sim 14}~$ 
can contribute to anomalous triple gauge
couplings while $~{\cal L}^{(4)}_{4,5,6,7,10}~$ do not. 
Among $~{\cal L}^{(2)\prime}~$ and
$~{\cal L}_{1,2,3,8,9,11\sim 14}~$, $~{\cal L}^{(2)\prime}~$ and
$~{\cal L}_{1,8}~$ can be constrained by the oblique corrections
[cf. (3.1.5) and (3.1.7)] so that we are left with seven operators
$~{\cal L}^{(4)}_{2,3,9,11,12,13,14}~$. There are just seven independent
relations in (3.1.9) by which the coefficients
$~\ell_{2,3,9,11,12,13,14}~$ can be independently determined in principle
if the seven TGC parameters $~g^Z_{1,4,5}~$ and $~\kappa^{Z,\gamma},~
\tilde{\kappa}^{Z,\gamma}~$ can all be measured. 
From (3.1.9), we derive
$$
\begin{array}{l}
\ell_2 = \displaystyle\frac{1}{e^2}
 \frac{s^2_{\rm w}c^2_{\rm w}}{c^2_{\rm w}-s^2_{\rm w}}\ell_0
 +\frac{c^2_{\rm w}}{c^2_{\rm w}-s^2_{\rm w}}\ell_1 +
 \displaystyle\frac{\Lambda^2}{f_\pi^2}
 \frac{s^2_{\rm w}c^2_{\rm w}}{e^2}
 (\Delta\kappa_\gamma -\Delta\kappa_Z) ~~,\\[0.4cm]
\ell_3 = -\displaystyle\frac{1}{e^2}
 \frac{s^2_{\rm w}c^2_{\rm w}}{c^2_{\rm w}-s^2_{\rm w}}\ell_0
 -\frac{c^2_{\rm w}}{c^2_{\rm w}-s^2_{\rm w}}\ell_1 +
 \displaystyle\frac{\Lambda^2}{f_\pi^2}
 \frac{s^2_{\rm w}c^2_{\rm w}}{e^2}\Delta g^Z_1 ~~,\\[0.4cm]
\ell_9 =\displaystyle\ell_8 + \frac{\Lambda^2}{f_\pi^2}
  \frac{s^2_{\rm w}c^2_{\rm w}}{e^2}\left(
 \Delta\kappa_Z+\frac{s^2_{\rm w}}{c^2_{\rm w}}\Delta\kappa_\gamma
 -\Delta g_1^Z\right) ~~,\\[0.4cm]
\ell_{11}=\displaystyle\frac{\Lambda^2}{f_\pi^2} 
 \frac{s^2_{\rm w}c^2_{\rm w}}{e^2}g^Z_5 ~~,\\[0.4cm]
\ell_{12}=-\displaystyle\frac{\Lambda^2}{f_\pi^2} 
 \frac{s^2_{\rm w}c^2_{\rm w}}{e^2}g^Z_4 ~~,\\[0.4cm]
 \ell_{13}=\displaystyle\frac{\Lambda^2}{f_\pi^2}
 \frac{s^2_{\rm w}c^2_{\rm w}}{e^2}
(\tilde{\kappa}_Z-\tilde{\kappa}_\gamma ) ~~,\\[0.4cm]
 \ell_{14}=-\displaystyle\frac{\Lambda^2}{f_\pi^2}
 \frac{s^2_{\rm w}c^2_{\rm w}}{e^2}
\left(\tilde{\kappa}_Z+\frac{s^2_{\rm w}}{c^2_{\rm w}}
      \tilde{\kappa}_\gamma \right)     ~~.
\end{array}
\eqno(3.1.10)                %(3.1.10)
$$
Inputting the experimentally measured $~S,~T,~U~$ and 
$~g^{Z,\gamma}_{1,4,5}~$, $\kappa_{Z,\gamma}$~ and
$~\tilde{\kappa}_{Z,\gamma}~$, we can derive constraints 
on all $\ell_n$~'s from (3.1.3) and (3.1.10).
For example, a recent global fit at LEP~\cite{TGC} gives the following 
$1\sigma$ (i.e. $68.27\%$ confidence level) bounds, for allowing 
only one TGC to be nonzero each time:
$$
\begin{array}{c}
-0.064 ~\leq ~ \Delta g^Z_1 ~\leq   ~-0.002 ~~,~~~
-0.070 ~\leq ~ \lambda_\gamma ~\leq ~-0.002 ~~,~~~
0.004 ~\leq ~\lambda_Z ~\leq ~ 0.094 ~~,\\
-0.046 ~\leq ~\Delta\kappa_Z ~\leq ~0.042 ~~,~~~
0 ~\leq ~ \Delta\kappa_\gamma ~\leq ~0.112  ~~.
\end{array}
\eqno(3.1.11)                    %(3.1.11)
$$
From the above result
 we can estimate the constraints on $~\ell_{2,3,9}~$ as
$$
\begin{array}{lllll}
-12.1\left(\frac{\Lambda}{\Lambda_0}\right)^2 & \leq & \ell_2 &
 \leq & 32.3\left(\frac{\Lambda}{\Lambda_0}\right)^2   ~~,\\[0.30cm]
-18.5\left(\frac{\Lambda}{\Lambda_0}\right)^2 & \leq & \ell_3 &
 \leq & 0.61\left(\frac{\Lambda}{\Lambda_0}\right)^2   ~~,\\[0.30cm]
-13.3\left(\frac{\Lambda}{\Lambda_0}\right)^2 & \leq & \ell_9 &
 \leq & 18.5\left(\frac{\Lambda}{\Lambda_0}\right)^2   ~~,
\end{array}
\eqno(3.1.12)                    %(3.1.12)
$$
where $~\Lambda \leq \Lambda_0 \equiv 4\pi f_\pi \simeq 3.1$~TeV.

At the FermiLab Tevatron, the TGCs can be directly measured at the tree
level instead of at the loop level. 
For instance, the CDF group gives,
at the $95\%$ confidence level (C.L.)~\cite{TGC}, 
$$
\begin{array}{cl}
-1.1~<~ \Delta\kappa_V ~<~1.3 ~~, &
~~~(~{\rm for}~\lambda_V=\Delta g^V_1 =0~)~,\\
-0.8~<~ ~~\lambda_V~      ~<~0.8 ~~, &
~~~(~{\rm for}~\Delta\kappa_V=\Delta g^V_1 =0~)~,\\
-1.2~<~ \Delta g^Z_1   ~<~1.2 ~~, &
~~~(~{\rm for}~\lambda_V=\Delta\kappa_V =0~)~,\\
\end{array}
\eqno(3.1.13)        %(3.1.13)
$$
where $~\Delta\kappa_\gamma =\Delta\kappa_Z~
~{\rm and} ~ \lambda_\gamma =\lambda_Z~$
are assumed. Thus we can estimate the $95\%$ C.L.
constraints on $~\ell_{3,9}~$  as
$$
\begin{array}{cc}
-346\left(\frac{\Lambda}{\Lambda_0}\right)^2 ~<~ \ell_3 ~<~
 346\left(\frac{\Lambda}{\Lambda_0}\right)^2 ~~,~~~~ &
-412\left(\frac{\Lambda}{\Lambda_0}\right)^2 ~<~ \ell_9 ~<~
 488\left(\frac{\Lambda}{\Lambda_0}\right)^2 ~~,
\end{array}
\eqno(3.1.14)         %(3.1.14)
$$
which gives, for $~\Lambda =2$~TeV~,
$$
\begin{array}{cc}
-145 ~<~ \ell_3 ~<~ 145 ~~,~~~ &
-173 ~<~ \ell_9 ~<~ 204 ~~,~~~(~\Lambda =2~{\rm TeV}~)~~.
\end{array}
\eqno(3.1.14a)         %(3.1.14a)
$$
As shown above, the indirect $1\sigma$ bounds from LEP/SLC allow  
$~\ell_{2,3,9}$  to be around of $O(10)$, 
and the direct $95\%$ C.L. bounds from 
Tevatron on $~\ell_{3,9}~$ are also too weak 
to be useful in discriminating different dynamical models
whose effects to these coefficients $\ell_n$~'s are theoretically 
expected to be of ~$O(1)$~\cite{georgi}. 

Since the operators $~{\cal L}^{(4)}_{4,5,6,7,10}~$
contain only quartic vertices, they cannot be constrained at tree level
by any low energy data. The current experiments can
only constrain these operators at one-loop level 
[i.e., of $~O(1/\Lambda^4)$].
By calculating the one-loop logarithmic contributions 
(with all the constant terms ignored) to the low energy data
from these operators, one can roughly estimate the indirect
experimental bounds on their coefficients~\cite{dv,ebo}.
Since the ignored constant terms are of the same order of magnitude
as the logarithmic contributions, we should keep in mind that 
some uncertainties (like a factor of $2$ or so) may naturally exist
in these estimated bounds.
It was found in Ref.~\cite{dv} that, at the $90\%$ C.L., 
the LEP data constraints
(allowing only one non-zero coefficient at a time) are
$$
\begin{array}{ll}
-11 ~<~ \ell_4 ~<~ 11 ~~,~~ &  -28 ~<~ \ell_5 ~<~ 26 ~~,
\end{array}
\eqno(3.1.15)               %(3.1.15)
$$
for the cut-off scale $~\Lambda = 2$~TeV. In another recent 
study~\cite{ebo}, for $~\Lambda = 2$~TeV and $m_t=170$~GeV, 
the following LEP constraints are derived at the $90\%$ C.L.:
$$
\begin{array}{c}
-3.97 ~\leq ~ \ell_4 ~\leq ~ 19.83 ~~,~~~
-9.91 ~\leq ~ \ell_5 ~\leq ~ 50.23 ~~,\\
-0.66 ~\leq ~ \ell_6 ~\leq ~ 3.50 ~~,~~~ 
-5.09 ~\leq ~ \ell_7 ~\leq ~ 25.78 ~~,~~~
-0.67 ~\leq ~ \ell_{10} ~\leq ~ 3.44 ~~.
\end{array}
\eqno(3.1.16)               %(3.1.16)
$$
The above results show that the low energy bounds on
the $SU(2)_C$-violating operators ${\cal L}_{6,10}$
are close to their theoretical expectation for $~\ell_n\sim O(1)~$,
which are stronger than that for ${\cal L}_{7}$ and
the $SU(2)_C$-conserving operators ${\cal L}_{4,5}$ 
(when turning off the $U(1)_Y$ gauge coupling).
Thus, ${\cal L}_{6,10}$ are more sensitive to the low energy data.
But these numerical values should not be taken too seriously
(except as a useful guideline) since 
all non-logarithmic contributions are ignored in the above estimates
and the correlations among different operators are not considered
for simplicity. We also note that the sensitivities of the 
$SU(2)_C$-violating operators to the low energy data
do {\it not} have a naive power-like dependence
on the number of $~{\cal T}$-operators. (3.1.16) shows that
the operators $~{\cal L}_{6,10}~$ (containing two and four 
$~{\cal T}$-operators, respectively) have quite similar sensitivities
to the low energy data and their bounds are much stronger than that for
$~{\cal L}_{7}~$ (containing two $~{\cal T}$~'s). 
When further looking at the LEP $68.27\%$~C.L. bounds (3.1.12)
for the triple gauge boson couplings (TGCs) from the $SU(2)_C$-violating 
operators $~{\cal L}_{2,9}~$, 
we find that they are similar and are both much weaker than the
$90\%$~C.L. bounds for quartic couplings from
$~{\cal L}_{6,10}~$ in (3.1.16) despite $~{\cal L}_{2,9;6,10}~$ containing
one, two, two and four $~{\cal T}$-operators, respectively.
Intuitively, it would be natural to expect that
the $SU(2)_C$-violating operators may get stronger bounds than the 
$SU(2)_C$-conserving ones [as implied in (3.1.16) for the quartic couplings  
of $~{\cal L}_{6,10}~$] when we consider the $SU(2)_C$
as a good approximate symmetry. But the real situation
is more involved. From the LEP bounds (3.1.12) and the Tevatron bounds
(3.1.14,14a), we see that, for TGCs, the $SU(2)_C$-violating operators 
$~{\cal L}_{2,9}~$ have weaker bounds than that of 
the $SU(2)_C$-conserving operator $~{\cal L}_3~$. 
Ref.~\cite{dv} also estimated the $90\%$~C.L. LEP bounds 
for $~{\cal L}_{2,3}~$ as $~~-47 < \ell_2 < 39~~$ 
and $~~-8 < \ell_3 < 11 ~~$ which impose stronger constraint on 
$~\ell_3~$.
[The relations $~L_{9R}=-2\ell_2~,~~L_{9L}=-2\ell_3~,~~L_5=\ell_4~,~~
L_4=\ell_5~$ have been used to translate the eq.~(39) of Ref.~\cite{dv} 
into our notations.]

In summary, the results in (3.1.12), (3.1.14,14a), (3.1.15) and (3.1.16) 
indicate that the current bounds on
~$\ell_{2,3,9}$~ and ~$\ell_{4,5,7}$~ are still too weak.
Concerning the $SU(2)_C$-violating operators, the bound on $~\ell_0~$
($T$) is most stringent while that on $~\ell_{1,8,6,10}~$ are all
around $~O(1)~$ (or larger) as shown in (3.1.7) and (3.1.16).
However, the updated constraints on other $SU(2)_C$-violating operators
$~{\cal L}_{2,9,7}~$ can be of $~O(10)~$ or larger, implying that
the current low energy tests do not well probe the 
$SU(2)_C$-violation effects from these operators.
Although LEPII and the upgraded Tevatron are expected to 
improve the current bounds somewhat, to further improve 
the precision on $\ell_n$~'s and to fully probe the EWSB sector require
finding the most sensitive
high energy scattering processes to {\it independently test}
{\it all} those coefficients in (3.1.2) at the LHC and the future 
linear colliders (LC).

\vspace{0.2cm}
\noindent
{\bf 3.2.  A Generalized Precise Electroweak Power Counting Rule}

We want to {\it separately} count the power dependences of the amplitudes
on the energy $E$, the cutoff scale 
$\Lambda$ of the EWCL and the Fermi scale (vacuum expectation value)
$f_\pi = 246$\,GeV ($\sim M_W, m_t$).\footnote{This is 
essentially different from the previous counting result in the literature
for the heavy Higgs SM ~\cite{hveltman} where only the {\it sum} 
of the powers of $E$ and $m_H$ has been counted.}~~
This is {\it crucial} for correctly 
estimating the order of magnitude of an amplitude
at any given order of perturbative calculation.
For instance, an amplitude of order 
$~\frac{E^2}{f_\pi^2}~$ differs by two orders
of magnitude from an amplitude of order $~\frac{E^2}{\Lambda^2}~$ 
in spite that they have the same
$E$-dependence. Also,  the amplitudes $~\frac{E^2}{f_\pi^2}$~ and 
~$\frac{E^2}{f_\pi^2}\frac{E^2}{\Lambda^2}$~ have the {\it same} sum
for the $E$ and $\Lambda$ powers,
but are clearly of different orders in magnitude. E.g.,
in the typical case $E=1$~TeV and $\Lambda\simeq 4\pi f_\pi\simeq 3.1~$TeV, 
they differ by a large factor $\sim 10$~.
Since the weak-boson mass ~$M_W=g f_\pi/2$~ and the fermion mass 
$~m_f=y_f f_\pi/\sqrt{2}~$, we can 
count them through powers of the coupling constants $g$ and $y_f$ 
and the vacuum expectation value $f_\pi$.
The $SU(2)$ weak gauge
coupling $g$ and the top quark Yukawa coupling $y_t$ are around of $O(1)$
and thus will not significantly affect the order of magnitude estimates.
The electromagnetic $U(1)_{\rm em}$ coupling 
$~e=g\sin\theta_W~$ is smaller than $g$ by
about a factor of $2$. The Yukawa couplings of all
light SM fermions other than the top quark are negligibly small.
In our following precise counting rule, the dependences on coupling
constants $~g,~g^{\prime} ({\rm or}~ e)~$ and $~y_t~$ are included, 
while all the light
fermion Yukawa couplings [ $y_f~(\neq y_t) \ll 1$ ] are ignored.

 Weinberg's power counting method  
was derived for only counting the energy dependence
in the un-gauged nonlinear $\sigma$-model as a description of low energy
QCD interaction~\cite{wei}. 
But some of its essential features are very general:~
{\bf (i).} The total dimension $~D_T~$ of 
an $S$-matrix element $~T~$ is determined 
by the number of external lines and the space-time dimension; 
{\bf (ii).}   Assume that all mass poles 
in the internal propagators of $~T~$ are much smaller
than the typical energy scale $E$ of $~T~$, then the total
dimension $~D_m~$ of the $E$-independent coupling constants included in 
$~T~$ can be directly counted 
according to the type of vertices contained. 
Hence, the total $E$-power $~D_E~$ for 
$~T~$ is given by $~~D_E= D_T -D_m~~$. 

Here, we shall make a natural generalization of 
Weinberg's power counting method 
for the EWCL in which, except the light SM
gauge bosons, fermions and would-be GB's, all possible heavy fields have
been integrated out.
It is clear that in this case the above conditions 
{\bf (i)} and {\bf (ii)} are satisfied. 
The total dimension of an $L$-loop $S$-matrix element $T$ is
$$
D_T = 4-e~~,
\eqno(3.2.1)                                          %(3.2.1)
$$
where $~~e=e_B+e_F~~$, and $e_B$ ($e_F$) is
 the number of external bosonic  
(fermionic) lines. Here the dimensions of the external spinor wave 
functions are already included in $D_T$. 
For external fermionic lines, we only 
count the SM fermions with masses 
$~~m_f\leq m_t \sim O(M_W)\ll E ~$. So the 
spinor wave function of each external fermion 
will contribute an energy factor
$~E^{1/2}~$ for $~~E\gg m_f~$, where the spinor wave functions are 
normalized as 
$~~ \bar{u}(p,s)u(p,s^\prime )=2m_f\delta_{ss^\prime}~$, etc. 

Let us label the different types of vertices by an index $n$. If
the vertex of type $n$ contains $b_n$ bosonic lines, 
$f_n$ fermionic lines and $d_n$ derivatives, 
then the dimension of the $~E$-independent 
effective coupling constant in $T$ is 
$$
D_m= \displaystyle\sum_n {\cal V}_n\left(4-d_n-b_n-\frac{3}{2}f_n\right) ~~,
\eqno(3.2.2)                              %(3.2.2)
$$
where ${\cal V}_n$ is the number of vertices of type $n$. 
Let $i_B$ and $i_F$ be the numbers of 
internal bosonic and fermionic lines, 
respectively. ( $i_B$ also includes possible internal ghost lines.)
Define  $~~i=i_B+i_F~~$,
we have, in addition, the following general relations
$$
\begin{array}{l}
\displaystyle\sum_n b_n{\cal V}_n =2i_B+e_B~~,~~~~
\displaystyle\sum_n f_n{\cal V}_n =2i_F+e_F ~~,~~~~
L=1+i-\displaystyle\sum_n {\cal V}_n ~~.
\end{array}
\eqno(3.2.3)
$$                                          %(3.2.3) 
These can further simplify the terms in (3.2.2).

Note that external vector-boson lines may cause extra contributions to the 
power of $~E~$ in $~D_E~$ due to the $E$-dependence of their polarization 
vectors since each longitudinal polarization vector $\epsilon_L^\mu$ 
is of $O(E/M_W)$ for $~E\gg M_W~$.
Thus, if we simply count all external $V_L$-lines directly,
the relation between $D_E,D_T$ and $D_m$ will become 
$~~~D_E=D_T-D_m+e_L-e_v~~~$, where $e_L$ 
and  $e_v$ denote the numbers of external 
$V_L$ and $v^a$ lines, respectively. [As shown in (2.1c), each external
$v^a$-line is a gauge-line $~V_\mu^a~$ suppressed by the factor
$~v^\mu = O(M_W/E)~$.]
However, when this relation is applied to the $V_L$-amplitudes with $D_m$
given in (3.2.2), it does not lead to the correct results. To see 
this, let us take the $V_LV_L \rightarrow V_LV_L$ scattering amplitude as an 
example, in which $e_L=4$ and $e_v=e_F=0$. 
To lowest order of the EWCL, the leading
powers of $E$ in the amplitudes $T[V^{a_1}_L, \cdots ,V^{a_4}_L]$,
$T[\pi^{a_1}, \cdots ,\pi^{a_4}]$ and the $B$-term [cf. (2.1)]
are $E^4$, $E^2$ and $E^0$, respectively. 
This is not consistent with the prediction of the ET (2.3). 
The reason for this inconsistency is that this naive power counting for the
$V_L$-amplitude only gives the leading $E$-power for individual Feynman
diagrams. It does not reflect the fact that gauge invariance
causes the cancellations of the $E^4$-terms between different diagrams, and 
leads to the final $E^2$-dependence of the whole $V_L$-amplitude.
Thus directly counting the external $V_L$-lines in the $V_L$-amplitudes for
$~D_E~$ does not give the correct answer. This problem can be elegantly
solved by implementing the ET identity (2.1). We see that the power counting 
of the GB-amplitude plus the $B$-term 
does give the correct $E$-dependence because, unlike in the $V_L$-amplitude, 
there is generally no large $E$-power cancellations in 
the GB-amplitudes and the $B$-term. 
Therefore based upon the ET identity (2.1),
the correct counting of the powers of $E$ for the $V_L$-amplitude
can be given by counting the corresponding GB-amplitude plus 
the $B$-term. Thus, in the following generalized power 
counting rule, we do not directly count 
the the external $V_L$-lines in a given diagram. Instead, they will be counted 
through counting the RHS of the ET identity (2.1). 
We shall therefore drop the $e_L$ term in the above relation between $D_E,D_T$
and $D_m$, and {\it make the convention that the number of external
vector-boson lines $e_V$ counts only the number of external $V_T$-lines and 
photon lines}. Then from (3.2.1), (3.2.2) and (3.2.3), the feasible formula 
for the leading energy power in $T$ is
$$
D_E  ~= D_T-D_m-e_v
     ~= 2L +2 + \sum_n {\cal V}_n
      \left(d_n +\frac{1}{2}f_n-2\right)- e_v ~~.
\eqno(3.2.4)                                                     %(3.2.4)
$$      
This is just the Weinberg's counting rule~\cite{wei} 
in its generalized form
with the gauge boson, 
ghost and fermion fields and possible $v_\mu$-factors 
included. (3.2.4) is clearly valid for
any gauge theory satisfying the above conditions 
{\bf (i)} and {\bf (ii)}.

To correctly estimate the magnitude of each given amplitude $T$,
besides counting the power of $E$, it is crucial to also {\it
separately} count the power dependences on the two typical mass scales
of the EWCL: the vacuum expectation value $f_\pi$ and the effective  
cutoff scale $\Lambda$. If the powers of $f_\pi$ and $\Lambda$ are not
separately counted, $\Lambda/f_\pi \simeq 4\pi >12$ will be mistakenly
counted as 1. This can make the estimated results off by orders of 
magnitudes.

Consider the $S$-matrix element $T$ at the $L$-loop order.
Since we are dealing with a spontaneously broken gauge theory which has
 a nonvanishing vacuum expectation value $f_\pi$,  
 $T$ can always be written as $~f_{\pi}^{D_T}~$ 
times some dimensionless function
of $~E,~\Lambda$, and $f_\pi~$, etc.  
The $E$-power dependence has been given by our generalized Weinberg
formula (3.2.4). 
We now  count the power of $\Lambda$.
The $~\Lambda$-dependence in $T$ can only come from two sources:
\begin{description}
\item{(i).} From tree vertices:
$T$ contains $~{\cal V}=\displaystyle\sum_n {\cal V}_n~$ vertices, 
each of which contributes a factor
$~1/\Lambda^{a_n}~$ so that the total factor from ${\cal V}$-vertices is
$~1/\left(\Lambda^{\sum_n a_n}\right)~$; 
\item{(ii).}  From loop-level: Since each loop brings
a factor $~(1/4\pi)^2 = (f_{\pi}/\Lambda_0 )^2~$, the total
$~\Lambda$-dependence from loop contribution is $~~1/\Lambda_0^{2L}~~$,
where $~\Lambda_0\equiv 4\pi f_\pi \geq \Lambda ~$.

\end{description}
Hence the total $~\Lambda$-dependence given by the above two sources
is $~~ 1/\left(\Lambda^{\sum_na_n}\Lambda_0^{2L}\right)~~$, which
reduces to $~~ 1/\left(\Lambda^{\sum_na_n +2L}\right)~~$ in the case
$~\Lambda \simeq \Lambda_0 =4\pi f_\pi ~$.
For generality, we shall explicitly keep the loop factor 
$~(1/4\pi )^{2L}= (f_{\pi}/\Lambda_0 )^{2L} ~$ in eq.~(3.2.5)
because $~\Lambda =\min (M_{SB}, \Lambda_0 )~$ can be 
somehow lower than $~\Lambda_0=4\pi f_\pi \approx 3.1$~TeV for strongly
coupled EWSB sector, as indicated by model buildings.
From the above discussion, we conclude the following
precise counting rule for $~T~$:
$$
\begin{array}{l}
T= c_T f_\pi^{D_T}\displaystyle 
\left(\frac{f_\pi}{\Lambda}\right)^{N_{\cal O}}
\left(\frac{E}{f_\pi}\right)^{D_{E0}}
\left(\frac{E}{\Lambda_0}\right)^{D_{EL}}
\left(\frac{M_W}{E}\right)^{e_v} H(\ln E/\mu)~~,\\[0.5cm]
N_{\cal O}=\displaystyle\sum_n a_n~,~~
%D_{E0}=2+\displaystyle\sum_n {\cal V}_n(d_n-2)+(i_F+\frac{1}{2}e_F)~, ~~ 
D_{E0}=2+\displaystyle\sum_n {\cal V}_n\left(d_n+\frac{1}{2}f_n-2\right)~, 
~~ D_{EL}=2L~,~~ \Lambda_0 =4\pi f_\pi ~, \\
\end{array}
\eqno(3.2.5)                                                  %(3.2.5)
$$ 
where the dimensionless coefficient $~c_T~$ contains possible powers of
gauge couplings ($~g,g^\prime~$) 
and Yukawa couplings ($~y_f~$) from the vertices in $~T~$. 
$~H$ is a function of $~\ln (E/\mu )~$ which arises from 
loop integrations in the standard 
dimensional regularization~\cite{app,georgi1}  and is insensitive to $E$. 
Here, $\mu$ denotes the relevant renormalization scale for loop corrections. 
In Ref.~\cite{georgi1}, it has been specially emphasized 
that the dimensional regularization supplemented 
by the minimal subtraction scheme is most
convenient for loop calculations in the effective Lagrangian formalism.

It is useful to 
give the explicit and compact form of $~D_{E0}~$ in (3.2.5) 
for the lowest order EWCL 
$~{\cal L}_{\rm G}+{\cal L}^{(2)}+{\cal L}_{\rm F}~$.\footnote{ 
~~It is straightforward to include the
higher order operators in the EWCL for
counting $D_{E0}$ although the possible vertices in this case are
more complicated.}~~ Expanding the interaction terms 
in $~{\cal L}_{\rm G}+{\cal L}^{(2)}+{\cal L}_{\rm F}~$, we find
$$
\begin{array}{l}
\sum_n {\cal V}_n  \equiv {\cal V}   =
 {\cal V}_F +{\cal V}_d^{(V)} + {\cal V}_\pi 
 + {\cal V}_4^{VVVV} + {\cal V}^{VV-\pi} 
             + {\cal V}^{c\bar{c}-\pi} ~~,\\[0.25cm]
\sum_n d_n{\cal V}_n = {\cal V}_d^{(V)} +2{\cal V}_{\pi} ~~,
\end{array}
\eqno(3.2.6)                                             %(3.2.6)
$$
with
$$
\begin{array}{l}
{\cal V}_F = {\cal V}_3^{F\bar{F}V} + {\cal V}^{F\bar{F}-\pi} ~~,\\[0.25cm]
{\cal V}_d^{(V)}\equiv 
{\cal V}_3^{\pi\pi V}+ 
\sum_{n=1}^{\infty}{\cal V}_{2n+2}^{V\pi^{2n+1}}
+{\cal V}_3^{VVV} 
            +{\cal V}^{c\bar{c}V}~~,\\[0.25cm]
{\cal V}_\pi     \equiv \sum_{n=2}^{\infty}{\cal V}_{2n}^{\pi} ~~.
\end{array}
\eqno(3.2.7)                                             %(3.2.7)
$$
In the above equations,
$~{\cal V}_\pi~$ denotes the number
of vertices with pure GB self-interactions;
$~{\cal V}^{F\bar{F}-\pi}~$ and $~{\cal V}^{c\bar{c}-\pi}~$
denote the numbers of fermion-GB vertices and ghost-GB vertices,
 respectively;  $~{\cal V}^{VV-\pi}~$ denotes the $~V$-$V$-$\pi^n~$
( $n\geq 1$ ) vertices; and $~{\cal V}_3^{F\bar{F}V}~$ 
denotes the three-point vertex $F$-$\bar{F}$-$V$, etc. 
(Note that $~{\cal V}^{c\bar{c}-\pi}~$  vanishes in the Landau gauge 
because of the decoupling of GB fields from ghost fields \cite{app}.)  
Hence, for $~{\cal L}_{\rm G}+{\cal L}^{(2)}+{\cal L}_{\rm F}~$, 
the $~D_{E0}~$ factor in (3.2.5) is
$$
D_{E0}=2- \left( {\cal V}_d^{(V)} +{\cal V}_F +2{\cal V}_4^{VVVV}+
           2{\cal V}^{VV-\pi} + 2{\cal V}^{c\bar{c}-\pi}\right) ~~.
\eqno(3.2.8)                                               %(3.2.8)
$$
This clearly shows that the leading energy-power dependence at $L$-loop
level ( $L\geq 0$ ) is always given by those diagrams with pure GB
self-interactions, i.e., $~~(D_E)_{max}=(D_{E0})_{max}+D_{EL}=2+2L~~$, 
because of  the negative contribution from 
$~~-( {\cal V}_d^{(V)} +{\cal V}_F +2{\cal V}_4^{VVVV}+
   2{\cal V}^{c\bar{c}-\pi} ) ~~$ in (3.2.8) which includes all 
types of vertices except the pure GB self-interactions.
This conclusion can be directly generalized to all higher order chiral 
Lagrangian operators such as $~{\cal L}_n~$'s in (3.1.2),  and
is easy to  understand since only pure GB 
self-interaction-vertices
contain the highest powers of the momenta in 
each order of the momentum expansion.
The same conclusion holds for pure $V_L$-scattering amplitudes 
since they can be decomposed into the corresponding 
GB-amplitudes plus the $~M_W/E$-suppressed $B$-term  [cf. (2.1)].
We finally conclude that {\it in the EWCL
$~(D_E)_{\max}=2L+2~$ which is  independent of
the number of external lines of a Feynman diagram.} 
To lowest order of EWCL and at the tree level (i.e. $L=0$~),
$~~(D_E)_{\max}=2~~$, which is in accordance with 
the well-known low energy theorem \cite{let}.
For example, by (3.2.5) and (3.2.8), the model-independent tree level
contributions to 
$~\pi^{a_1}+\pi^{a_2}\rightarrow\pi^{a_3}+\cdots +\pi^{a_n}~$ and
$~V^{a_1}_T +\pi^{a_2}\rightarrow\pi^{a_3}+\cdots +\pi^{a_n}~$ 
( $n\geq 4$ ) are estimated as
$$
\begin{array}{ll}
T_0[\pi^{a_1},\cdots ,\pi^{a_n}]
=\displaystyle O\left(\frac{E^2}{f^2_\pi}f^{n-4}_\pi\right) ~,~~~~&
B_0^{(0)} = g^2f^{n-4}_\pi ~~;\\[0.4cm]
T_0[V_T^{a_1},\pi^{a_2},\cdots ,\pi^{a_n}]
=\displaystyle O\left(g\frac{E}{f_\pi}f^{n-4}_\pi\right) ~,~~~~&
B_0^{(1)}= O\left( g^2\displaystyle\frac{M_W}{E}f^{n-4}_\pi\right) ~,
\end{array}
\eqno(3.2.9)        %(3.2.9)
$$
where $B_0^{(0)}$ and $B_0^{(1)}$ 
are the leading order $B$-terms contained in the
corresponding $V_L$-amplitudes for the above two processes.  
(3.2.9) also coincides with the lowest order explicit calculations in 
Appendix A.

\vspace{0.5cm}
\noindent
{\bf 4. Classification of sensitivities at the 
level of $S$-matrix elements}
\vspace{0.3cm}

Armed with the above counting rule (3.2.5), 
we can conveniently estimate  contributions 
from all effective operators in the EWCL to  
any high energy scattering process.  
In the literature (cf. Ref.~\cite{wwlhc}), what usually done 
was to study
only a small subset of all effective operators for simplicity. But,
to discriminate different underlying theories for a complete test of the
EWSB mechanism, it is necessary to measure all these
operators via various high energy processes.
As the first step global study,
our electroweak power counting analysis 
makes it possible to quickly grasp the overall physical picture 
which provides a useful guideline
for selecting relevant operators and scattering processes to perform
further detailed numerical studies. 
In this and the next sections, we shall systematically classify all
possible NLO effective operators for both the
$S$-matrix elements and the LHC event rates.

We concentrate on the high energy weak-boson fusion and 
quark-anti-quark annihilation processes.
As shown in Refs.~\cite{wwlhc,mike-ww}, for the non-resonance case, 
the most important fusion process 
for probing the EWSB sector is the same-charged channel: 
$~W^{\pm}W^{\pm}\rightarrow W^{\pm}W^{\pm}~$, which gets 
dominant contributions from the 4-GB vertices in the EWCL.
In Tables~1a and 1b we estimate the contributions from the lowest order 
(model-independent) operators in 
$~{\cal L}_{\rm MI}\equiv 
{\cal L}_{\rm G} +{\cal L}_{\rm F} +{\cal L}^{(2)}~$ up to one-loop 
and from all the 
NLO (model-dependent) bosonic operators in (3.1.2) at the tree-level
for $~W^{\pm}W^{\pm}\rightarrow W^{\pm}W^{\pm}~$.
The contributions of different operators to a given amplitude are
different due to their different structures.
For instance, the commonly discussed operators $~{\cal L}_{4,5}~$
contribute the model-dependent leading term of
$~~O\left(\frac{E^2}{f_\pi^2}\frac{E^2}{\Lambda^2}\right)~~$ 
to the $~T[4W_L]~$ amplitude,
and the sub-leading term of 
$~~O\left(g\frac{E}{f_\pi}\frac{E^2}{\Lambda^2}\right)~~$ to 
the $~T[3W_L,W_T]~$ amplitude, while $~{\cal L}_{3,9}~$ give their largest
contributions to $~T[3W_L,W_T]~$ rather than $~T[4W_L]~$ at high energies.
The model-independent
and model-dependent contributions to various $B$-terms 
are summarized in Tables~2a and 2b, in which
$B^{(i)}_{\ell}$ ($i=0,\cdots ,3;~\ell =0,1,\cdots$)
denotes the $B$-term from $V_L$-amplitudes containing $i$ external 
$V_T$-lines with $B^{(i)}_{0}$ obtained from the leading order and
$B^{(i)}_1$ from the NLO calculations.  
We see that the largest $B$-term is  
$B^{(0)}_0$ from the $4W_L$ amplitudes, as given in (2.5).
The term $B^{(0)}_0$ [of $O(g^2)$], is 
a model-independent constant containing only 
the SM gauge coupling constants.
All the other $B$-terms are further suppressed 
by a factor of $M_W/E$ or $(E/\Lambda )^2$, or their product.

For all the $~q\bar{q}^{(\prime )}\rightarrow V^aV^b~$ 
processes (with $q$ or ${q}^{(\prime)}$ being light quarks except
the top),  which get dominant contributions from $s$-channel diagrams 
(containing the $~V_T$-GB-GB vertices), the
model-independent and the model-dependent contributions are estimated
in Tables 3 and 4, respectively. 
Note that the tree-level $~q\bar{q}\rightarrow ZZ~$ 
annihilation process has no model-dependent NLO contribution, 
therefore to probe new physics in the EWSB sector, we have to study
$~q\bar{q}^{(\prime )}\rightarrow W^+W^-,~W^\pm Z~$ annihilations.
As shown in Tables~4a and 4b, the operators $~{\cal L}_{2,3,9}~$
give the leading contributions, 
of $~O\left(g^2\frac{E^2}{\Lambda^2}\right)~$, to 
$~q\bar{q}\rightarrow W^+W^-~$ via
$~T_1[q\bar{q};W^+_LW^-_L]~$ channel\footnote{We note that
the contributions from $~{\cal L}_2~$ 
(and also $~{\cal L}_{1,13}~$)  are always associated
with a suppressing factor $~\sin^2\theta_W \sim {1\over 4}$~.},~
and the operators $~{\cal L}_{3,11,12}~$
give the same leading contributions 
to $~q\bar{q}^{\prime}\rightarrow W^\pm Z~$
via $~T_1[q\bar{q}^{\prime};W^\pm_L Z_L]~$ channel.
But $~{\cal L}^{(2)\prime}~$ does not 
contribute any positive $E$-power term to any of the
 $~V^aV^b$ final states via
tree-level quark-anti-quark annihilations.
 Tables~3 and 4 also show that the largest $B$-term is
$~B=B_0^{(1)}=O\left(g^2\frac{M_W}{E}\right)~$ which is model-independent
and comes from $~T_0[q\bar{q}^{(\prime )};V_T,v]~$, a part of 
the $~T_0[q\bar{q}^{(\prime )};V_T,V_L]~$ amplitude (cf. Table~3a).
All model-dependent $B$-terms, as listed in Tables~4a and 4b,
are either constant terms of 
$~O\left((g^4,e^2g^2)\frac{f_\pi^2}{\Lambda^2}\right)~$ or further
suppressed by negative $E$-power(s) and are thus negligibly small.

From Tables 1-2, we  further classify in Table~5
the sensitivities to all the bosonic  operators for probing
the EWSB sector either
directly (from pure GB interactions) or indirectly (from interactions
suppressed by the SM gauge coupling constants).
The same classification for 
all $~q\bar{q}^{(\prime )}\rightarrow V^aV^b~$
annihilation processes is separately given in Table~6.
The classifications in Tables~5 and 6 
are based upon the following hierarchy in the power counting:
$$
\frac{E^2}{f_\pi^2}\gg \frac{E^2}{f_\pi^2}\frac{E^2}{\Lambda^2}, 
~g\frac{E}{f_\pi} \gg g\frac{E}{f_\pi}\frac{E^2}{\Lambda^2}, ~g^2 
\gg g^2\frac{E^2}{\Lambda^2}, ~g^3\frac{f_\pi}{E} 
\gg g^3\frac{Ef_\pi}{\Lambda^2},~g^4\frac{f^2_\pi}{E^2} 
\gg g^4\frac{f_\pi^2}{\Lambda^2}~~.
\eqno(4.1)                               %(4.1)
$$
In the typical TeV region, for 
$~~E\in (750\,{\rm GeV},~1.5\,{\rm TeV})$,
this gives:
$$
\begin{array}{c}
(9.3,37)\gg (0.55,8.8),(2.0,4.0)\gg (0.12,0.93),(0.42,0.42) \gg \\
(0.025,0.099),(0.089,0.045)\gg (5.3,10.5)\times 10^{-3},
(19.0,4.7)\times 10^{-3}\gg (1.1,1.1)\times 10^{-3} ~,
\end{array}
\eqno(4.2)                             %(4.2)
$$
where $E$ is taken to be the invariant mass of the $VV$ pair.
The numerical values in (4.2) convincingly show the existence of 
the power counting hierarchy in (4.1). 
This governs the order of magnitude
of the results from  detailed numerical calculations. 
This  hierarchy makes it possible to conveniently and globally 
classify the sensitivities of various scattering processes to
the complete set of the effective operators in the EWCL. 
The construction of this power counting hierarchy is based upon the
property of the chiral perturbation expansion and  
can be understood as follows.
The leading term $~\frac{E^2}{f_\pi^2}~$ in (4.1) comes from the 
model-independent lowest order $4V_L$ ($\neq 4Z_L$) scatterings.
 Starting from this leading term, 
(4.1) is built up by {\it increasing either
the number of derivatives 
(i.e. the power of $E/\Lambda$) or the number of
external transverse gauge bosons 
(i.e. the power of gauge coupling constants).}
The NLO contributions from the derivative expansion are 
always suppressed
by $~E^2/\Lambda^2~$ relative to the model-independent leading term.
Also, for each given process, when an external $V_L$-line is replaced 
by a corresponding $V_T$-line, 
a factor $\frac{E}{f_\pi}$ in the amplitude would be replaced by 
a gauge coupling $g$ (or $g^\prime$).\footnote{
The counting on the amplitudes $T_0[4W_T]$ and 
$T_0[q\bar{q}^{(\prime )};V_TV_T]$ are exceptions of this rule
since they have a contribution from the tree-level pure
Yang-Mills gauge term $~{\cal L}_{\rm G}~$.
These two similar exceptions can be found at the second line of 
Tables~1a and 3a, respectively.}~  
This explains why the power counting hierarchy takes the form of (4.1).

Tables~5 and 6 are organized in accordance 
with the power counting hierarchy 
given in (4.1) for all $VV$-fusion and 
$q\bar{q}^{(\prime )}$-annihilation amplitudes.
It shows the {\it relevant} effective new physics operators and the 
corresponding physical processes for probing the EWSB sector
when calculating the scattering amplitudes to the required precision.
For instance, according to the classification of  Table~5, 
the model-independent operator $~{\cal L}_{\rm MI}~$ can be probed
via studying the leading tree-level scattering amplitude
$~T_0[4V_L]~(\neq T_0[4Z_L])~$ which is of 
$~O\left(\frac{E^2}{f_\pi^2}\right)~$. A sensitive probe of 
$~{\cal L}_{\rm MI}~$ via this amplitude requires $~T_0\gg B_0~$, i.e.,
$~O\left(\frac{E^2}{f_\pi^2}\right)\gg O(g^2)~$ or 
$~~\left(2M_W/E\right)^2\ll 1~~$ 
which can be well satisfied in the high energy
region $~E\geq 500$~GeV. We note that the test of the leading order 
operator ${\cal L}_{\rm MI}$ will first distinguish
the strongly interacting EWSB sector from the weakly interacting one.
To test the model-dependent operators $~{\cal L}_{4,5,6,7,10}~$ demands
a higher precision than the leading tree level contribution by a factor
of $~\frac{E^2}{\Lambda^2}~$. As an example, 
in order to sensitively test the 
$~{\cal L}_{4,5}$ operators with coefficients of $O(1)$ 
via the $4V_L$-processes, the criterion (2.4) requires
$~~O\left(\frac{E^2}{f_\pi^2}\frac{E^2}{\Lambda^2}\right)\gg O(g^2)~$,~
or, $~~\left(0.7{\rm TeV}/E\right)^4\ll 1~~$. 
This indicates that sensitively
probing $~{\cal L}_{4,5}~$ via the $4V_L^\pm$-scatterings
requires $~E\geq 1$~TeV. Thus, we find that,
in the TeV region, the $~4V_L$ scatterings can 
sensitively probe $~{\cal L}_{4,5}~$; 
while, similarily, $~{\cal L}_{6,7}~$ can
be probed via $~2W_L+2Z_L~$ or $~4Z_L~$ scattering and $~{\cal L}_{10}~$ 
can only be tested  via $~4Z_L~$ scattering. 
As shown in Table~3, to probe the operators
$~{\cal L}_{2,3,9,11;12}~$, one has to detect the $~3V_L+V_T$
scatterings, which are further suppressed by a factor $~\frac{M_W}{E}~$
relative to the leading model-dependent contributions from the
$~{\cal L}_{4,5}~$ and $~{\cal L}_{6,7,10}~$ via $4V_L$ processes.
Since the model-independent
leading order $2V_T+2V_L$ and $4V_T$ amplitudes 
(from ${\cal L}_{\rm MI}$) and the largest constant
$B$-term $\left( B^{(0)}_0\right)$  are all around of the same order, 
i.e. $~O\left(g\frac{E}{f_\pi}\frac{E^2}{\Lambda^2},g^2\right)~$ 
[cf. (4.2)],\footnote{
They can in principle be separated if the polarization of
the external $V$-lines are identified.
For the final state $V$'s, one can study the angular 
distribution of the leptons from $V$-decay. 
For the incoming $V$'s, one can use forward-jet
tagging and central-jet vetoing to select 
longitudinal $V$'s \cite{wwww}.}~ 
it requires  a significantly higher precision 
to sensitively probe the operators $~{\cal L}_{2,3,9,11;12}~$ 
which can only contribute
the $g$-suppressed indirect EWSB information
and therefore are more difficult to be tested. Here the ratio
$~B_0/T_1\sim g^2/\left[g\frac{E}{f_\pi}\frac{E^2}{\Lambda^2}\right]~$ 
gives 
$~\left(1.15{\rm TeV}/E\right)^3\simeq 0.45\not\ll 1~$ for $E=1.5$~TeV,
which shows the probe of these operators is at most marginally
sensitive when their coeficients $~\ell_n=O(1)~$.
Finally, the operators $~{\cal L}_{1,8;13,14}~$ can be probed 
via the amplitude
$~T_1[2V_L,2V_T] ~(\neq T_1[2Z_L,2Z_T])~$ which is of 
$~O\left(g^2\frac{E^2}{\Lambda^2},g^3\frac{f_\pi}{E}\right)~$ 
and numerically much smaller [cf. (4.2)] in comparison with the
leading $B$-term in (2.5). 
Therefore, $~{\cal L}_{1,8;13,14}~$ should be effectively probed 
via scattering processes other than the $VV$-fusions.

We then look at Table~6 for $q\bar{q}^{(\prime )}$-annihilations.
For the lowest order Lagrangian 
$~{\cal L}_{\rm MI}= 
  {\cal L}_{\rm G}+{\cal L}^{(2)}+{\cal L}_{\rm F}~$,
the model-independent operators 
$~{\cal L}^{(2)}~$ and $~{\cal L}_{\rm G}~$
can be probed via tree-level amplitudes [of $~O(g^2)~$]
with $~V_LV_L~$ and ~$V_TV_T$~ final states, respectively.\footnote{
$~{\cal L}^{(2)}~$ just gives the low energy theorem results
and thus denotes the model-independent part of the EWSB sector, while
$~{\cal L}_{\rm G}~$ is the standard tree-level Yang-Mills gauge term
which is irrelevant to the EWSB mechanism.}~
Thus, the contribution of $~{\cal L}^{(2)}~$ to $V_LV_L$ final state is
not enhanced by any $E$-power in the high energy region, in contrast to
the case of $VV$-fusions (cf. Table~1a). 
So, the leading order $~T_0[q\bar{q}^{(\prime )};V_LV_L]~$ amplitude,
similar to the $~T_0[q\bar{q}^{(\prime )};V_TV_T]~$ amplitude,
is not sensitive to the strongly coupled EWSB sector. 
We then discuss the contributions of model-dependent 
NLO operators to the $q\bar{q}^{(\prime )}$-annihilations.
We first note that the operators $~{\cal L}_{4,5,6,7,10}~$ cannot
contribute to $q\bar{q}^{(\prime )}$-annihilations at $~1/\Lambda^2$-order
and thus should be best probed via $VV$-fusions (cf. Table~5).
Among all other NLO operators, 
the probe of $~{\cal L}_{2,3,9}~$ 
are most sensitive via $~q\bar{q}\rightarrow W^+_LW^-_L$ amplitude
and the probe of $~{\cal L}_{3,11,12}~$ are best via
$~q\bar{q}^{\prime}\rightarrow W^\pm_LZ_L$ amplitude. For operators 
$~{\cal L}_{1,8;13,14}~$, the largest amplitudes are
$~T_1[q\bar{q}^{(\prime )};W^+_LW^-_T/W^+_TW^-_L]~$ and
$~T_1[q\bar{q}^{\prime};W^\pm_LZ_T/W^\pm_TZ_L]~$, which are
at most of $~O\left(g^3\frac{Ef_\pi}{\Lambda^2}\right)~$ and are
suppressed by a factor $~\frac{f_\pi}{E}~$ relative to the 
model-dependent leading amplitudes of 
$~O\left(g^2\frac{E^2}{\Lambda^2}\right)~$.

In summary, applying the power counting technique allows us to
conveniently estimate contributions of various operators 
to any scattering amplitude.
For a given scattering process, this result tells us which 
operators can be sensitively probed. Similarly, the same result can 
also tell us which process would be most sensitive for probing 
new physics via a given effective operator.
In the next section, we shall examine the important
$W^\pm W^\pm \rightarrow W^\pm W^\pm$ fusion and 
$~q\bar{q}^{\prime}\rightarrow W^\pm Z~$ annihilation
processes at the LHC to illustrate how to use 
the electroweak power counting method
to estimate the event rates and 
how to use the ET as a theoretical criterion
to classify the sensitivities of these typical scattering processes
to the NLO bosonic operators in the EWCL.

\vspace{0.5cm}
\noindent
{\bf 5. Probing EWSB Mechanism at the LHC via Weak-Boson Fusions and
Quark-Anti-quark Annihilations}
\vspace{0.3cm}

\vspace{0.2cm}
\noindent
{\bf 5.1. Preliminaries}

In this section, we estimate the production rates for both
weak boson fusions and quark-anti-quark annihilations at the 
LHC (a ${\rm p}{\rm p}$ collider with $\sqrt{S}=14$\,TeV 
and an integrated luminosity of $~100\,{\rm fb}^{-1}~$).
The gauge-boson fusion process $W^+W^+ \rightarrow W^+W^+$  and the
quark-anti-quark annihilation process $~q\bar{q}^{\prime}\rightarrow W^+Z~$ 
will be separately studied.
To calculate the event rates for gauge-boson fusions, we multiply
the luminosity of the incoming weak-boson pair $VV$
(by the effective-$W$ approximation (EWA)~\cite{effective-W}) 
and the constituent cross section of the weak-boson scattering
(from the amplitude estimated by the power counting analysis in 
the last section). Note that the validity of the EWA
requires the $VV$ invariant mass $~~M_{VV}\gg 2M_W~~$ \cite{effective-W},
which coincides with the condition in (2.3a) for the validity of
the ET.\footnote{ Here, we 
reasonably take the typical energy scale $E$ of the $VV$ scattering  
to be $M_{VV}$ for estimating the event rates.}~~
Thus, the EWA and the ET 
have similar precisions in computing the
event rate from $V_LV_L$ fusion process in hadron collisions.
As $M_{VV}$ increases, they become more accurate. 
It is known that the EWA is less accurate for
sub-processes involving initial transverse gauge 
boson(s)~\cite{gunion,mike-EWA}.  Nevertheless, the EWA has
been widely used in the literature for computing event rates from 
gauge-boson (either transversely or longitudinally
polarized) fusion processes because it is easy to implement and can be
used to reasonably estimate event rates before any exact calculation is 
available. As to be shown shortly, our power counting results
agree well to the existing detailed calculations  
within about a factor of $~2$. Hence,
it is appropriate to apply the power counting analysis
together with the EWA for estimating the event rates from weak-boson 
fusions at the LHC.  
The coincidence for the case of the $q\bar{q}^{(\prime )}$-annihilation 
is even better, where the EWA is not needed.

For the purpose of systematically analyzing the 
sensitivity to each bosonic operator in (3.1.2), 
in what follows, we separately compare the rates contributed by each 
individual operator with that by the $B$-term.
The actual experiments contain the contributions from {\it all} 
possible operators and are thus more complicated. For simplicity and
clearness, we follow the well-known naturalness 
assumption (i.e., contributions from different operators do not 
accidently cancel each other) and estimate the 
contributions from each operator separately. 

Let us denote the production rate for the scattering 
process $~W_{\alpha}^+W_{\beta}^+\rightarrow W_{\gamma}^+W_{\delta}^+~$
as $~R_{\alpha\beta\gamma\delta (\ell)}~$,
where $\alpha ,\beta ,\gamma ,\delta =L,T$ label the polarizations
of the $W$-bosons 
and $~\ell=0,1$ indicates contributions from
leading order and next-to-leading order, respectively. 
Up to the one-loop level, we define
$$
\begin{array}{lll}
R_{\alpha\beta\gamma\delta}
& = & R_{\alpha\beta\gamma\delta (0)} +R_{\alpha\beta\gamma\delta (1)} ~~,
\\[0.2cm]
 R_{\alpha\beta\gamma\delta (\pm )} 
& = & R_{\alpha\beta\gamma\delta (0)} \pm 
|R_{\alpha\beta\gamma\delta (1)}| ~~.
\end{array}
\eqno(5.1a,b)
$$
Also, $~R_B~$ denotes the rate contributed by the largest $B$-term.
For convenience, we use the subscript ``$_S$'' to stand for 
summing up the polarizations of the corresponding gauge boson. For the
$~q\bar{q}^{(\prime)}\rightarrow V_\alpha V_\beta~$ processes, the rates are 
denoted by $~R_{\alpha\beta(\ell )}~$, and correspondingly we define
$~R_{\alpha\beta}=R_{\alpha\beta(0)}+R_{\alpha\beta(1)}~$ and
$~R_{\alpha\beta (\pm )}=R_{\alpha\beta(0)}\pm |R_{\alpha\beta(1)}|~$.
We also note that when applying the power counting analysis,
we have ignored the angular dependence in the scattering 
amplitudes (cf. Tables~1$\sim$2)
because it does not affect the order of magnitude estimates
for the total cross sections (or the event rates). 

To check the reliability of our power counting method,
we have compared our numerical results for the 
$~W^+W^+\rightarrow W^+_LW^+_L$ fusion and 
the $q\bar{q}\rightarrow W^+_LW^-_L$ annihilation with 
those in Fig.~8 and Fig.~5 of Ref.~\cite{BDV} 
in which the above constituent amplitudes were explicitly calculated and
the polarizations of the initial weak-bosons were summed over 
for the fusion process.
As shown in Figs.~1a and 1b, both results coincide
well within about a factor of $2$ or better.
These are two typical examples showing that the correct physical picture
can be consistently and quickly grasped by our power counting analysis. 

We conclude this subsection by briefly commenting on  
a recent paper~\cite{dobad}. The sole purpose of Ref.~\cite{dobad} 
was to avoid using the ET, but still within the EWA, 
to increase the calculation precision and to extend the results 
to lower energy regions.
This approach is, however, inconsistent  because the validity of
the EWA also requires the {\it same high energy condition
$~E\gg M_W~$} as that of the ET.
To study  the operators ${\cal L}_{4,5}$ (dominated by pure $V_L$-modes),
the approach in Ref.~\cite{dobad} cannot really get 
higher precision than previous studies~\cite{BDV,wwlhc} 
(using the ET combined with the EWA) except 
making unnecessary complications and confusions in their calculation.
To study other operators like ${\cal L}_{1,2,3,8,9,11}$, 
the $V_T$-modes must be included, for which the EWA is much 
worse~\cite{gunion,mike-EWA}. 
Hence, to get the {\it consistent and precise results}
for these operators, one must go beyond the EWA for full calculations.
This was not contained in Ref.~\cite{dobad}. On the contrary,
our present global power counting analysis
provides the first complete and consistent estimate for all 
NLO operators in the EWCL (3.1.2).

\vspace{0.2cm}
\noindent
{\bf 5.2. Analyzing the Model-Independent Contributions 
          to the Events Rates}

In Fig.~2a we give the power counting estimates 
for the production rates
of the $W^+_L W^+_L$ pairs at the LHC 
from the initial state $W$-bosons with different polarizations.
In this figure, setting the renormalized coefficients 
$~\ell_{0\sim 14}$ to be zero, we include only the model-independent 
contributions up to one-loop.\footnote{It is understood 
that the divergent pieces 
from one-loop calculations have been absorbed by the
coefficients of the corresponding 
NLO effective operators~\cite{app,georgi}.}~~
As clearly shown in Fig.~2a, the rate from $4W_L$ scattering
dominates over the rate from $W_T+3W_L$ scattering. The latter
is lower by about
an order of magnitude for large $M_{WW}$ in spite of the fact that
the $W_T W_L$ luminosity is larger than the $W_LW_L$
luminosity in the initial state.
Also separately shown in the same figure is the 
event rate $~|R_B|~$ contributed by the 
largest $B$-term [cf. (2.1) and (2.5)]
which is even significantly lower than that from the $W_T+3W_L$ scattering
by a factor of ~$2\sim 7$~ for $~M_{WW}>500$\,GeV.
However, the rate from $W_TW_T$ initial state is lower than 
that from the $B$-term in the $4W_L$ amplitude as $~E\geq 600$\,GeV.
This implies that if the contribution from $W_TW_T$ initial state
is to be included in calculating the total production rate
of the  $W_LW_L$ pair,
the contribution from the $B$-term in the $4W_L$ amplitude 
should also be included because
they are of the same order in magnitude. 
If, however, only the pure Goldstone boson amplitude
$~T[\pi^+ \pi^+ \rightarrow \pi^+ \pi^+]~$
is used to calculate the $4W_L$-amplitude 
(with the $B$-term ignored) 
the contribution from $~T[W^+_T W^+_T \rightarrow W_L^+ W_L^+]~$
should also be consistently ignored for computing the 
total rate of $W^+_L W^+_L$ pair production 
via the weak-boson fusion mechanism.

As shown in Ref.~\cite{wwlhc}, it is possible to statistically,
though not on the event-by-event basis,
choose event with longitudinally polarized 
$W$-bosons in the initial state 
by applying the techniques of 
forward-jet tagging and central-jet vetoing.
In this work we do not intend to study the details of the 
event kinematics, and 
we shall sum over all the initial state polarizations 
for the rest of discussions. 
Let us first compare the rates for different polarizations in the final 
state. Fig.~2b
shows that the rate of $W_LW_L$ final state dominates, while the
rate of $B$-term and the
rates of $W_LW_T$ and $W_TW_T$ final states are of the same order, 
and all of them are about an $O(10)$ to $O(10^2)$  
lower than the rate of $W_LW_L$ final state in the energy region 
$~E=M_{WW} > 500\,{\rm GeV}$. 
Therefore, if one wants to increase the precision
in calculating the total event rates by 
including the small contribution from 
the $B$-term in pure $4W^+_L$ scattering,
then the contributions from $W^+_SW^+_S\rightarrow W^+_TW^+_T$
and $W^+_SW^+_S\rightarrow W^+_LW^+_T$ scatterings should also be 
consistently included. Otherwise, they must be neglected all together.
From Figs.~2a and 2b, we conclude that the scattering process
$W^+_L W^+_L \rightarrow W^+_L W^+_L$ dominates the $W^+W^+$-pair
productions when the model-dependent coefficients $~\ell_{0\sim 14}$ 
in (3.1.2) are set to be zero.

Similarly, we apply the power counting analysis to 
estimate the model-independent contribution to the rates of 
polarized $W^+Z^0$ pair produced from 
$q {\bar{q}}'$ fusion up to one-loop order. 
 The results are plotted in Fig.~3. 
It shows that the $~W_T^+Z^0_T~$ and $~W_L^+Z^0_L~$ rates (i.e.,
$R_{TT}$ and $R_{LL}$) are of the same order. They are much larger
than the rate $~R_{LT}~$ (from the $~W^+_LZ^0_T~$ and $~W^+_TZ^0_L~$
final states) and the rate $~|R_B|~$ (from the largest $B$-term
contained in the $~q\bar{q}'\rightarrow W^+_LZ^0_T,W^+_TZ^0_L~$
amplitudes). The rate $~R_{LT}~$ is only slightly above the $~|R_B|~$ 
because the leading GB-amplitude $~T_0[q\bar{q}';\pi^+ Z_T/\pi^0 W_T^+]~$ 
is of the same order as
the term $~B^{(1)}_0~$ (cf. Table~3a). Here we see that for the
lowest order total signal rate, both $~R_{TT}~$ and $~R_{LL}~$ have to be
included since they are of the same order and larger than the
NLO model-dependent contributions. If one wants
to further include $~R_{LT}~$, then $~R_B~$ should also be included.

\vspace{0.2cm}
\noindent
{\bf 5.3. Estimating Sensitivities for Probing 
          the Model-Dependent Operators}

In this section, we classify the sensitivities to all the 
NLO bosonic operators at the LHC. Without knowing the values of the 
{\it model-dependent coefficients} ($\ell_n$~'s), we shall take them to
vary from $O(1)$ to $O(10)$  except that $~\ell_0,
~\ell_1$ and $\ell_8$ are bounded to be of $O(1)$ 
by the low energy data [cf. (3.1.7)].

We first consider the scattering process $W^+W^+ \rightarrow W^+W^+$.
Our theoretical criterion for discriminating different sensitivity levels
({\it sensitive, marginally sensitive}, or {\it insensitive})
to probe a particular operator via the production of $W^+W^+$ pairs
is to compare its contribution to the event rate 
( $|R_{\alpha\beta\gamma\delta (1)}|$ ) with that from the
largest model-independent contribution of the LNI $B$-term ( $|R_B|$ ),
according to the Sec.~2. 
In Figs.~4-7, we show the results
for varying $~|\ell_n|~$ from $~O(1)$ to $~O(10)$ (except $\ell_0,
~\ell_1$ and $\ell_8$). Here, the polarizations of the initial and the final 
states have been summed over. 
In Figs.~4a and 4b, we consider the coefficients ( $\ell_n$'s ) 
to be naturally of ~$O(1)$~ according to 
the naive dimensional analysis~\cite{georgi}.
Fig.~4a shows that the event rates/($100$\,${\rm fb}^{-1}$GeV) from
operators $~{\cal L}_{4,5}~$ are larger than that from the $B$-term 
when $~E=M_{WW} >600~$GeV, while the rates from operators
$~{\cal L}_{3,9,11;12}~$ can exceed $~|R_B|~$ only if $~E=M_{WW}>860$~GeV.
As $M_{WW}$ increases, the 
rates contributed by $~{\cal L}_{4,5}~$ remain flat,
while the rates by $~{\cal L}_{3,9,11;12}~$  and the $B$-term decrease.
The ratio of the event rates from $~{\cal L}_{4,5}~$ 
to $~|R_B|~$ is $~5.0$ at $~E=M_{WW}=1$\,TeV, and rapidly increases 
to $~19.6$ at $~E=M_{WW}=1.5~$\,TeV.
In contrast, the ratio between the rates from
$~{\cal L}_{3,9,11;12}~$  and the $B$-term 
only varies from $~1.4$ to $3.0$ for $~E=M_{WW}=1\sim 1.5$\,TeV.
Fig.~4b shows that for the coefficients of $~O(1)~$, the event rates
contributed by operators $~{\cal L}^{(2)\prime}~$ and 
$~{\cal L}_{1,2,8;13,14}~$  are all below $~|R_B|~$ for a wide 
region of energy up to about $~2$\,TeV, 
so that they cannot be sensitively probed in this case. 
Especially, the contributions from
$~{\cal L}_{1,13}~$  are about two orders of magnitude 
lower than that from the $B$-term. 
This suggests that $~{\cal L}_{1,13}~$   must be tested via
other processes~\cite{LCHKY}.
In Figs.~5a and 5b, different event rates are compared for 
the coefficients (except $\ell_0,~ \ell_1,~\ell_8$) to be of $O(10)$. 
Fig.~5a shows that 
the rates from $~{\cal L}_{3,9,11;12}~$ could significantly dominate 
over $~|R_B|~$ by an order of magnitude for 
$E=M_{WW}\sim 1$\,TeV
if their coefficients are increased by a factor
of $~10$ relative to the natural size of $~O(1)~$.
Fig.~5b shows that the rates from 
$~{\cal L}_{13}~$ is still lower than $~|R_B|~$ by about an order of 
magnitude,
while the rate from $~{\cal L}_{2}~$ agrees with
 $~|R_B|~$  within a factor of $~2$. 
The contribution from $~{\cal L}_{14}~$ exceeds 
$~|R_B|~$ by about a factor $2\sim 3$ at $~E=M_{WW}=1$\,TeV and a
factor of $3\sim 5$ at $~E=M_{WW}=1.5$\,TeV when its coefficients is of 
$~O(10)~$.

As discussed above, the cutoff scale $\Lambda$ can be lower than
$~\Lambda_0=4 \pi f_\pi \simeq 3.1~$TeV if there is any new 
heavy resonance below $\Lambda_0$.
In that case, the signal rates $|R_1|$ will be higher than 
that reported in Figs.~4 and 5 by about a factor of 
$~\left(\frac{\Lambda_0}{\Lambda}\right)^2~$.
For comparison, we repeat the above calculations for 
$\Lambda=2~$TeV in Figs.~6 and 7.
As indicated in Fig.~6a and 7b, the sensitivities to probing 
$~\ell_{3,9,11,12}\sim O(1)~$ and $~\ell_{14,2}\sim O(10)~$ 
via $W^+W^+ \rightarrow W^+W^+$ 
process increase for a lower  ~$\Lambda$~ value.

From the above analyses, we conclude that studying the 
$~W^+W^+\rightarrow W^+W^+~$ process can sensitively probe
the operators $~{\cal L}_{4,5}~$, but is only marginally sensitive for
probing $~{\cal L}_{3,9,11;12}~$ and insensitive for 
$~{\cal L}^{(2)\prime}~$ and $~{\cal L}_{1,2,8;13,14}~$, if their 
coefficients are naturally of $~O(1)~$. 
In the case where these coefficients are of $~O(10)~$, the probe of 
$~{\cal L}_{14}~$ (for lower $\Lambda$) and $~{\cal L}_{3,9,11;12}~$
could become sensitive and that of $~{\cal L}_{2}~$ (for lower $\Lambda$)
could become marginally sensitive, while 
 $~{\cal L}_{13}~$ still cannot be sensitively or marginally sensitively
measured.

Moreover, we note that the operators $~{\cal L}_{6,7,10}~$, 
which violate the custodial $SU(2)_C$ symmetry,
do not contribute to the $W^+W^+$ pair productions up to
$~O(1/\Lambda^2)~$.
They can however contribute to the other scattering channels such as
$~WZ\rightarrow WZ~$, $~WW\rightarrow ZZ~$,
$~ZZ\rightarrow WW~$ and $~ZZ\rightarrow ZZ~$,
cf. Table~3. (Here, $~\,{\cal L}_{10}$ only contributes 
to $~ZZ\rightarrow ZZ~$ channel.)
By our order of magnitude estimates,
we conclude that they will give the similar kind of contributions
to the $WZ$ or $ZZ$ channel as
$~{\cal L}_{4,5}~$ give to the $W^+W^+$ channel.
This is because all these operators
contain four covariant derivatives [cf. (3.1.2)]
and thus become dominant in the high energy $VV$-fusion processes. 

Let us consider the $W^-W^- \rightarrow W^-W^-$ production process.
At the LHC, in the TeV region,
the luminosity of $W^-W^-$ is 
typically smaller than that of $W^+W^+$ by a factor of $3 \sim 5$.
This is because in the TeV region, where the fraction
of momentum ($x$) of proton carried by the quark (which 
emitting the initial state $W$-boson) is large 
(for $x=\frac{E}{\sqrt{S}}\sim 0.1$), the parton luminosity is
dominated by the valence quark contributions.
Since in the large-$x$ region, the probability of finding a down-type 
valence quark in the proton is smaller than finding an up-type 
valence quark, 
the luminosity of $W^-W^-$ is smaller than that of $W^+W^+$.
However, as long as there are enough $W^-W^-$ pairs detected,
which requires a large integrated luminosity of the machine and a 
high detection efficiency of the detector, conclusion similar to probing 
the effective operators for the $W^+W^+$ 
channel can also be drawn for this channel.
For ~$M_{WW} > 1.5 $~TeV, the $W^-W^-$ production rate becomes 
about an order of magnitude smaller than the $W^+W^+$ rate for any given 
operator. Thus, this process will not be sensitive to probing
the NLO operators when ~$M_{WW} > 1.5 $~TeV.

Next, we examine the $W^+Z^0$ production
 rates in the $~q\bar{q}^{\prime}\rightarrow W^+ Z^0~$
channel. As shown in Figs.~8, for
~$\Lambda=3.1~$TeV, when the coefficients are of
~$O(1)~$, the probe of ~${\cal L}_{3,11,12}~$ is sensitive 
when $~E>750~$GeV, while that of ~${\cal L}_{8,9,14}~$ is 
marginally sensitive if ~$E>950$~GeV. 
The probe of ~${\cal L}^{(2)\prime}$~ 
becomes marginally sensitive if ~$E>1.4$~TeV, and
that of ${\cal L}_{1,2,13}$ is insensitive for ~$E<1.9~$TeV. When the
coefficients other than $~\ell_0,~\ell_1,~\ell_8~$ are of ~$O(10)$~, 
~${\cal L}_{3,11,12,9,14}$~ 
could all be sensitively probed when $~E>500~$GeV, 
and the probe of ~${\cal L}_{2,13}$~ could be sensitive when ~$E>1.2~$TeV. 
For ~$\Lambda=2~$TeV, the sensitivities are increased overall by a
factor $~\left(\frac{\Lambda_0}{\Lambda}\right)^2\simeq 2.4~$,
as shown in Fig.~9.   The event rate for
$~q\bar{q}^{\prime}\rightarrow W^- Z^0~$ is slightly lower than that of
$~q\bar{q}^{\prime}\rightarrow W^+ Z^0~$ by 
only about a factor of $1.5$ due
to the lower luminosity for producing $W^-$ bosons in 
${\rm p}{\rm p}$ collisions. Hence, the 
above conclusion also holds for the 
$~q\bar{q}^{\prime}\rightarrow W^- Z^0~$ process.

We note that the $~q\bar{q}^{\prime}\rightarrow W^\pm Z^0~$ annihilation
provides {\it complementary} information on probing 
the EWSB sector, in comparison with 
the ~$W^\pm W^\pm\rightarrow W^\pm W^\pm$~ fusion. 
In the former, $~{\cal L}_{3,11,12}$~ can be 
unambiguously probed, while in the latter, 
~${\cal L}_{4,5}~$ can be sensitively probed.
Furthermore, as shown in Table~5 and 6, the operators
$~{\cal L}_{6,7}~$ can be probed from either 
~$T_1[2 W_L, 2 Z_L]~$ or $~T_1[4 Z_L]$~, and ~${\cal L}_{10}$~ can 
only be probed from ~$T_1[4 Z_L]$~, while ~${\cal L}_{2,9}$~
can be tested from $~T_1[q\bar{q};W^+_LW^-_L]~$.
It is therefore necessary and useful to measure all the 
the gauge boson fusion and quark-anti-quark annihilation
processes for completely exploring the EWSB sector.

\vspace{0.5cm}
\noindent
{\bf 6. Conclusions}
\vspace{0.3cm}

In this work, based upon our recent study 
on the intrinsic connection between
the longitudinal weak-boson scatterings and probing the EWSB sector, we
first formulate the physical content of the ET as a criterion for
discriminating processes which are sensitive/insensitive to probing
the EWSB mechanism [cf. Eqs.~(2.3)$\sim$(2.5)]. 
Then, we develop a precise power counting rule (3.2.5) for the EWCL, 
from a natural generalization of Weinberg's counting method for 
the ungauged non-linear sigma model. For completeness
and for other possible applications, 
in Appendix B, we also generalize our power 
counting rule for a linearly realized effective Lagrangian~\cite{linear}
which is often studied in the literature. 
The renormalizable SM with a light Higgs boson is
included in the linear effective Lagrangian formalism at the lowest order.

Armed with this powerful counting rule and using the ET as the theoretical
criterion for probing the EWSB sector, we further systematically classify
the sensitivities of various scattering processes to 
the complete set of bosonic operators at the level of $S$-matrix
elements (cf. Tables~1-6). 
The power counting hierarchy in (4.1) governs the 
order of magnitude of all relevant scattering amplitudes.

Finally, based on the above power counting analysis combined with the EWA, 
we study the phenomenology for probing the EWSB sector
at the LHC via the $~W^\pm W^\pm\rightarrow W^\pm W^\pm~$ fusion 
and the $~q\bar{q}'\rightarrow W^\pm Z^0~$ annihilation processes.
In this simple power counting analysis,
our numerical results for the production rates agree, 
within about a factor of $2$ (cf. Fig.~1.),
with the explicit calculations performed in the literature
in which only a small subset of the NLO operators were studied.
This indicates that our power counting analysis conveniently and reasonably
grasps the overall physical picture.
With this powerful tool, we perform the first complete survey on the
sensitivities~\footnote{Cf. footnote-$b$.}~ 
of all fifteen next-to-leading order 
$CP$-conserving and $CP$-violating effective operators at the LHC
via $W^\pm W^\pm$-fusions and 
$q\bar{q}'\rightarrow W^\pm Z^0~$ 
annihilations. The results are shown in Figs.~4-7 and Fig.~8-9, 
respectively.
We find that, for $W^+W^+$-channel, when the  coefficients $\ell_n$'s 
are naturally of $O(1)$,
$~{\cal L}_{4,5}~$ are most sensitive, 
$~{\cal L}_{3,9,11;12}~$ are marginally sensitive, and
$~{\cal L}^{(2)\prime}~$ and $~{\cal L}_{1,2,8;13,14}~$ are insensitive.
For the case where
the coefficients other than ~$\ell_0,~\ell_1,~\ell_8$~ are of $~O(10)~$, 
the probe of $~{\cal L}_{14}~$ (for lower $\Lambda$) and 
$~{\cal L}_{3,9,11;12}~$ could become sensitive and that of 
$~{\cal L}_{2}~$ (for lower $\Lambda$) could become 
marginally sensitive. However,
$~{\cal L}_{13}~$ cannot be sensitively
probed via this process so that it must be 
measured via other processes.\footnote{We note that 
$~{\cal L}_{13}~$ (and $~{\cal L}_{14}~$) can be sensitively probed via
$~e^- \gamma \rightarrow \nu_e W^-_L Z^0_L \, ~{\rm or}~ \,
e^- W^-_L W^+_L~$ processes at the future TeV 
linear collider~\cite{LCHKY}.}~~ A similar conclusion holds for the 
$W^-W^-$ channel except that the event rate is lower by about a factor 
of $3 \sim 5$ in the TeV region because the quark 
luminosity for producing a $W^-W^-$ pair is smaller 
than that for a $W^+W^+$ pair in pp collisions. 
 Up to the next-to-leading order,
the $SU(2)_C$-violating operators $~{\cal L}_{6,7,10}~$
do not contribute to the $W^\pm W^\pm$ channel. They, 
however, can be probed 
via the $WZ \rightarrow WZ$,~ $WW \rightarrow ZZ$,~
$ZZ \rightarrow WW$,  and  $ZZ \rightarrow ZZ$ processes~\cite{LCHKY}.

For the ~$q\bar{q}^{\prime}\rightarrow W^\pm Z^0$~ process,
the conclusion is quite different. 
The operators $~{\cal L}_{4,5,6,7,10}~$ do not contribute at the tree level. 
Using this process, $~{\cal L}_{3,11,12}~$ can be 
sensitively probed in the high energy range ($E>750~$GeV), 
and the probe of $~{\cal L}_{8,9,14}~$ can be marginally sensitive 
for $~E>950$~GeV if their coefficients are of $~O(1)~$ and that of
~${\cal L}_{9,14}$~ can be sensitive if their coefficients are of ~$O(10)$~. 
The results are plotted in Figs.~8-9.
We conclude that the $VV$-fusion and the 
$q {\bar q}^{(\prime )}$-annihilation processes are 
{\it complementary} to each other 
for probing the complete set of the 
NLO effective operators in the electroweak chiral Lagrangian (3.1.2).
Extensions of our analysis to the future linear colliders
are given in Ref.~\cite{LCHKY}.

%\newpage
\vspace{1.0cm}
\noindent
{\bf Acknowledgements}~~~
We thank Sally Dawson for useful conversations on the EWA used in 
Ref.~\cite{BDV} and Mike Chanowitz for helpful discussions on the validity
of the EWA~\cite{mike-EWA}. 
We are grateful to William Bardeen, John Donoghue, Tao Han 
and Peter Zerwas for valuable discussions and reading the manuscript.
We also thank Ulrich Baur for useful discussions on the current 
LEP and Tevatron bounds for TGCs given in Ref.~\cite{TGC}.
H.J.H. is supported by the AvH of Germany and the U.S. DOE.
Y.P.K. is supported by the NSF of China 
and the FRF of Tsinghua University;
C.P.Y. is supported in part by the U.S. NSF.

%%%%%%%%%%%%%%%%%%%%%%%%%%%%%%%%%%%%%%%%%%%%%%%%%%%%%%%%%%%%%%%%%%%%%%%%%
%%%%%%%%%%%%%%%%%%%%%%%%%%%%%%%%%%%%%%%%%%%%%%%%%%%%%%%%%%%%%%%%%%%%%%%%%

\vspace{1cm}
%\newpage
\noindent
{\bf  Appendix A. Validity of the ET in some special kinematic regions}
\vspace{0.3cm}

Here we examine the validity of the ET in some special kinematic regions
and its physical implication in probing the EWSB, which often cause
confusion in the literature.
It is known that there are kinematic regions
in which the Mandelstam variables 
~$t$~ or ~$u$~ is small or even vanishing
despite the fact that $~\sqrt{s}\gg M_W~$ for high energy scatterings.
Therefore, the amplitude that contains a~$t$- or $u$-channel diagram 
with massless photon field can generate
a kinematic singularity when the scattering angle $\theta$ 
approaches to $0^{\circ}$ or $180^{\circ}$.
In the following, we study in such special kinematic regions
whether the $B$-term [cf. (2.1)] can be safely ignored to validate
the ET and its physical consequence to probing the EWSB sector.

For illustration, let us consider the tree level 
$~W^+_LW^-_L\rightarrow W^+_LW^-_L~$ scattering
in the chiral Lagrangian formalism.
Generalization to loop orders is obvious since the kinematic problem
 analyzed here only concerns the one-particle-reducible (1PR) internal
$W$, $Z$ or photon line in the $t$-channel (or $u$-channel) diagram.
Both the tree level $~W^+_LW^-_L\rightarrow W^+_LW^-_L~$ and
$~\pi^+\pi^-\rightarrow \pi^+\pi^-~$ 
amplitudes in the chiral Lagrangian formalism 
contain contact diagrams, $s$-channel $Z$-exchange and
photon-exchange diagrams, and $t$-channel $Z$-exchange and
photon-exchange diagrams. 
In the C.M. frame, the precise tree-level amplitudes 
$~T[W_L]~$ and $~T[{\rm GB}]~$ are:
$$
\begin{array}{l}
T[W_L]=
 ig^2\left[ -(1+\kappa )^2 \sin^2\theta 
+ 2\kappa (1+\kappa )(3\cos\theta -1)
-{\rm c}_{\rm w}^2
   \displaystyle\frac{4\kappa (2\kappa +3)^2\cos\theta}{4\kappa +3
  -{\rm s}_{\rm w}^2{\rm c}_{\rm w}^{-2}}\right.\\[0.4cm]
  \left.  +{\rm c}_{\rm w}^2
   \displaystyle\frac{8\kappa (1+\kappa )(1-\cos\theta )(1+3\cos\theta )
   +2[(3+\cos\theta )\kappa +2][(1-\cos\theta )\kappa -\cos\theta ]^2}
   {2\kappa (1-\cos\theta )+{\rm c}_{\rm w}^{-2}} \right]\\[0.4cm]
 +ie^2 \left[ -\displaystyle\frac{\kappa (2\kappa +3)^2\cos\theta}
    {\kappa +1} +4(1+\kappa )(1+3\cos\theta )+
   \displaystyle\frac{[(3+\cos\theta )\kappa +2][(1-\cos\theta )\kappa 
       -\cos\theta ]^2}{\kappa (1-\cos\theta )} \right]~~,
\end{array}
\eqno(A1a)             %(A1a)
$$  \\
$$
\begin{array}{l}
T[{\rm GB}] = 
ig^2 \left[\displaystyle\frac{(1+\cos\theta )}{2}\kappa 
           +\frac{1}{3}
+\displaystyle\frac{({\rm c}_{\rm w}^2-{\rm s}_{\rm w}^2)^2}
{2{\rm c}_{\rm w}^2}\left(-\displaystyle\frac{2\kappa\cos\theta}
{4\kappa +3-{\rm s}_{\rm w}^2{\rm c}_{\rm w}^{-2}}
+\displaystyle\frac{(3+\cos\theta )\kappa +2}{2(1-\cos\theta )\kappa
+ {\rm c}_{\rm w}^{-2}}\right)\right]\\[0.4cm]
+ie^2\left[ -\displaystyle\frac{4\kappa\cos\theta}{4\kappa +1}
+\displaystyle\frac{(3+\cos\theta )
   \kappa +2}{(1-\cos\theta )\kappa}\right]~~, \\[0.4cm]
\end{array}
\eqno(A1b)            %(A1b)
$$\\[0.10cm]
where $~\kappa \equiv p^2/M_W^2~$ with $~p~$ equal to the C.M. momentum;
$~{\rm s}_{\rm w}\equiv \sin\theta_{\rm W}~$,
      $~{\rm c}_{\rm w}\equiv \cos\theta_{\rm W}~$ with $\theta_{\rm W}$
equal to the weak mixing angle; 
and $\theta$ is the scattering angle.
In (A1a) and (A1b) the terms without a momentum factor in the denominator
come from contact diagrams, terms with denominator independent of
scattering angle come from $s$-channel diagrams and terms with denominator
containing a factor $~1-\cos\theta~$ 
are contributed by $t$-channel diagrams.
Let us consider two special kinematic regions defined below.

\noindent
(i). In the limit of ~$\theta \rightarrow 0^{\circ}$:

As $~\theta\rightarrow 0^{\circ}~$, the $t$-channel photon propagator 
has a kinematic pole, but  
both $W_L$ and GB amplitudes have the {\it same} pole structure, i.e.
$$
\begin{array}{ll}
(T[W_L]-T[{\rm GB}])_{\rm pole~term}
& =-ie^2\left[\displaystyle
\frac{[(3+\cos\theta )\kappa +2]\cos^2\theta}{(1-\cos\theta )\kappa} -
\frac{(3+\cos\theta )\kappa +2}{(1-\cos\theta )\kappa}\right]\\[0.45cm] 
& =-ie^2(1+\cos\theta )(3+\cos\theta +2\kappa^{-1}) = O(g^2)~~,
\end{array}
\eqno(A2)                      %(A2)
$$
which is finite.\footnote{
This conclusion can be directly generalized to other $t$ or $u$
channel processes.}~~ 
Hence, {\it the B-term, 
which is defined as the difference $~T[W_L]-T[{\rm GB}]~$,
is finite at $~\theta =0^{\circ}~$, and is of $~O(g^2)~$.}
This means that when $~\theta~$ is close to the $t$-channel photon pole,
the $B$-term is negligibly small relative to the GB-amplitude
so that (2.3b) is satisfied and
the ET works. More explicitly,  in the limit of 
$~\theta =0^{\circ}~$ (i.e. $~t=0~$), and from (A1a,b), 
the $W_L$ and GB amplitudes are
$$
\begin{array}{ll}
T[W_L] & = i\left[ 4(3-8{\rm c}_{\rm w}^2+8{\rm c}_{\rm w}^4)
\displaystyle\frac{p^2}{f_\pi^2}
+2e^2\left( 2+\displaystyle\frac{M_W^2}{p^2}\right)\frac{1}{1-c_0}\right]
 +O(g^2)~~,\\[0.45cm]
T[{\rm GB}] & = i\left[ 4(3-8{\rm c}_{\rm w}^2+8{\rm c}_{\rm w}^4)
\displaystyle\frac{p^2}{f_\pi^2}
+2e^2\left( 2+\displaystyle\frac{M_W^2}{p^2}\right)
\displaystyle\frac{1}{1-c_0}\right]
 +O(g^2)~~,\\[0.4cm]
T[W_L] & = T[{\rm GB}] + O(g^2) ~~,
\end{array}
\eqno(A3)                              %(A3)
$$
where $~~c_0\equiv \lim_{\theta\rightarrow 0}\cos\theta~$.~
Notice that in this case one cannot make 
the $M_W^2/t~$ expansion\footnote{This
expansion is {\it unnecessary} for the validity of the ET, 
cf. (2.3) and (2.3a,b).}~~ because ~$t$ vanishes identically. Since 
both $W_L$ and GB amplitudes have exactly the same kinematic
singularity and the $B$-term is much smaller than $T[{\rm GB}]$,
{\it the ET still holds} in this special kinematic region.
We also emphasize that {\it in the 
kinematic regions where $t$ or $u$ is not much larger than $M_W^2$,
the $t$-channel or $u$-channel 
internal gauge boson lines must be included
according to the precise formulation of the ET} 
[cf. (2.3) and (2.3a,b)].\footnote{
This does not imply, in any sense, a violation of the ET 
since the ET, cf. (2.3) and (2.3a,b), does not require
either $t \gg M_W$ or $u \gg M_W$.}~~

\noindent
(ii). In the limit of ~$\theta \rightarrow 180^{\circ}$: 

In the kinematic region with $~s,~t\gg M_W^2~$, (A1a) and (A1b) yield
$$
\begin{array}{ll}
T[W_L] & =i\displaystyle\left[ 2(1+\cos\theta )\frac{p^2}{f_\pi^2}
                   +O(g^2)\right] ~~,\\[0.4cm]
T[{\rm GB}] & = i\displaystyle\left[ 2(1+\cos\theta )\frac{p^2}{f_\pi^2}
                   +O(g^2)\right]  ~~,\\[0.55cm]
T[W_L] & = T[{\rm GB}] + O(g^2) ~~,
\end{array}                                     %(A4)
\eqno(A4)                                
$$
where the $~O(g^2)~$ term is the largest term 
we ignored which denotes the
order of the $B$-term [cf. (2.5)]; all other terms we ignored 
in $(A4)$ are of $~O(M_W^2/p^2)~$ or $~O(e^2)~$
which are smaller than $O(g^2)$  and thus will not affect 
the order of magnitude estimate of the $B$-term.
For $~s,~t\gg M_W^2~$, the $W_L$ and GB amplitudes are dominated
by the $p^2$-term in (A4), which is actually proportional to $u$ for
this process.  
When the scattering angle $\theta$ is close to $180^{\circ}$,
$u$ becomes small and 
thus this leading $p^2$ term is largely suppressed
so that both the $W_L$ and GB amplitudes can be 
as small as the $B$-term, i.e.
of  $O(g^2)$. In this case our condition (2.3a) is satisfied
while (2.3b) is not, which means that the EWSB sector
cannot be sensitively probed for this kinematic region. Since
the total cross section of this process 
is not dominated by this special kinematic region and
is mainly determined by the un-suppressed
leading large $p^2$-term, 
so {\it the kinematic dependence of the amplitude
will not affect the order of magnitude of the total cross section}.
Hence, {\it our application of the power counting analysis in Sec.~5 for 
computing the total event rates remains valid even
though we have ignored the angular dependence in
estimating the magnitude of the scattering amplitudes.}
Neglecting the angular dependence in the amplitude may 
cause a small difference in the event rate 
as compared to that from a precise calculation.
For the processes such as
$~W^\pm_L W^\pm_L\rightarrow W^\pm_L W^\pm_L ~$ and
$~W_L^+ W^-_L \rightarrow Z_LZ_L~$, the leading $p^2$-term
is proportional to $~s/f_\pi^2~$ with no angular dependence, so that
the angular integration causes {\it no difference} between our power
counting analysis and the exact calculation for the leading 
$p^2$-term contribution.\footnote{
The small difference (a factor of 1.4) in Fig.~1 
mainly comes from  neglecting  the tree level sub-leading terms 
in our order of magnitude estimate for the amplitudes.}~~
In the above example for $~W^+_L W^-_L\rightarrow W^+_L W^-_L ~$ channel
[cf. (A4)], the leading amplitude is proportional to $~-u/f_\pi^2~$.
When applying the power counting method, 
we ignore the $\theta$-dependence and estimate it as $~s/f_\pi^2~$.
In computing the total rate, we integrate out the scattering angle.
This generates a difference from the precise one:
$$
{ {\int_{-1}^{1} u^2 \, {\rm d}\! \cos \theta} \over  
{\int_{-1}^{1} s^2 \,   {\rm d}\! \cos \theta} } = \frac{1}{3} ~~,
$$
which, as expected, is only a factor of $3$ and does not affect our order
of magnitude estimates.

Finally, we make a precise numerical analysis on the equivalence between
the $W_L$ and the GB amplitudes to show how well the ET works in different
kinematic regions and its implication to probing the EWSB sector.
We use the full expressions (A1a,b) for $W_L$ and GB amplitudes 
as required by the ET, cf. (2.3) and (2.3a,b).
In Fig.~10a, we plot the ratio $~|B/g^2|~$ for scattering angle
$~\theta = 2^{\circ}, 10^{\circ}, 45^{\circ}, 
90^{\circ}, 100^{\circ}, 120^{\circ}, 135^{\circ}, 
150^{\circ}, 180^{\circ}~$.
It shows that the LNI $B$-term is {\it always of $~O(g^2)~$
in the whole kinematic region,} and thus is irrelevant to
the EWSB sector,  in accordance
 with our general physical analysis in Sec.~2. 
Hence, to have a sensitive probe of the EWSB mechanism, condition
(2.3b) or (2.4) must be satisfied.
Fig.~10b shows that
for $~0^{\circ}\leq \theta \leq 100^{\circ}~$, 
the ratio $~~|B/T[W_L]|\leq 10\% ~~$ when
$~M_{WW}\geq 500$\,GeV. 
For $~\theta \geq 120^{\circ}~$, this ratio becomes
large and reaches $~O(1)~$ when $~\theta~$ is close to $180^{\circ}$.
This is because the kinematic factor
$(1+\cos\theta )$, associated with the leading $p^2$ term [cf. (A4)],
becomes small. This, however, will
not alter the conclusion that for $4W_L$-scattering the total cross
section from $T[{\rm GB}]$ is much larger than that from the $B$-term
as $~M_{WW}\geq 500$\,GeV.\footnote{
In practice, this is even more true after applying the necessary kinematic
cuts to require the final state $W$-bosons to be in the central rapidity 
region of the detector for detecting the signal event~\cite{wwlhc}, so that
the $\theta$ angle cannot be close to either 
$180^{\circ}$ or $0^{\circ}$.}~~  
Note that in Fig.~10b, for  $~\theta \leq 10^{\circ}~$, i.e. close to the
$t$-channel photon pole,
 the ratio $~|B/T[W_L]|~$ is below $~1\% ~$ and thus the ET holds
very well.
In Fig.~10c, we plot both the $W_L$ and GB amplitudes for
$~\theta = 10^{\circ},45^{\circ}, 100^{\circ}, 
150^{\circ}~$. The solid lines denote the 
complete $W_L$ amplitude and the dotted lines denote the GB amplitude.
We find that when $~\theta \leq 100^{\circ}~$, the GB amplitude is almost
indistinguishable from the $W_L$ amplitude. For $~\theta =150^{\circ}~$,
the $W_L$ amplitude is of the same order as the $B$-term, i.e. of 
$~O(g^2)~$, when $~M_{WW}< 1$\,TeV. In this case the $W_L$ or GB
amplitude is
too small and the strongly coupled EWSB sector cannot be sensitively
probed. As the energy $E$ increases, we see that the $W_L$ and GB
amplitudes rapidly dominate over the $B$-term and agree 
better and better even for large scattering angles.
   Finally, using the effective-$W$ method~\cite{effective-W},
we compare the LHC production rates in Figs.~10d and 10e for 
the invariant mass ($M_{WW}$) and the polar angle ($\cos\theta$) 
distributions, repectively. 
To avoid the $t$-channel photon singularity (at
$~\theta =0^{\circ}~$) in the phase space integration, 
we add an angular cut $~-0.1\leq \cos\theta  \leq 0.8~$
(i.e., $~36.9^{\circ}\leq \theta\leq 180^{\circ}~$).
Fig.~10d shows that the total cross sections computed from the
$W_L$ and 
the $GB$ amplitudes [cf. eq.~$(A1)$] indeed agree with each other
very well. From Fig.~10e, we see that the difference clearly appears 
only for the region of large scattering angle (i.e., $\cos\theta <-0.6~$ or
$~\theta > 127^{\circ}~$) where both the leading 
$W_L$ and GB amplitudes are suppressed by the kinematic factor 
$~1+\cos\theta~$ and thus the event rates are too low to be
sensitive to the EWSB sector. Hence, the difference from the large
$\theta$ region has only {\it negligible} effects on the total 
cross sections, as clearly shown in Fig.~10d. This also agrees with
our conclusion from Figs.~10a-c.
We have also made the comparison with symmetric angular cuts (such as
$~|\cos\theta | \leq 0.8~$) and found similar good agreement to
that in Fig.~10d. This is clear since in the large $\theta$ region 
the event rates become much lower and are close to their 
difference, i.e., of $~O(|R_B|)~$.

The above conclusions hold for the tree level contributions from
the lowest order operators in 
$~{\cal L}_{\rm G} +{\cal L}^{(2)}+{\cal L}_{\rm F}~$, cf. (3.1.2).
However, independent of the kinematic region considered,
not all the contributions from the NLO effective operators
can dominate the $B$-term and satisfy the condition (2.3b) [or (2.4)]. 
This is why the condition (2.3b) [or (2.4)] can serve as the 
criterion for classifying the sensitivities of these NLO operators 
in probing the EWSB sector for a given scattering process.

We therefore conclude that 
for the process considered here {\it the B-term, as 
defined in (2.1), can be at most of $O(g^2)$ 
for all kinematic regions} (cf. Fig.~10a), 
and is insensitive to the EWSB 
mechanism, in accordance with our general analysis in Sec.~2.
{\it When ~$t$ or ~$u$
is not large, the ~$t$- or ~$u$-channel 
internal  lines must be included}.
We find that in certain kinematic region {\it even  $t$ (or $u$) is close to 
zero, the ET still works well} [cf. Eq.~(A3) and Fig.~10b].
This is because
the validity of the ET does not require either
$~t \gg M_W^2$ or $~u \gg M_W^2$ [cf. (2.3) and (2.3a,b)]. 
 For some scattering processes,
there may be special kinematic regions in which the GB and the $W_L$
amplitudes are largely suppressed\footnote{This large suppression can also 
arise from the polarization effects of the in/out states.}~~
 so that the EWSB sector cannot be sensitively  probed 
in these special kinematic regions (cf. Figs.~10b,c and 10e). 
But, as shown in this work, 
measuring the total event rates from these processes
can still be used to sensitively probe the EWSB sector 
(cf. Fig.~10d and Figs.~2-9).

\vspace{1cm}
%\newpage
\noindent
{\bf  Appendix B. Electroweak Power Counting Rule for Linearly Realized
Effective Lagrangians}
\vspace{0.3cm}

For completeness and 
for other possible applications, 
we also generalize Weinberg's power counting method
to another popular effective Lagrangian formalism~\cite{linear} for
the weakly coupled EWSB sector, which is usually called as the 
{\it decoupling scenario.} In this formalism, the lowest order Lagrangian
is just the linear SM with a relatively light Higgs boson and all higher
order new physics effective operators must have dimensions larger than
$4$ and are suppressed by the effective cutoff scale $~\Lambda~$.
Even if a relatively light scalar
is found in future colliders, it remains important to know
whether such a scalar particle trivially serves 
as the SM Higgs boson or orginates from a more complicated dynamics.
For instance, the possible new physics effects 
parametrized in $(B1)$ should be probed 
in details for discriminating the SM Higgs boson
from the non-SM Higgs boson at the LHC and the future linear colliders.  

Following Ref.~\cite{linear}, we can generally write
the $SU(2)_L\otimes U(1)_Y$ linear effective Lagrangian as follows
$$
{\cal L}_{\rm eff}^{\rm linear} ~=~\displaystyle 
{\cal L}_{\rm SM}+\sum_{n} \frac{\ell_n}{\Lambda^{d_n-4}}{\cal O}_n
\eqno(B1)   %(B1)
$$
where $~d_n~(\geq 5)~$ is the dimension of the effective operator 
$~{\cal O}_n~$.
In $(B1)$, the lowest order Lagrangian $~{\cal L}_{\rm SM}~$ 
is just the SM Lagrangian with a relatively light
Higgs boson. The interesting high energy region considered here is
$$
\begin{array}{ccccc}
M_W, m_H, m_t & \ll & E & < & \Lambda 
\end{array}
\eqno(B2)
$$
in which $~m_H = \sqrt{2\lambda}f_\pi~$ denotes the Higgs boson mass.

Since the field content of $~{\cal L}_{\rm eff}^{\rm linear}~$ in
$(B1)$ is the same as that of the SM and the masses of 
all the known fields are much lower than the typical high
energy scale $~E~$ under consideration [cf. eq.~($B2$)], it is clear
that all the essential features of Weinberg's counting method hold
for this linear case.
Following the same reasoning as done in Sec.~3.2 [cf. 
eqs.~(3.2.1)-(3.2.4)], 
we find that that, for a given $S$-matrix element $~T~$, 
the counting formula for the linear case is very similar to Eq.~(3.2.5):
$$
\begin{array}{l}
T= c_T f_\pi^{D_T}\displaystyle 
\left(\frac{f_\pi}{\Lambda}\right)^{N_{\cal O}}
\left(\frac{E}{f_\pi}\right)^{D_{E0}}
\left(\frac{E}{\Lambda_0}\right)^{D_{EL}}
\left(\frac{M_W}{E}\right)^{e_v} H(\ln E/\mu)~~,\\[0.5cm]
N_{\cal O}=\displaystyle\sum_n (d_n -4)~,~~
%D_{E0}=2+\displaystyle\sum_n {\cal V}_n(d_n-2)+(i_F+\frac{1}{2}e_F)~, ~~ 
D_{E0}=2+\displaystyle\sum_n {\cal V}_n\left(d_n+\frac{1}{2}f_n-2\right)~, 
~~ D_{EL}=2L~,~~ \Lambda_0 =4\pi f_\pi ~, \\
\end{array}
\eqno(B3)                                                  %(B3)
$$ 
where the only difference is that $~N_{\cal O}~$ is now determined
by the canonical dimensional counting in $(B1)$ instead of the naive
dimensional analysis (NDA)~\cite{georgi} for the
non-decoupling scenario discussed in Sec.~3.

For $~{\cal L}_{\rm SM}~$ (i.e., for $\ell_n=0$ in $(B1)$~),
the counting for $~D_E~$ defined in $(B3)$ or (3.2.4) 
can be further simplified since we know that the 
total $E$-power dependence of the SM contributions
 will not increase as the loop number $L$ increases 
because of the perturbative unitarity of the light Higgs SM. 
We shall show that, due to the renormalizable feature of 
$~{\cal L}_{\rm SM}~$, the $D_{EL}$ term, $~2L~$, 
in $(B3)$ will be cancelled by a counter term 
from the vertex-contribution in the $D_{E0}$ term. 

In $~{\cal L}_{\rm SM}~$, there are 
only  $3$-point and $4$-point vertices. 
Due to the renormalizability of the
SM, all the $4$-point vertices do not contain 
partial derivatives, while each $3$-point vertex may contain at most one 
partial derivative. So, we have $d_n=0 \, ~{\rm or}~ \, 1~$. Thus,
$$
\sum_n {\cal V}_n d_n ={\cal V}_d~~,~~~ 
{\cal V}_d \equiv {\cal V}_3^{VVV}+{\cal V}_3^{ssV}+{\cal V}_3^{c\bar{c}V}~~,
\eqno(B4)
$$                                    %(B4)
where $~~{\cal V}_d~~$ is the number of all vertices containing one partial
derivative and $~~{\cal V}_n^{\chi_1\cdots\chi_n}~~$ is the total number
of $n$-point  vertices of type $~~\chi_1$-$\chi_2$-$\cdots$-$\chi_n~~$
($\chi$ denotes any possible field in the theory). 
The symbol $~s~$ denotes scalar fields 
( Higgs or GB ), $~c~(\bar{c})~$ denotes (anti-)ghost field and 
$~F~(\bar{F})~$ denotes (anti-)fermion field. 
Furthermore, in $~{\cal L}_{\rm SM}~$,
$$
\begin{array}{l}
{\cal V} = \sum_n {\cal V}_n = {\cal V}_3 + {\cal V}_4  ~~,~~~
\\[0.25cm]
{\cal V}_3 \equiv {\cal V}_d + {\cal V}_F + \bar{\cal V}_3 ~~,~~~
{\cal V}_F\equiv {\cal V}_3^{sF\bar{F}}+{\cal V}_3^{VF\bar{F}} ~~,~~~
\bar{\cal V}_3 \equiv 
  {\cal V}_3^{sVV}+{\cal V}_3^{sc\bar{c}}+{\cal V}_3^{sss}~~.
\\[0.25cm]
{\cal V}_4 \equiv {\cal V}_4^{ssss}+{\cal V}_4^{ssVV}+{\cal V}_4^{VVVV} ~~,
\end{array}
\eqno(B5)                                       %(B5)
$$
Substituting $(B4)$, $(B5)$ and the SM relation $~{\cal V}_F = 2i_F +e_F~$
into $(B3)$, we obtain
$$
\begin{array}{ll}
D_E^{\rm SM}~=~ D_{E0}^{\rm SM}+D_{EL}^{\rm SM} 
& =~ 2L+2 -2{\cal V} + {\cal V}_d +{\cal V}_F -e_v \\
    & =~ 2L+2 - ({\cal V}_d +{\cal V}_F +2\bar{\cal V}_3 +2{\cal V}_4)-e_v~~,
\end{array}
\eqno(B6) 
$$                                              %(B6)                   
which, with the aid of another SM relation 
$$
\begin{array}{l}
3{\cal V}_3 + 4{\cal V}_4 = e + 2i  ~~~,~~~~{\rm or},\\[0.25cm]
{\cal V}_d + {\cal V}_F = 4{\cal V}-\bar{\cal V}_3 -2i-e
= 2-2L + 2{\cal V}-\bar{\cal V}_3-e~~~,
\end{array}
\eqno(B7)                                       %(B7)
$$                                              
can be further simplified as
$$
D_E^{\rm SM}~  =~ 4 -e -e_v - \bar{\cal V}_3 ~~,
\eqno(B8)                                           %(B8)
$$
where $~~\bar{\cal V}_3 \equiv 
  {\cal V}_3^{sVV}+{\cal V}_3^{sc\bar{c}}+{\cal V}_3^{sss}~~$.
Note that the loop-dependence term  $~2L~$ is indeed canceled by 
the counter term from the vertex-contribution [cf. $(B6)$ and $(B7)$] 
as expected. This is 
the unique feature of the renormalizable SM and this feature 
is absent in the EWCL with the derivative expansion which has been fully
studied in Sec.~3. 
In summary, for $~{\cal L}_{\rm SM}~$,
a Feynman diagram with its external lines fixed 
can have the leading energy dependence 
if it does not contains the trilinear vertices 
$~s$-$V$-$V$, $~s$-$c$-$\bar c$~ and
~$s$-$s$-$s$~, and  the $v_\mu$-factor. Equivalently, 
$(B6)$ shows in another way that at a given $L$-loop level the leading energy 
behavior of a diagram corresponds to the minimal $~~({\cal V}_d +{\cal V}_F 
+2\bar{\cal V}_3 +2{\cal V}_4)~~$ and the vanishing $~e_v~$. 

The NLO linear operators $~{\cal O}_n~$ in $(B1)$
have been fully compiled in Ref.~\cite{linear}. Here are a few typical
dimension-$6$ effective operators:
$$
\begin{array}{ll}
{\cal O}_W ~=~ -i4
{\rm Tr}({\bf W}_\mu~^\nu{\bf W}_\nu~^\rho{\bf W}_\rho~^\mu )
~~,~~~~~ & 
{\cal O}_{\partial\phi}~=~ 
\displaystyle{1\over 2}\partial_\mu (\phi^{\dagger}\phi )
                       \partial^\mu (\phi^{\dagger}\phi ) ~~,\\[0.38cm]
{\cal O}_{qq}^{(1,1)}~=~ 
\displaystyle{1\over 2}(\bar{q}\gamma_\mu q)(\bar{q}\gamma^\mu q)
~~,~~~~~ &
{\cal O}_{qq}^{(1,3)}~=
~\displaystyle{1\over 2}(\bar{q}\gamma_\mu\tau^aq)
                        (\bar{q}\gamma^\mu\tau^aq)
~~,\\[0.38cm]
{\cal O}_{\phi W}~=~\displaystyle (\phi^{\dagger}\phi )
{\rm Tr}({\bf W}_{\mu\nu}{\bf W}^{\mu\nu})~~,~~~~~ &
{\cal O}_{\phi B} ~=~ (\phi^{\dagger}{\bf W}_{\mu\nu}\phi )B^{\mu\nu}~~,
\\[0.38cm]
{\cal O}_{\phi}^{(1)}~=~\displaystyle{1\over 2}
(\phi^{\dagger}\phi )(D_\mu\phi^{\dagger}D^{\mu}\phi )~~,~~~~~ &
{\cal O}_{\phi}^{(3)}
~=~(\phi^{\dagger}D^\mu\phi )[(D_\mu\phi)^{\dagger}\phi]~~,\\[0.38cm]
\end{array}
\eqno(B9)
$$
where $~\phi~$ denotes the Higgs doublet which conatins the 
linearly realized Higgs field ($H$) and three would-be Goldstone bosons
($\pi^\pm ,~\pi^0$).

It is straightforward to apply our power counting rule $(B3)$ [and $(B8)$]
for estimating various scattering amplitudes contributed by $(B1)$.
Some typical examples are in order. 
First, we count the model-independent
contributions from $~{\cal L}_{\rm SM}~$ to some $~2\rightarrow 2$~
scattering processes:
$$
\begin{array}{ll}
T[V_T^{a_1}V_T^{a_2}\rightarrow V_T^{a_3}V_T^{a_4}]\displaystyle
~=~O(g^2)+O\left({g^4\over 16\pi^2}\right)~, &\\[0.42cm]
T[\pi^{a_1}\pi^{a_2}\rightarrow \pi^{a_3}\pi^{a_4}]\displaystyle
~=~ O(g^2,\lambda )+O\left({g^4,\lambda^2\over 16\pi^2}\right)~, &
B^{(0)}_0~=~O\left((g^2,\lambda )\frac{M_W^2}{E^2}\right)~,\\[0.35cm]
T[V_T^{a_1}\pi^{a_2}\rightarrow \pi^{a_3}\pi^{a_4}]\displaystyle
~=~O\left((g^2,\lambda )\frac{M_W}{E}\right)
   +O\left(\frac{g^4}{16\pi^2}\right)~~,~~~ & 
B_0^{(1)}~=~O\left(g^2\frac{M_W}{E}\right)~~.
\end{array}
\eqno(B10)
$$
Second, we count the NLO model-dependent contributions
from $(B9)$ to some tree-level 
high energy processes at the $~O(1/\Lambda^2)~$:
$$
\begin{array}{l}
T_1[V_T^{a_1}V_T^{a_2}\rightarrow V_T^{a_3}V_T^{a_4}]({\cal O}_W)
~=~\displaystyle O\left(\ell_W \frac{gE^2}{\Lambda^2}\right)~~,\\[0.4cm]
T_1[\pi^{a_1}\pi^{a_2}\rightarrow \pi^{a_3}\pi^{a_4}]({\cal O}_W)
\displaystyle ~=~ 0~~,~~~~~
B^{(0)}_1~=~O\left(\ell_W\frac{gM_W^2}{\Lambda^2}\right)~~,\\[0.42cm]
T_1[\pi^{a_1}\pi^{a_2}\rightarrow HH]({\cal O}_{\partial\phi})~=~
\displaystyle
O\left(\ell_{\partial\phi}\frac{E^2}{\Lambda^2}\right)~~,~~~~~
B^{(0)}_1~=~ O\left(\frac{M_W^2}{\Lambda^2}\right)~~,\\[0.42cm]
\displaystyle
T_1[q\bar{q}\rightarrow q\bar{q}]({\cal O}_{qq}^{(1,1)}) ~=~
O\left(\ell^{(1,1)}_{qq}\frac{E^2}{\Lambda^2}\right)~~.\\[0.4cm]
\end{array}
\eqno(B11)
$$
The above examples illustrate, 
in the linear effective Lagrangian formalism, 
how to conveniently apply our power counting rule $(B3)$ 
to determine the high energy behavior of any given
amplitude and estimate its order of magnitude.

%\newpage
\vspace{1.2cm}
%%%%%%%%%%%%%%% RRRRRRRRRRRRRRRRRRRRRRRRRRRR %%%%%%%%%%%%%%%%%%%%%%

\newpage
\noindent
{\bf Table Captions}
\vspace{0.2cm}

\noindent
{\bf Table 1.}  Estimates of amplitudes 
for $~W^\pm W^\pm \rightarrow W^\pm W^\pm ~$
scattering.\\[0.35cm]
{\bf Table 1a.} Model-independent contributions from 
$~{\cal L}_G +{\cal L}_F +{\cal L}^{(2)}~$.\\[0.20cm]
{\bf Table 1b.} Model-dependent contributions from the 
next-to-leading order operators.

\noindent
{\bf Table 2.} Order estimates of $B$-terms for 
$~W^\pm W^\pm \rightarrow W^\pm W^\pm ~$ scattering.\\[0.35cm]
{\bf Table 2a.}  Model-independent contributions.\\[0.20cm]
{\bf Table 2b.}  Relevant operators for model-dependent 
                 contributions.$^{(a)}$

\noindent
{\bf Table 3.}~ Estimates of amplitudes for $~q\bar{q}^{(\prime )}   
      \rightarrow V^aV^b~$: Model-independent contributions.\\[0.35cm]
{\bf Table 3a.} ~For $~q\bar{q}^{(\prime )}\rightarrow W^{+}W^{-}~,
                ~W^{\pm}Z ~$.\\[0.20cm] 
{\bf Table 3b.} ~For $~q\bar{q}\rightarrow ZZ~$.

\noindent
{\bf Table 4.}~ 
 Estimates of amplitudes for $~q\bar{q}^{(\prime )}\rightarrow V^aV^b~$: 
 Model-dependent contributions.$^{(a)}$\\[0.35cm] 
{\bf Table 4a.} ~For $~q\bar{q}\rightarrow W^{+}W^{-}~ $.\\[0.20cm]
{\bf Table 4b.} ~For $~q\bar{q}^{\prime}\rightarrow W^{\pm}Z~ $.

\noindent
{\bf Table 5.} Global classification of sensitivities 
to probing direct and indirect EWSB information from effective 
operators at the level of $S$-matrix elements (I).$^{(a)}$\\
Notes:\\ {\footnotesize
$^{(a)}$ The contributions from
${\cal L}_{1,2,13}$ are {\it always} associated 
with a factor of $\sin^2\theta_W$, unless specified otherwise. 
Also, for contributions to the $B$-term in 
a given $V_L$-amplitude, we list them
separately with the $B$-term specified.\\ 
$^{(b)}$ MI $=$ model-independent, MD $=$ model-dependent.\\
$^{(c)}$ There is no contribution when all the external lines are 
electrically neutral.\\
$^{(d)}$ $B_0^{(1)}\simeq T_0[2\pi ,v,V_T]~(\neq T_0[2\pi^0 ,v^0,Z_T])$,~
$B_0^{(3)}\simeq T_0[v,3V_T]~(\neq T_0[v^0,3Z_T])$.\\
$^{(e)}$  $T_1[2V_L,2V_T]=T_1[2Z_L,2W_T],~T_1[2W_L,2Z_T]$, ~or
$~T_1[Z_L,W_L,Z_T,W_T]$.\\
$^{(f)}$ ${\cal L}_2$ only contributes to $T_1[2\pi^\pm ,\pi^0,v^0]$ and
$T_1[2\pi^0,\pi^\pm ,v^\pm ]$ at this order; 
${\cal L}_{6,7}$ do not contribute
to $T_1[3\pi^\pm ,v^\pm ]$.\\
$^{(g)}$  ${\cal L}_{10}$ contributes only 
to $T_1[\cdots ]$ with all the external
lines being electrically neutral.\\
$^{(h)}$ $B_0^{(2)}$ is dominated by $~T_0[2V_T,2v]~$ since 
$~T_0[\pi,2V_T,v]~$ contains a suppressing factor $\sin^2\theta_W$ 
as can be deduced from $~T_0[\pi,3V_T]~$ (cf. Table~1a) 
times the factor $~v^\mu =O\left(\frac{M_W}{E}\right)~$.\\
$^{(i)}$ Here, $T_1[2W_L,2W_T]$ contains a coupling 
$e^4=g^4\sin^4\theta_W$.\\
$^{(j)}$ ${\cal L}_2$ only contributes to $T_1[3\pi^\pm ,v^\pm ]$.\\
$^{(k)}$ ${\cal L}_{1,13}$ do not contribute to 
$T_1[2\pi^\pm ,2v^\pm ]$.  }

\noindent
{\bf Table 6.} 
Global classification of sensitivities to probing direct and indirect EWSB\\ 
information from effective operators at the level of $S$-matrix elements (II). 
$^{(a)}$

%\newpage
\vspace{1.8cm}
\noindent
{\bf Figure Captions}
\vspace{0.2cm}

\noindent
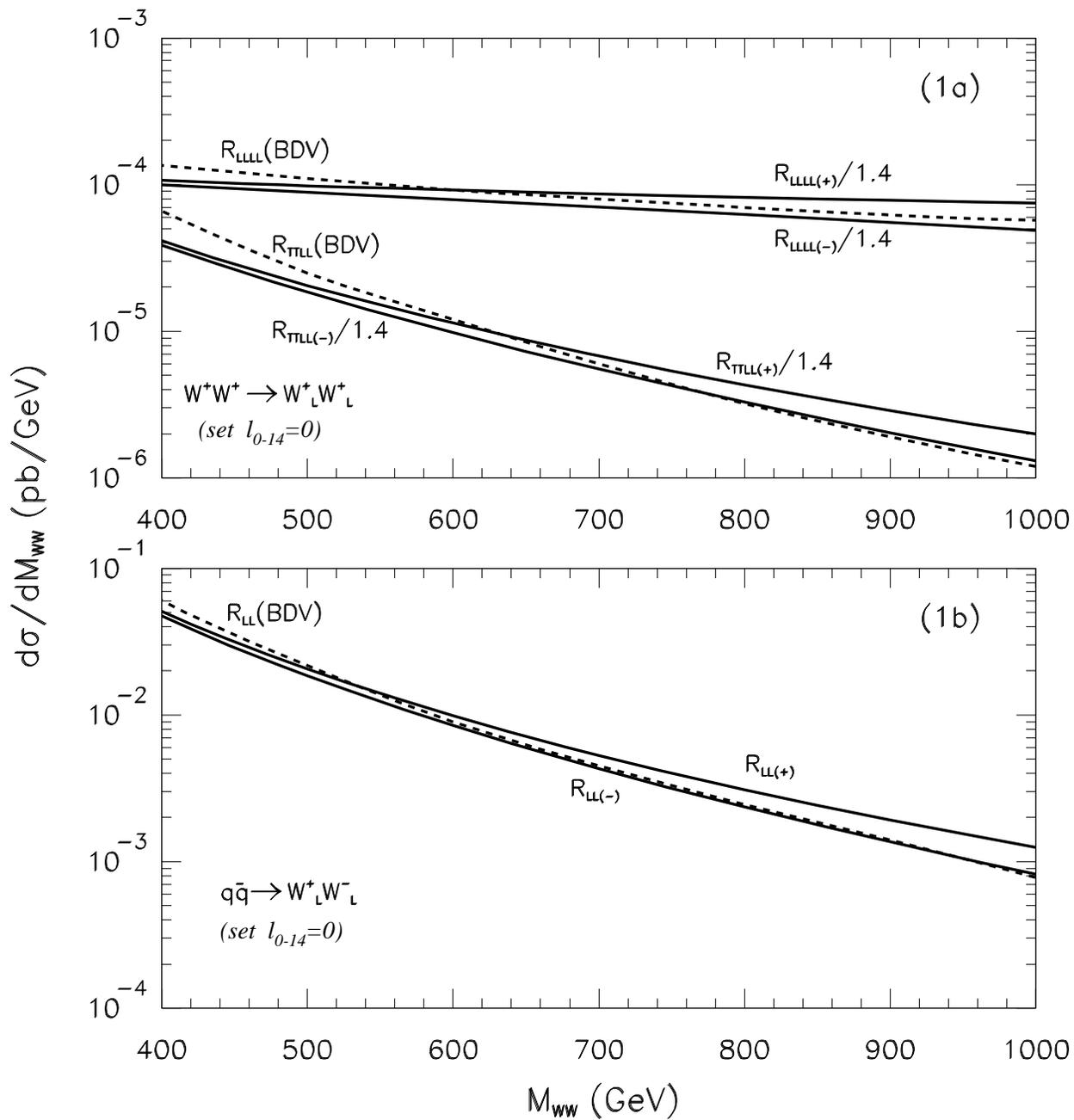
{\bf Fig.~1.}  Comparison of the power counting predictions with the
corresponding ones in Fig.~8 of Ref.~\cite{BDV} up to one-loop for 
a pp collider with
$\sqrt{S}= 40$\,TeV. The solid lines are given by our power counting 
analysis; the dashed lines are from Ref.~\cite{BDV}. [The meanings of the 
rates $R_{\alpha\beta\gamma\delta}$'s and $R_{\alpha\beta}$'s
 are defined in eq.~(5.1a,b) and below.]
\\[0.20cm]
{\bf (1a).}  $W^+W^+\rightarrow W^+_L W^+_L$.\\
{\bf (1b).}  $q\bar{q}^{\prime}\rightarrow W^+_L W^-_L$.

\noindent
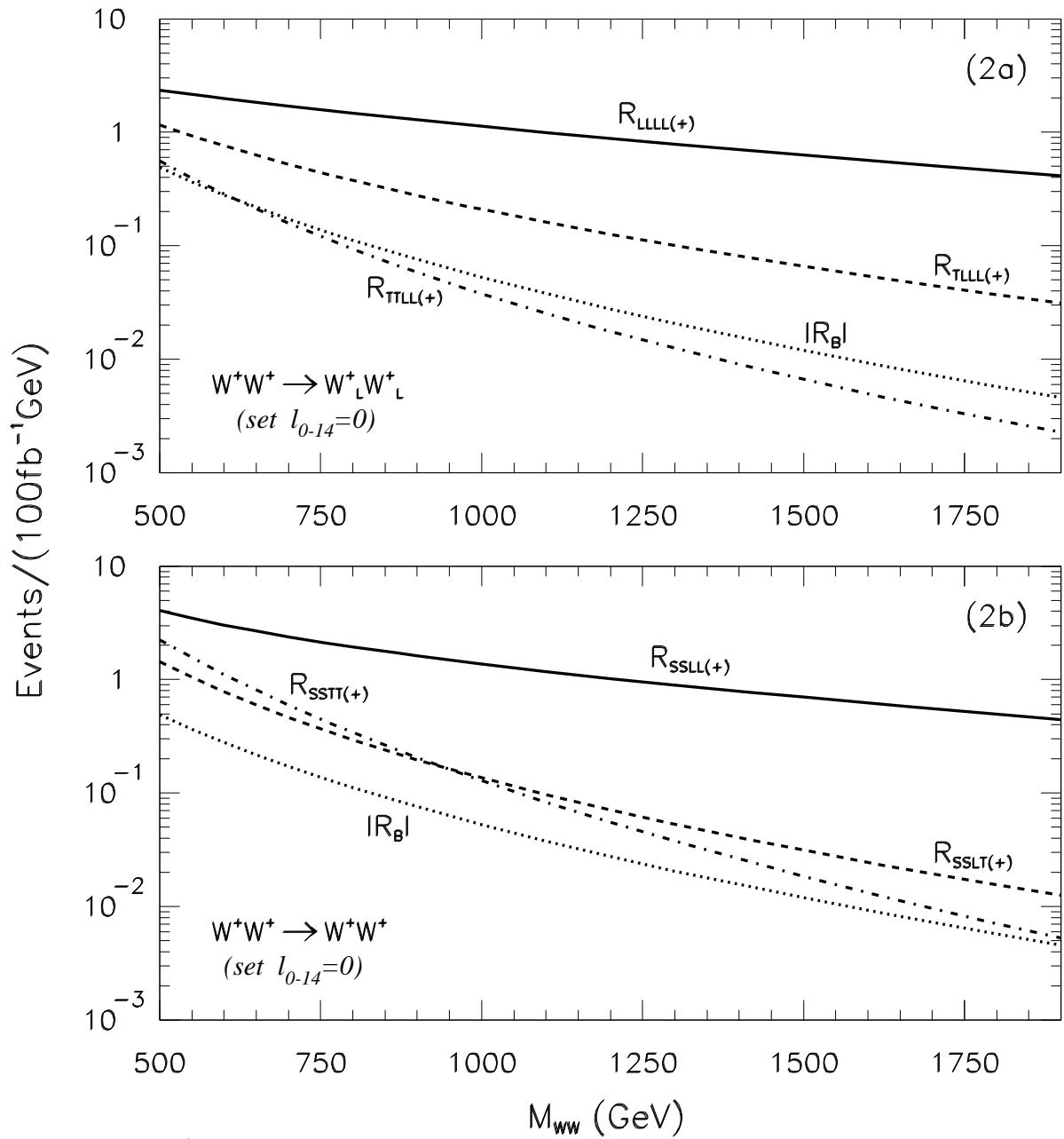
{\bf Fig.~2.}\\ 
{\bf (2a).} Comparison of the $W^+_LW^+_L$ production rates up to 
one-loop (for $\ell_{0-14}=0$)
with $W^+_LW^+_L$, $W^+_LW^+_T$ and $W^+_TW^+_T$ initial states,
at the  $14$\,TeV LHC.\\
{\bf (2b).} Comparison of the production rates for
different final-state polarizations up to one-loop (for $\ell_{0-14}=0$)
after summing over the polarizations of the initial states, at the 
$14$\,TeV LHC.

\noindent
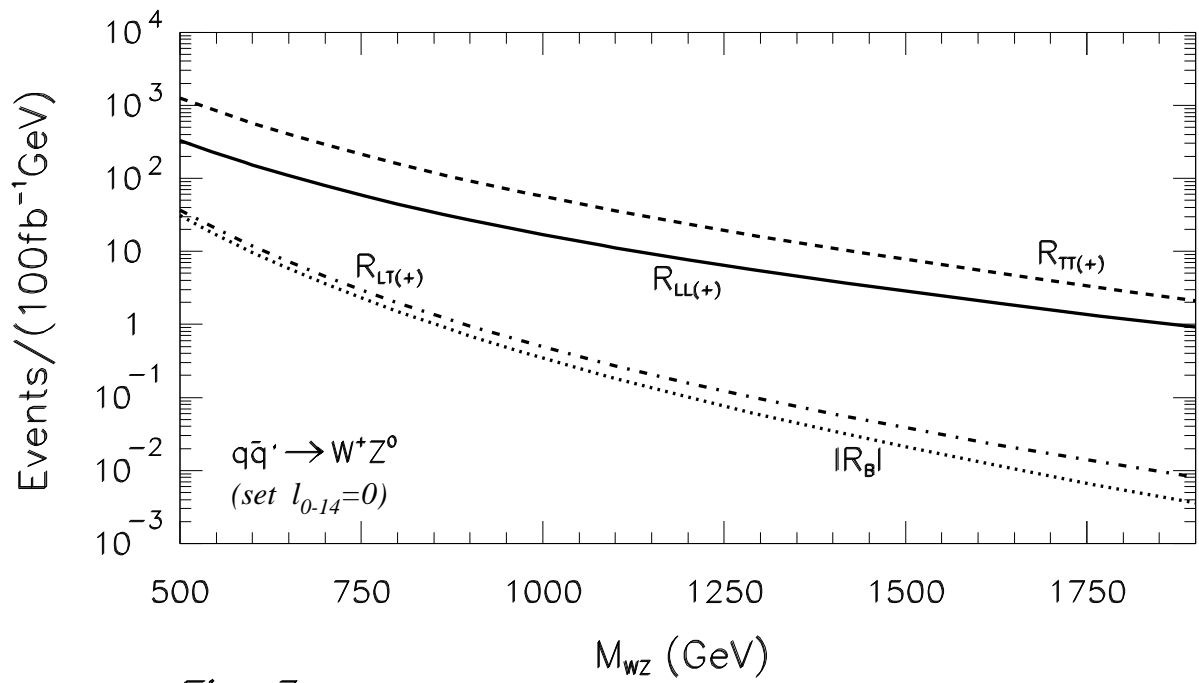
{\bf Fig.~3.} Comparison of the production rates for different
final-state polarizations up to one-loop (for $\ell_{0-14}=0$)
via $q\bar{q}^{\prime}\rightarrow W^+ Z^0$ at the  $14$\,TeV LHC.

\noindent
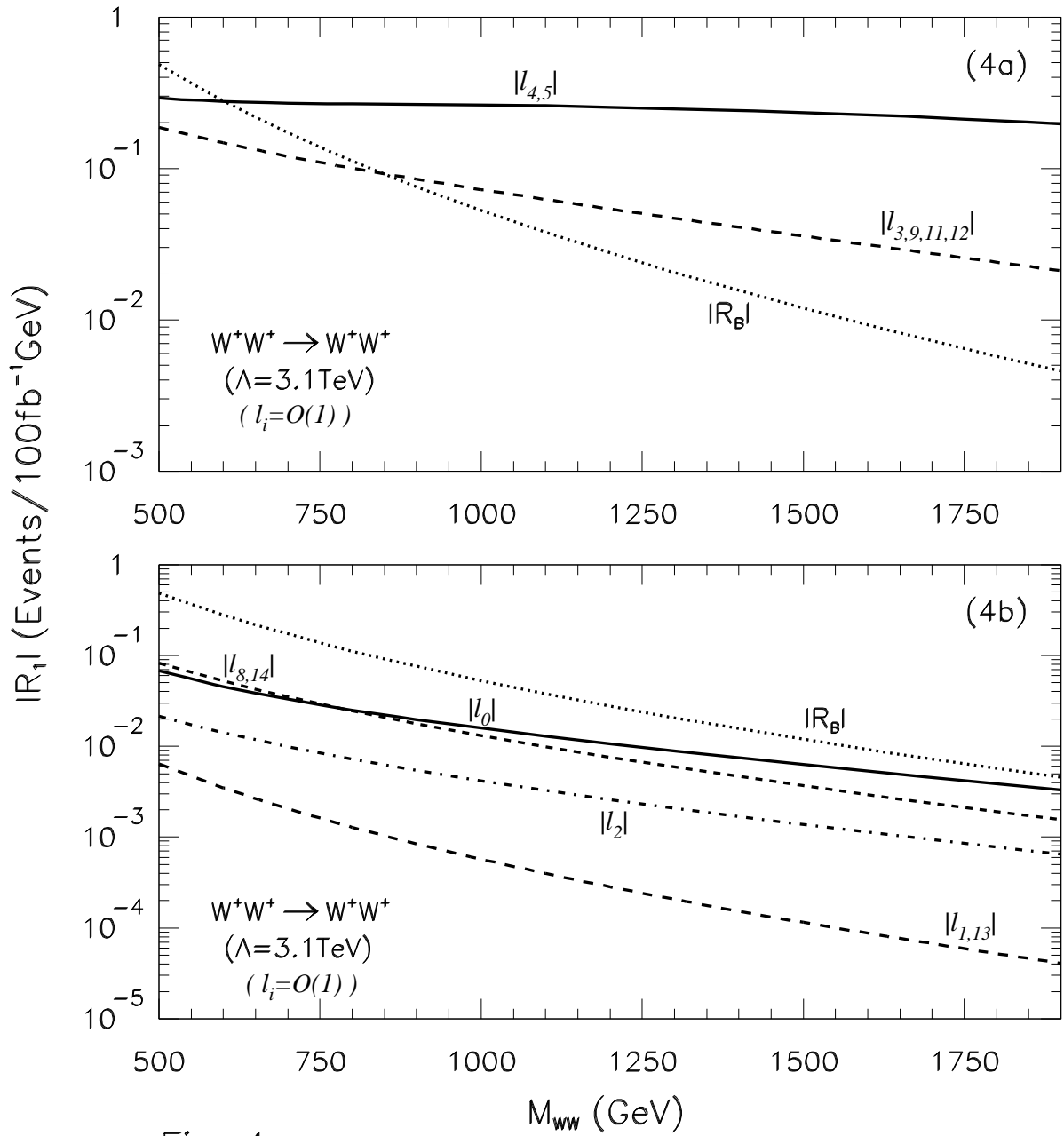
{\bf Fig.~4.} Sensitivities of the operators ${\cal L}^{(2)\prime}$ and 
${\cal L}_{1\sim 14}$ at the $14$ TeV LHC with $\Lambda=3.1~$TeV. 
The coefficients $\ell_n$'s are
taken to be of $O(1)$,\\[0.20cm]
{\bf (4a).} For operators ${\cal L}_{3,4,5,9,11,12}~$.\\
{\bf (4b).} For operators 
${\cal L}^{(2)\prime}$ and ${\cal L}_{1,2,8,13,14}~$.

\noindent
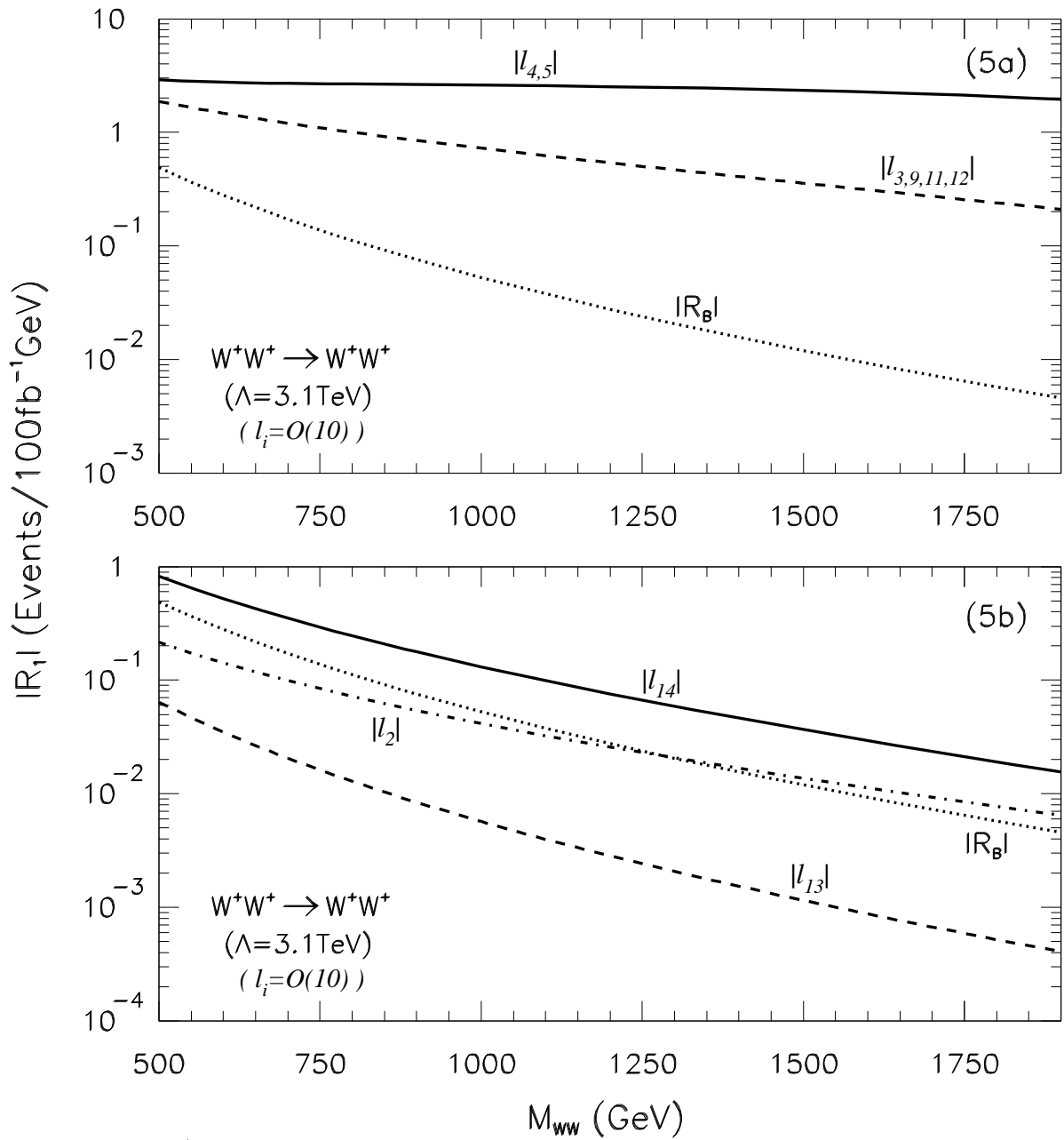
{\bf Fig.~5.}  Same as Fig.~4, but the coefficients $\ell_n$'s are 
taken to be of $O(10)$ except $\ell_{0,1,8}$ which are already 
constrained by low energy data to be of $O(1)$.\\[0.20cm]
{\bf (5a).} For operators ${\cal L}_{3,4,5,9,11,12}~$.\\
{\bf (5b).} For operators ${\cal L}_{2,13,14}~$.

\noindent
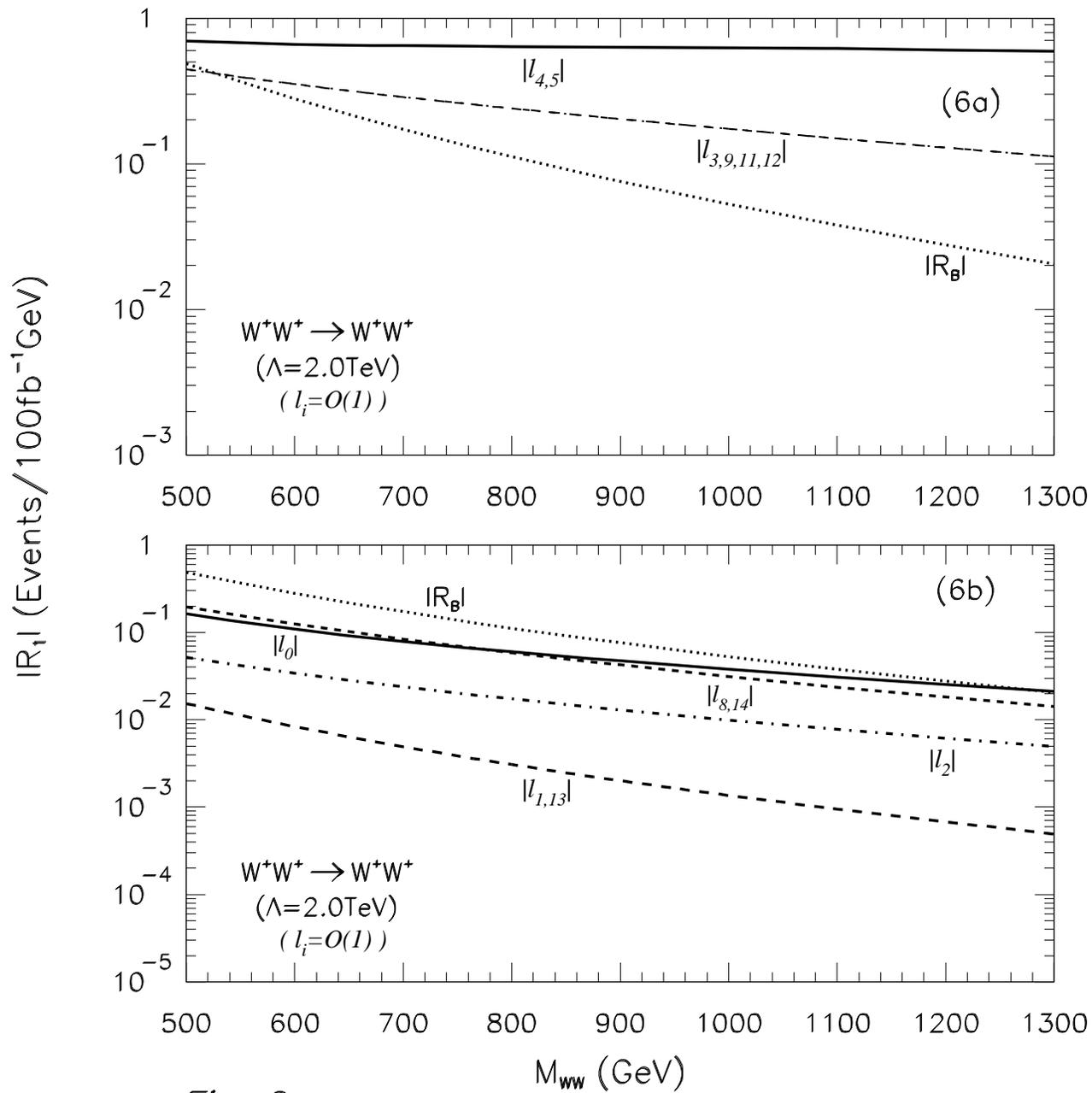
{\bf Fig.~6} Same as Fig.~4, but with $\Lambda=2.0~$TeV~.

\noindent
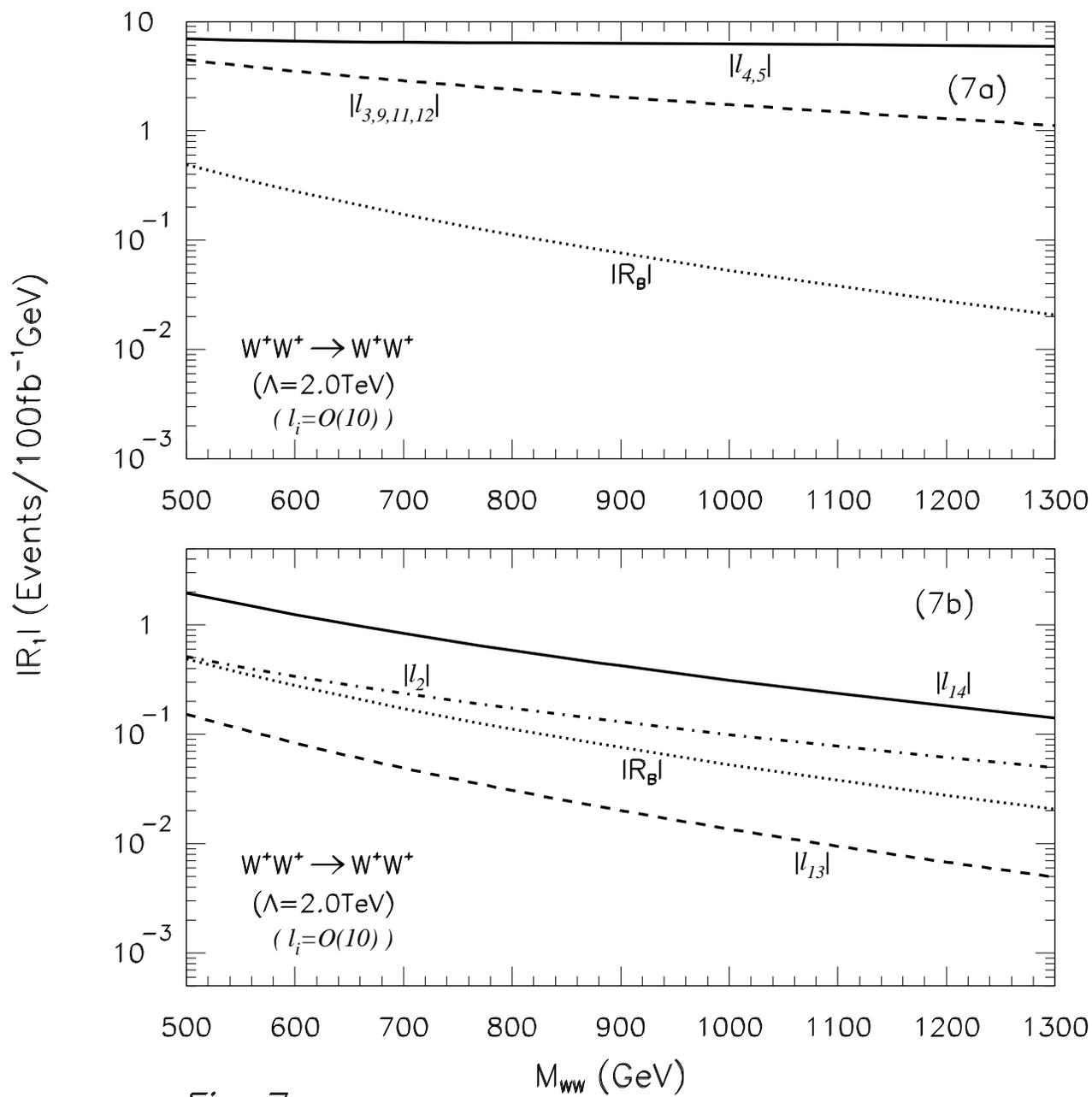
{\bf Fig.~7.} Same as Fig.~5, but with $\Lambda=2.0~$TeV~.

\noindent
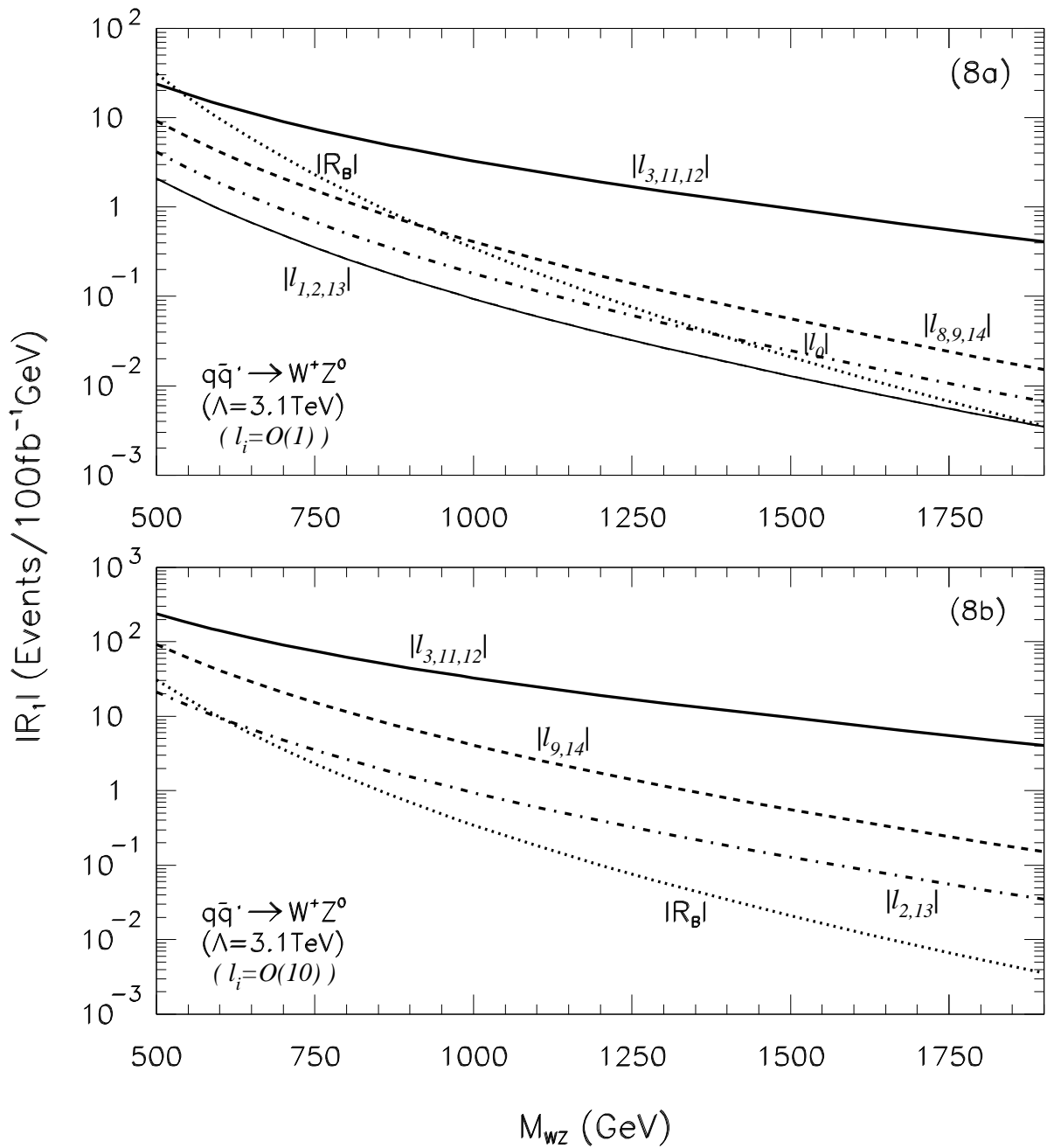
{\bf Fig.~8.} Sensitivities of the operators ${\cal L}^{(2)\prime}$ and
${\cal L}_{1-14}$ in $q\bar{q}^{\prime}\rightarrow W^+ Z^0$ at the 
$14~$TeV LHC with $\Lambda=3.1~$TeV~.
\\[0.20cm]
{\bf (8a).}  The coefficients $\ell_n$~'s are taken to be of $O(1)$.\\
{\bf (8b).}  The coefficients $\ell_n$~'s are taken to be of $O(10)$
except $\ell_{0,1,8}$ which are already 
constrained by low energy data to be of $O(1)$.

\noindent
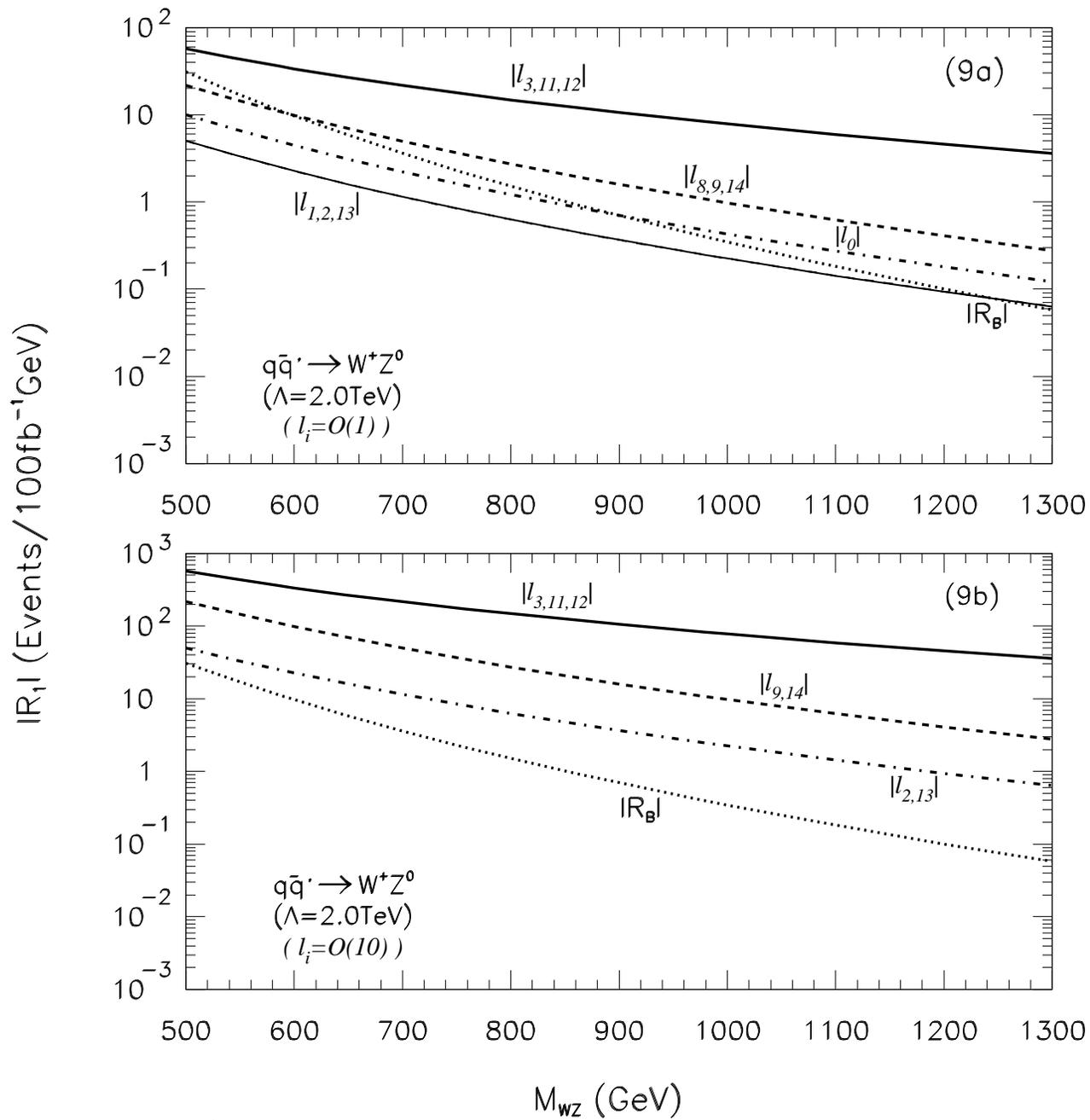
{\bf Fig.~9.} Same as Fig.~8, but with $\Lambda=2.0~$TeV~.

\noindent
{\bf Fig.~10.} Examination on the kinematic dependence and the 
validity of the ET
for the $~W^+_LW^-_L \rightarrow W^+_LW^-_L ~$ scattering 
process.   \\[0.20cm]
{\bf (10a).}  The ratio $~|B/g^2|~$ for $~\theta 
= 2^{\circ},~10^{\circ},~45^{\circ},~
90^{\circ},~100^{\circ},~120^{\circ},~ 135^{\circ},~ 150^{\circ},~ 
180^{\circ}~$. \\
{\bf (10b).}  Same as (10a), but for the ratio $~|B/T[W_L]|~$.\\
{\bf (10c).}  Comparison of the $W_L$-amplitude (solid lines) and 
the corresponding GB-amplitude (dotted lines) for 
$~\theta = 10^{\circ},~45^{\circ},~100^{\circ},~150^{\circ}~$. 
Here, $~B\,[150^{\circ}]~$ denotes
the $B$-term at $~\theta =150^{\circ}~$.\\
{\bf (10d).} Comparison of the LHC production rates 
contributed by the exact $W_L$-amplitude (solid line) and the
GB-amplitude (dotted line) from eq.~(A1). \\
{\bf (10e).} Same as Fig.~10d, but for the angular distributions.

%%%%%%%%%%%%%%%%%%%%%%%%%%%%%%%%%%%%%%%%%%%%%%%%%%%%%%%%%%%%%

\addtolength{\textwidth}{2.0cm}
\addtolength{\oddsidemargin}{1.1cm}
\addtolength{\oddsidemargin}{-0.9cm}
\evensidemargin=\oddsidemargin
\addtolength{\textheight}{-0.5cm}
%\addtolength{\topmargin}{-0.8cm}
\addtolength{\topmargin}{1.0cm}

\newpage

%%%%%%%%%%%%%%%%%%%%%%%%%%%%%%%%%%%%%%%%%%%%%%%%%%%%%%%%%%%%%%%%%%%%%%%%%%%
%%%%%      **********[ Table 1A; 2A,B; 3; 3-pages ]*************      %%%%%
%%%%%%%%%%%%%%%%%%%%%%%%%%%%%%%%%%%%%%%%%%%%%%%%%%%%%%%%%%%%%%%%%%%%%%%%%%%
\renewcommand{\baselinestretch}{1}

%%%%%% Beginning of Table 1. %%%%%%%%%%%%%%%%%%%%%%%%%%%%%%%%%%%%%%%%%%%%%

\begin{table}[t]  
\begin{center}

{\bf Table 1.}  Estimates of
amplitudes for $~W^{\pm}W^{\pm}\rightarrow W^{\pm}W^{\pm}~$ scattering.
\vspace{0.8cm}

%%%%%%Table 1A.%%%%%%%%%%%%%%%%%%%%%%%%%%%%%%%%%%%%%%%%%%%%%%%%%%%%%%%%%%
{\bf Table 1a.} 
~Model-independent contributions from 
$~{\cal L}_G+{\cal L}_F +{\cal L}^{(2)}~$.
\vspace{0.5cm}

\small

\begin{tabular}{||c||c|c|c|c|c||} 
%\hline\hline
& & & & &  \\
${\cal L}_G +{\cal L}_F + {\cal L}^{(2)}$  
&  $~~~T_{\ell}[4\pi]~~~     $  
&  $~T_{\ell}[3\pi,W_T]~ $  
&  $~T_{\ell}[2\pi,2W_T]~$  
&  $~T_{\ell}[\pi,3W_T]~ $  
&  $~T_{\ell}[4W_T]~     $  \\ 
& & & & &  \\
\hline\hline
& & & & &\\
Tree-Level
&  $ \frac{E^2}{f_{\pi}^2} $
&  $ g\frac{E}{f_\pi} $
&  $ g^2 $
&  $ e^2g\frac{f_\pi}{E} $ 
&  $g^2$ \\
( ${\ell}=0$ ) & & & & &\\
& & & & &\\
\hline
& & & & &\\
One-Loop 
& $\frac{E^2}{f_{\pi}^2}\frac{E^2}{\Lambda_0^2}$ 
& $g\frac{E}{f_{\pi}}\frac{E^2}{\Lambda_0^2}$ 
& $g^2\frac{E^2}{\Lambda_0^2}$  
& $g^3\frac{f_{\pi}E}{\Lambda_0^2}$
& $g^4\frac{f_{\pi}^2}{\Lambda_0^2}$   \\
( ${\ell}=1$ ) & & & & &\\
& & & & & \\
%\hline\hline 
\end{tabular}
\end{center}
\end{table}

\vspace{0.8cm}

%%%%%%%%%%%%%%%%%%%%%%%%%%   End of TABLE 1A. %%%%%%%%%%%%%%%%%%%%%%%%%%%

\newpage
%%%%%% Beginning of Table 1b. %%%%%%%%%%%%%%%%%%%%%%%%%%%%%%%%%
\begin{table}[t]  
\begin{center}

{\bf Table 1b.} Model-dependent contributions from the 
next-to-leading order  operators.
\vspace{0.8cm}

\renewcommand{\baselinestretch}{1.0}
\small
\begin{tabular}{||c||c|c|c|c|c||} 
%\hline\hline
& & & & &  \\
Operators 
& $ T_1[4\pi] $
& $ T_1[3\pi,W_T] $
& $ T_1[2\pi,2W_T] $
& $ T_1[\pi,3W_T] $
& $ T_1[4W_T] $ \\
& & & & &  \\
\hline\hline
& & & & &  \\
$ {\cal L}^{(2)\prime} $
& $ \ell_0 ~\frac{E^2}{\Lambda^2} $
& $ \ell_0~g\frac{f_\pi E}{\Lambda^2} $ 
& $ \ell_0~g^2\frac{f_\pi^2}{\Lambda^2} $
& $ \ell_0~g^3\frac{f_\pi^3}{E \Lambda^2} $
& /  \\
& & & & &  \\
\hline
& & & & &  \\
$ {\cal L}_{1,13} $
&  /
& $ \ell_{1,13}~e^2g\frac{f_\pi E}{\Lambda^2} $
& $ \ell_{1,13}~e^4\frac{f_\pi^2}{\Lambda^2} $
& $ \ell_{1,13}~e^2g\frac{f_\pi E}{\Lambda^2} $
& $ \ell_{1,13}~e^2g^2\frac{f_\pi^2}{\Lambda^2} $  \\   
& & & & &  \\
\hline
& & & & &  \\
$ {\cal L}_2 $
& $ \ell_2~e^2\frac{E^2}{\Lambda^2} $ 
& $ \ell_2~e^2g\frac{f_\pi E}{\Lambda^2} $
& $ \ell_2~e^2\frac{E^2}{\Lambda^2} $
& $ \ell_2~e^2g\frac{f_\pi E}{\Lambda^2} $
& $ \ell_2~e^2g^2\frac{f_\pi^2}{\Lambda^2} $ \\
& & & & &  \\
\hline
& & & & &  \\
$ {\cal L}_3 $
& $ \ell_3~g^2\frac{E^2}{\Lambda^2} $
& $ \ell_3~g\frac{E}{f_\pi}\frac{E^2}{\Lambda^2} $
& $ \ell_3~g^2\frac{E^2}{\Lambda^2} $
& $ \ell_3~g^3\frac{f_\pi E}{\Lambda^2} $
& $ \ell_3~g^4\frac{f_\pi^2}{\Lambda^2} $ \\
& & & & &  \\
\hline
& & & & &  \\
$ {\cal L}_{4,5} $
& $ \ell_{4,5}~\frac{E^2}{f_\pi^2}\frac{E^2}{\Lambda^2} $
& $ \ell_{4,5}~g\frac{E}{f_\pi}\frac{E^2}{\Lambda^2} $
& $ \ell_{4,5}~g^2\frac{E^2}{\Lambda^2} $
& $ \ell_{4,5}~g^3\frac{f_\pi E}{\Lambda^2} $ 
& $ \ell_{4,5}~g^4\frac{f_\pi^2}{\Lambda^2} $ \\
& & & & &  \\
\hline
& & & & & \\
$ {\cal L}_{6,7,10} $
& /
& /
& /
& /
& /  \\
& & & & &  \\
\hline
& & & & &  \\
$ {\cal L}_{8,14} $
& / 
& $ \ell_{8,14}~g^3\frac{f_\pi E}{\Lambda^2} $
& $ \ell_{8,14}~g^2\frac{E^2}{\Lambda^2} $
& $ \ell_{8,14}~g^3\frac{f_\pi E}{\Lambda^2} $
& $ \ell_{8,14}~g^4\frac{f_\pi^2}{\Lambda^2} $ \\
& & & & & \\
\hline
& & & & & \\
$ {\cal L}_9 $
& $ \ell_9~g^2\frac{E^2}{\Lambda^2} $
& $ \ell_9~g\frac{E}{f_\pi}\frac{E^2}{\Lambda^2} $ 
& $ \ell_9~g^2\frac{E^2}{\Lambda^2} $
& $ \ell_9~g^3\frac{f_\pi E}{\Lambda^2} $
& $ \ell_9~g^4\frac{f_\pi^2}{\Lambda^2} $ \\
& & & & & \\
\hline
& & & & & \\
$ {\cal L}_{11,12} $
& /
& $ \ell_{11,12}~g\frac{E}{f_\pi}\frac{E^2}{\Lambda^2} $
& $ \ell_{11,12}~g^2\frac{E^2}{\Lambda^2} $
& $ \ell_{11,12}~g^3\frac{f_\pi E}{\Lambda^2} $
& $ \ell_{11,12}~g^4\frac{f_\pi^2}{\Lambda^2} $ \\
& & & & & \\
%\hline\hline 
\end{tabular}
\end{center}
\end{table}
%%%%%%%%%%%% End of Table 1b. %%%%%%%%%%%%%%%%%%%%%%%%%%%%%%%%%

\newpage

\renewcommand{\baselinestretch}{1}

%%%%%% Beginning of Table 2. %%%%%%%%%%%%%%%%%%%%%%%%%%%%%%%%%%%%%%%%%
\begin{table}[t]   
\begin{center}

{\bf Table 2.} Order estimates of $B$-terms 
for $~W^{\pm}W^{\pm}\rightarrow W^{\pm}W^{\pm}~$  scattering. 
\vspace{0.8cm}

%%%%%%Table 2A.%%%%%%%%%%%%%%%%%%%%%%%%%%%%%%%%%%%%%%%%%%%%%%%%%%%%%
{\bf Table 2a.} Model-independent contributions.
\vspace{0.5cm}

\small

\begin{tabular}{||c||c|c|c|c||} 
%\hline\hline
& & & &  \\
${\cal L}_G +{\cal L}_F + {\cal L}^{(2)}$  
&  $~~~B_{\ell}^{(0)}~~~     $  
&  $~~~B_{\ell}^{(1)}~~~     $  
&  $~~~B_{\ell}^{(2)}~~~     $  
&  $~~~B_{\ell}^{(3)}~~~     $  \\ 
& & & &   \\
\hline\hline
& & & & \\
Tree-Level
&  $ g^2$
&  $ g^2\frac{M_W}{E} $
&  $ g^2\frac{M_W^2}{E^2} $
&  $ g^2\frac{M_W}{E} $\\
 ( ${\ell}=0$ ) & & & &\\
& & & & \\
\hline
& & & & \\
One-Loop 
& $g^2\frac{E^2}{\Lambda_0^2}$ 
& $g^3\frac{Ef_\pi}{\Lambda_0^2}$ 
& $ g^4\frac{f_{\pi}^2}{\Lambda_0^2}$  
& $g^4\frac{f_{\pi}^2}{\Lambda_0^2}\frac{M_W}{E} $\\
( ${\ell}=1$ ) & & & &\\
& & & &  \\
%\hline\hline 
\end{tabular}

\vspace{1.5cm}

%%%%%% Beginning of Table 2b. %%%%%%%%%%%%%%%%%%%%%%%%%%%%%%%%%%%%%%%%%%%%

{\bf Table 2b.} 
Relevant operators for model-dependent contributions.$^{(a)}$ 

\vspace{0.8cm}

\small

\begin{tabular}{||c|c|c|c||} 
%\hline\hline
& & &   \\
    $O(g^2\frac{E^2}{\Lambda^2})$
&    $O(g^3\frac{Ef_\pi}{\Lambda^2})$
&    $O(g^2\frac{f_{\pi}^2}{\Lambda^2})$ 
&    $O(g^4\frac{f_{\pi}^2}{\Lambda^2})$ \\
    ( from $B^{(0)}_1$ )
&   ( from $B^{(1)}_1$ )
&   ( from $B^{(0)}_1$ )  
&   ( from $B^{(2)}_1 ~{\rm or} ~B^{(0)}_1 $ )\\
 & & &   \\
\hline
 & & & \\
  ${\cal L}_{3,4,5,9,11,12}$  
& ${\cal L}_{2,3,4,5,8,9,11,12,14}$ 
& ${\cal L}^{(2)\prime}$ 
& \parbox[t]{4.28cm}{
${\cal L}_{1\sim 5,8,9,11\sim 14} ~~(B^{(2)}_1)$ \\
${\cal L}_{1,2,8,13,14} ~~(B^{(0)}_1)$ \\ 
${\cal L}_{2\sim 5,8,9,11,12,14} ~~(B^{(0)}_1)~^{(b)}$ } \\
& & & \\
%\hline\hline 
\end{tabular}
\end{center}
\end{table}

\begin{table}[t]
\begin{center}
\vspace{0.4cm}
\begin{tabular}{l}
{\footnotesize 
$^{(a)}$ We list the relevant operators for each order of $B$-terms.} \\
{\footnotesize 
$^{(b)}$ Here $B^{(0)}_1$ is contributed by $ T_1[2\pi^\pm ,2v^\pm ] $.}\\
\end{tabular}
\end{center}
\end{table}

%%%%%% End of Table 2b. %%%%%%%%%%%%%%%%%%%%%%%%%%%%%%%%%%%%%%%%%%%%%%%

\newpage
%%%%%%%%%%%%%%%%%%%%%%%%%% Beginning of Table-3 %%%%%%%%%%%%%%%%%%%%%%%%%%%%%%%
\begin{table}[t]  
\begin{center}
{\bf Table 3.}~ Estimates of amplitudes for $~q\bar{q}^{(\prime )}
                \rightarrow V^aV^b~$: Model-independent contributions.
\vspace{0.8cm}

%%%%%%Table 3A.%%%%%%%%%%%%%%%%%%%%%%%%%%%%%%%%%%%%%%%%%%%%%%%%%%%%%%%%%%%%%%%%
{\bf Table 3a.} ~For $~q\bar{q}^{(\prime )}\rightarrow W^{+}W^{-}~,
                ~W^{\pm}Z ~$. 
\vspace{0.5cm}

\small

\begin{tabular}{||c||c|c|c||c|c||} 
%\hline\hline
& & & & & \\
$~~{\cal L}_G +{\cal L}_F + {\cal L}^{(2)} ~~$  
&  $~~T_{\ell}[q\bar{q}^{(\prime )}\rightarrow \pi\pi ]~~    $  
&  $~T_{\ell}[q\bar{q}^{(\prime )}\rightarrow \pi V_T]~ $  
&  $~T_{\ell}[q\bar{q}^{(\prime )}\rightarrow V_T V_T ]~$ 
&  $B^{(0)}_{\ell}$
&  $B^{(1)}_{\ell}$  \\ 
& & & & & \\
\hline\hline
& & & & & \\
Tree-Level ( $\ell = 0$ )
&  $ g^2 $
&  $ e^2g \frac{f_\pi}{E} $
&  $ g^2 $
&  $ g^2\frac{M_W^2}{E^2} $
&  $ g^2\frac{M_W}{E}$  \\
& & & & & \\
\hline
& & & & & \\
One-Loop ( $\ell = 1$ )
& $g^2\frac{E^2}{\Lambda_0^2} $
& $g^3\frac{f_\pi E}{\Lambda_0^2} $
& $g^4\frac{f_\pi^2}{\Lambda_0^2} $
& $g^3\frac{f_\pi M_W}{\Lambda_0^2}$ 
& $g^4\frac{f_\pi^2}{\Lambda^2_0}\frac{M_W}{E}$  \\
& & & & & \\
%\hline\hline 
\end{tabular}

\vspace{1.8cm}

%%%%%%Table 3B.%%%%%%%%%%%%%%%%%%%%%%%%%%%%%%%%%%%%%%%%%%%%%%%%%%%%%%%%%%%%%
{\bf Table 3b.} ~For $~q\bar{q}\rightarrow ZZ~$. 
\vspace{0.5cm}
\small

\begin{tabular}{||c||c|c|c||c|c||} 
%\hline\hline
& & & & &  \\
$~~{\cal L}_G +{\cal L}_F + {\cal L}^{(2)} ~~$  
&  $~~T_{\ell}[q\bar{q}\rightarrow \pi\pi ]~~     $  
&  $~T_{\ell}[q\bar{q}\rightarrow \pi Z_T]~ $  
&  $~T_{\ell}[q\bar{q}\rightarrow Z_T Z_T ]~$ 
&  $B^{(0)}_{\ell}$
&  $B^{(1)}_{\ell}$  \\ 
& & & & &  \\
\hline\hline
& & & & & \\
Tree-Level ( $\ell = 0$ )
&  /
&  /
&  $g^2$
&  $g^2\frac{M_W^2}{E^2}$
&  $g^2\frac{M_W}{E}$  \\
& & & & & \\
\hline
& & & & &\\
One-Loop ( $\ell = 1$ )
& $g^2\frac{E^2}{\Lambda_0^2} $
& $g^3\frac{f_\pi E}{\Lambda_0^2} $
& $g^4\frac{f_\pi^2}{\Lambda_0^2} $
& $g^3\frac{f_\pi M_W}{\Lambda_0^2}$ 
& $g^4\frac{f_\pi^2}{\Lambda^2_0}\frac{M_W}{E}$  \\
& & & & & \\
%\hline\hline 
\end{tabular}
\end{center}
\end{table}

\vspace{0.8cm}

%%%%%% End of Table 3B. %%%%%%%%%%%%%%%%%%%%%%%%%%%%%%%%%%%%%%%%%%%%%%%%%%%%%

%%%%%%%%%%%%%%%%%%%%%%%%%%   End of TABLE 3. %%%%%%%%%%%%%%%%%%%%%%%%%%%%%%%%%

\newpage
%%%%%%%%%%%%%%%%%%%%%%%%%% Beginning of Table-4 %%%%%%%%%%%%%%%%%%%%%%%%%%%%%

\addtolength{\oddsidemargin}{-2.5cm}

\begin{table}[t]  
\begin{center}
{\bf Table 4.}~ 
 Estimates of amplitudes for $~q\bar{q}^{(\prime )}\rightarrow V^aV^b~$: 
 Model-dependent contributions.$^{(a)}$ 

\vspace{0.8cm}

%%%%%%Table 4A.%%%%%%%%%%%%%%%%%%%%%%%%%%%%%%%%%%%%%%%%%%%%%%%%%%%%%%%%%%%%%%%%
{\bf Table 4a.} ~For $~q\bar{q}\rightarrow W^{+}W^{-}~ $.
\vspace{0.5cm}

\small

\begin{tabular}{||c||c|c|c||c|c||} 
%\hline\hline
& & & & & \\
~Operators~
&  $~~T_1[q\bar{q}\rightarrow \pi\pi ]~~ $  
&  $~T_1[q\bar{q}\rightarrow \pi V_T]~ $  
&  $~T_1[q\bar{q}\rightarrow V_T V_T ]~$ 
&  $B^{(0)}_{1}$
&  $B^{(1)}_{1}$  \\  
& & & & & \\
\hline\hline
& & & & & \\
   $ {\cal L}^{(2)\prime} $ 
&  $\ell_0 ~g^2\frac{f_\pi^2}{\Lambda^2} $
&  $\ell_0 ~g^3\frac{f_\pi^3}{E\Lambda^2} $
&  / 
&  $g^2\frac{f_\pi^2}{\Lambda^2}\frac{M_W^2}{E^2}$
&  /  \\
& & & & & \\
\hline
& & & & & \\
   $ {\cal L}_{1,13} $ 
& / 
&  $\ell_{1,13}~ e^2g\frac{f_\pi E}{\Lambda^2} $ 
&  $\ell_{1,13}~ e^2g^2\frac{f_\pi^2}{\Lambda^2} $ 
&  $e^2g^2\frac{f_\pi^2}{\Lambda^2}$
&  $e^2g^2\frac{f_\pi^2}{\Lambda^2}\frac{M_W}{E}$\\
& & & & & \\
\hline
& & & & & \\
$ {\cal L}_2 $ 
& $\ell_2~ e^2\frac{E^2}{\Lambda^2} $
& $\ell_2~ e^2 g\frac{f_\pi E}{\Lambda^2} $ 
& $\ell_2~ e^2g^2\frac{f_\pi^2}{\Lambda^2} $ 
&  $e^2g^2\frac{f_\pi^2}{\Lambda^2}$
&  $e^2g^2\frac{f_\pi^2}{\Lambda^2}\frac{M_W}{E}$\\
& & & & & \\
\hline
& & & & & \\
$ {\cal L}_3 $ 
& $\ell_3~ g^2\frac{E^2}{\Lambda^2} $
& $\ell_3~ g^3\frac{f_\pi E}{\Lambda^2} $ 
& $\ell_3~ g^4\frac{f_\pi^2}{\Lambda^2} $  
&  $g^4\frac{f_\pi^2}{\Lambda^2}$
&  $g^4\frac{f_\pi^2}{\Lambda^2}\frac{M_W}{E}$\\
& & & & & \\
\hline
& & & & & \\
$ {\cal L}_{8,14} $ 
& /
& $\ell_{8,14}~ g^3\frac{f_\pi E}{\Lambda^2} $ 
& $\ell_{8,14}~ g^4\frac{f_\pi^2}{\Lambda^2} $  
&  $g^4\frac{f_\pi^2}{\Lambda^2}$
&  $g^4\frac{f_\pi^2}{\Lambda^2}\frac{M_W}{E}$ \\
& & & & & \\
\hline
& & & & & \\
$ {\cal L}_9 $ 
& $\ell_9~ g^2\frac{E^2}{\Lambda^2} $ 
& $\ell_9~ g^3\frac{f_\pi E}{\Lambda^2} $ 
& $\ell_9~ g^4\frac{f_\pi^2}{\Lambda^2} $ 
&  $g^4\frac{f_\pi^2}{\Lambda^2}$
&  $g^4\frac{f_\pi^2}{\Lambda^2}\frac{M_W}{E}$ \\
& & & & & \\
\hline
& & & & & \\
$ {\cal L}_{11,12} $ 
& /
& $\ell_{11,12} ~g^3\frac{f_\pi E}{\Lambda^2} $ 
& $\ell_{11,12} ~g^4\frac{f_\pi^2}{\Lambda^2} $  
&  $g^4\frac{f_\pi^2}{\Lambda^2}$
&  $g^4\frac{f_\pi^2}{\Lambda^2}\frac{M_W}{E}$\\
& & & & & \\
%\hline\hline 
\end{tabular}
\end{center}
\end{table}

\vspace{1.0cm}
\begin{table}[t]
\begin{center}
\begin{tabular}{l}
{\footnotesize 
$^{(a)}$ Here we only consider the
 light quarks ( $q \neq t$ ) whose Yukawa coupling $~y_q \approx 0~$.
At tree level,}\\
{\footnotesize  $~q\bar{q}\rightarrow ZZ~$ contains no 
model-dependent contribution and the operators $~{\cal L}_{4,5,6,7,10}~$ 
do not contribute to }\\
{\footnotesize   $~q\bar{q}^{(\prime )}\rightarrow W^+W^-,~W^{\pm}Z~$. }  
\end{tabular}
\end{center}
\end{table}

\addtolength{\oddsidemargin}{2cm}
\newpage
\addtolength{\oddsidemargin}{1.6cm}
\addtolength{\topmargin}{1.0cm}

%%%%%%%%%%%%%%%%%%%%%%%%%% Beginning of Table-4B %%%%%%%%%%%%%%%%%%%%%%%%%%%%%%
\begin{table}[t]  
\begin{center}

%%%%%%Table 4B.%%%%%%%%%%%%%%%%%%%%%%%%%%%%%%%%%%%%%%%%%%%%%%%%%%%%%%%%%%%%%%%%
{\bf Table 4b.} ~For $~q\bar{q}^{\prime}\rightarrow W^{\pm}Z~ $.
\vspace{0.5cm}

\vspace{0.8cm}
\small

\begin{tabular}{||c||c|c|c||c|c||} 
%\hline\hline
& & & & &  \\
~Operators~
&  $~~T_1[q\bar{q}^{\prime}\rightarrow \pi^{\pm}\pi^0 ]~~$  
&  $~T_1[q\bar{q}^{\prime}\rightarrow \pi V_T]~ $  
&  $~T_1[q\bar{q}^{\prime}\rightarrow W_T^{\pm} Z_T ]~$ 
&  $B^{(0)}_{1}$
&  $B^{(1)}_{1}$  \\  
& & & & &  \\
\hline\hline
& & & & &  \\
   $ {\cal L}^{(2)\prime} $ 
&  $\ell_0 ~g^2\frac{f_\pi^2}{\Lambda^2} $
&  $\ell_0 ~g^3\frac{f_\pi^3}{E\Lambda^2} $
& /  
&  $g^2\frac{f_\pi^2}{\Lambda^2}\frac{M_W^2}{E^2}$
&  / \\
& & & & &  \\
\hline
& & & & &  \\
   $ {\cal L}_{1,13} $ 
&  / 
&  $\ell_{1,13}~ e^2g\frac{f_\pi E}{\Lambda^2} $ 
&  $\ell_{1,13}~ e^2g^2 \frac{f_\pi^2}{\Lambda^2} $ 
&  $e^2g^2\frac{f_\pi^2}{\Lambda^2}$
&  $e^2g^2\frac{f_\pi^2}{\Lambda^2}\frac{M_W}{E}$\\
& & & & &  \\
\hline
& & & & &  \\
$ {\cal L}_2 $ 
& /
& $\ell_2~ e^2g \frac{f_\pi E}{\Lambda^2} $ 
& $\ell_2~ e^2g^2\frac{f_\pi^2}{\Lambda^2} $ 
&  $e^2g^2\frac{f_\pi^2}{\Lambda^2}$
&  $e^2g^2\frac{f_\pi^2}{\Lambda^2}\frac{M_W}{E}$\\
& & & & &  \\
\hline
& & & & &  \\
$ {\cal L}_3 $ 
& $\ell_3~ g^2\frac{E^2}{\Lambda^2} $
& $\ell_3~ g^3\frac{f_\pi E}{\Lambda^2} $ 
& $\ell_3~ g^4\frac{f_\pi^2}{\Lambda^2} $  
&  $g^4\frac{f_\pi^2}{\Lambda^2}$
&  $g^4\frac{f_\pi^2}{\Lambda^2}\frac{M_W}{E}$\\
& & & & &  \\
\hline
& & & & &  \\
$ {\cal L}_{8,14} $ 
& /
& $\ell_{8,14}~ g^3\frac{f_\pi E}{\Lambda^2} $ 
& $\ell_{8,14}~ g^4\frac{f_\pi^2}{\Lambda^2} $  
&  $g^4\frac{f_\pi^2}{\Lambda^2}$
&  $g^4\frac{f_\pi^2}{\Lambda^2}\frac{M_W}{E}$\\
& & & & &  \\
\hline
& & & & &  \\
$ {\cal L}_9 $ 
& /
& $\ell_9~ g^3\frac{f_\pi E}{\Lambda^2} $ 
& $\ell_9~ g^4\frac{f_\pi^2}{\Lambda^2} $  
&  $g^4\frac{f_\pi^2}{\Lambda^2}$
&  $g^4\frac{f_\pi^2}{\Lambda^2}\frac{M_W}{E}$\\
& & & & &  \\
\hline
& & & & &  \\
$ {\cal L}_{11,12} $ 
& $\ell_{11,12}~ g^2\frac{E^2}{\Lambda^2} $ 
& $\ell_{11,12} ~g^3\frac{f_\pi E}{\Lambda^2} $ 
& $\ell_{11,12} ~g^4\frac{f_\pi^2}{\Lambda^2} $ 
&  $g^4\frac{f_\pi^2}{\Lambda^2}$
&  $g^4\frac{f_\pi^2}{\Lambda^2}\frac{M_W}{E}$  \\
& & & & &  \\
%\hline\hline 
\end{tabular}
\end{center}
\end{table}
%%%%%%%%%%%%%%%%%%%%% End of Table-4 %%%%%%%%%%%%%%%%%%%%%%%%%%%%%%%%%%%%%%%%%%

%%%%%%%%%%%%%%%%%%%%%%%%%%%%%%%%%%%%%%%%%%%%%%%%%%%%%%%%%%%%%%%%%%%%%%%

\addtolength{\textheight}{0.5cm}
\addtolength{\topmargin}{-1.5cm}
\addtolength{\oddsidemargin}{-0.6cm}
\addtolength{\oddsidemargin}{-0.8cm}
\evensidemargin=\oddsidemargin
\newpage

%%%%%% Beginning of Table 5. %%%%%%%%%%%%%%%%%%%%%%%%%%%%%%%%%%%%%%%%%%

\tabcolsep 1pt
\begin{table}[t]  
\begin{center}
{\bf Table 5.} 
Global classification of sensitivities to probing direct and indirect EWSB\\ 
information from effective operators at the level of $S$-matrix elements (I). 
$^{(a)}$

\vspace{0.5cm}

\small

\begin{tabular}{||c||c|c|c||} 
%\hline\hline
& & &  \\
Required Precision
&  Relevant Operators
&  Relevant Amplitudes
&  MI or MD $^{(b)}$ \\
& & & ? \\
\hline\hline
%& & & \\
   $O\left(\frac{E^2}{f_{\pi}^2}\right)$ 
&  ${\cal L}_{\rm MI}~(\equiv 
   {\cal L}_{\rm G}+{\cal L}_{\rm F}+{\cal L}^{(2)}) $
&  $ T_0[4V_L] (\neq T_0[4Z_L]) $
&  MI \\
%& & & \\
\hline
%& & & \\
\parbox[t]{2.2cm}{ 
~~\\
~~\\
$O\left(\frac{E^2}{f_\pi^2}
 \frac{E^2}{\Lambda^2},~g\frac{E}{f_\pi}\right)$\\
~~\\
~~ }
&  \parbox[t]{2.8cm}{ 
  ${\cal L}_{4,5}$\\
  ${\cal L}_{6,7}$\\
  ${\cal L}_{10}$\\
  ${\cal L}_{\rm MI}$\\
  ${\cal L}_{\rm MI}$ }
&  \parbox[t]{5.0cm}{ 
  $T_1[4V_L]$\\
  $T_1[2Z_L,2W_L],~T_1[4Z_L]$\\
  $T_1[4Z_L]$\\
  $T_0[3V_L,V_T] ~(\neq T_0[3Z_L,Z_T])$\\
  $T_1[4V_L]$ }
&  \parbox[t]{0.8cm}{ 
  MD\\
  MD\\
  MD\\
  MI\\
  MI }\\
%& & & \\
\hline
%& & & \\
\parbox[t]{2.2cm}{ 
~~\\
~~\\
$O\left(g\frac{E}{f_\pi}\frac{E^2}{\Lambda^2},~g^2\right)$\\
~~\\
~~\\
~~  }
& \parbox[t]{2.8cm}{ 
  $ {\cal L}_{3,4,5,9,11,12} $\\
  $ {\cal L}_{2,3,4,5,6,7,9,11,12} $\\
  $ {\cal L}_{3,4,5,6,7,10} $\\
  $ {\cal L}_{\rm MI} $\\ 
  $ {\cal L}_{\rm MI} $\\
  $ {\cal L}_{\rm MI} $} 
&  \parbox[t]{5.8cm}{ 
  $T_1[3W_L,W_T]$\\
  $T_1[2W_L,Z_L,Z_T],~T_1[2Z_L,W_L,W_T]$\\
  $T_1[3Z_L,Z_T]$\\
  $T_0[2V_L,2V_T],~T_0[4V_T]~~^{(c)}$\\
  $T_1[3V_L,V_T]$\\
  $B^{(0)}_0 \simeq T_0[3\pi ,v]~(\neq T_0[3\pi^0 ,v^0])$  }
&   \parbox[t]{0.8cm}{ 
  MD\\
  MD\\
  MD\\
  MI\\
  MI\\
  MI }\\
%& & & \\
\hline
%& & & \\
  $O\left(\frac{E^2}{\Lambda^2}\right)$
& ${\cal L}^{(2)\prime}$ 
& $T_1[4W_L],~T_1[2W_L,2Z_L]$
& MD \\
%& & & \\
\hline
%& & & \\
   \parbox[t]{2.5cm}{
        ~~\\
~~\\
~~\\
$O\left(g^2\frac{E^2}{\Lambda^2},~g^3\frac{f_\pi}{E}\right)$ \\ 
~~\\          
~~\\
~~ }
&  \parbox[t]{3.0cm}{ 
   ${\cal L}_{\rm MI}$\\
   ${\cal L}_{2,3,9}$\\
   ${\cal L}_{3,11,12}$\\
   ${\cal L}_{2,3,4,5,8,9,11,12,14}$\\
   ${\cal L}_{1\sim 9,11\sim 14}$ \\
   ${\cal L}_{4,5,6,7,10}$ \\
   ${\cal L}_{{\rm MI},2,3,4,5,6,7,9\sim 12}$ }
&  \parbox[t]{6.3cm}{  
   $T_0[V_L,3V_T],T_1[2V_L,2V_T], B^{(1,3)}_0~~^{(c,d)} $\\
   $T_1[4W_L]$\\
   $T_1[2Z_L,2W_L]$\\
   $T_1[2W_L,2W_T]$\\
   $T_1[2V_L,2V_T]~~~^{(e)}$\\
   $T_1[2Z_L,2Z_T]$\\
{\footnotesize  $B^{(0)}_1 \simeq T_1[3\pi ,v]~~^{(f,g)}$}  }
&  \parbox[t]{2.0cm}{ 
  MI\\
  MD\\
  MD\\
  MD\\
  MD\\
  MD\\
  MI $+$ MD  }\\
%& & & \\
\hline
%& & & \\
\parbox[t]{2.75cm}{
~~\\
~~\\
  $O\left(g^3\frac{Ef_\pi}{\Lambda^2},~g^4\frac{f_\pi^2}{E^2}\right)$\\
~~\\
~~}
& \parbox[t]{3.3cm}{ 
${\cal L}_{{\rm MI},1,2,3,8,9,11\sim 14}$ \\
${\cal L}_{4,5}$\\  
${\cal L}_{6,7,10}$ \\
${\cal L}_{2\sim 5,8,9,11,12,14}$\\
${\cal L}_{\rm MI}$   }
& \parbox[t]{5.5cm}{ 
$T_1[V_L,3V_T]~(\neq T_1[Z_L,3Z_T])$\\
$T_1[V_L,3V_T]$\\
$T_1[V_L,3V_T]~(\neq T_1[W_L,3W_T])~~^{(g)}$ \\
{\footnotesize $B^{(1)}_1\simeq T_1[2\pi ,V_T,v]$} \\
{\footnotesize $B^{(2)}_0\simeq T_0[2V_T,2v]$}$~~^{(c,h)}$  }
& \parbox[t]{2.0cm}{ MI$+$MD \\
                     MD\\ 
                     MD\\
                     MD\\
                     MI }\\
\hline
%& & & \\
\parbox[t]{2.0cm}{ 
~~\\
~~\\
~~\\
~~\\
~~\\
  $O\left((g^2,g^4)\frac{f_\pi^2}{\Lambda^2}\right)$\\
~~\\
~~\\
~~\\
~~\\
~~}
& \parbox[t]{3.1cm}{ 
  ${\cal L}^{(2)\prime} $ \\
  ${\cal L}_1$\\
  ${\cal L}_{{\rm MI},1\sim 5,8,9,11\sim 14}$\\
  ${\cal L}_{{\rm MI},1\sim 9,11\sim 14}$\\
  ${\cal L}_{{\rm MI},1,4,5,6,7,10}$\\
  ${\cal L}_{1,2,8,13,14}$\\
  ${\cal L}_{{\rm MI},1\sim 9,11\sim 14}$\\
  ${\cal L}_{{\rm MI},4,5,6,7,10}$\\
  ${\cal L}_{{\rm MI},1\sim 5,8,9,11\sim 14}$\\
  ${\cal L}_{{\rm MI},1\sim 9,11\sim 14}$\\
  ${\cal L}_{{\rm MI},4,5,6,7,10}$ } 
& 
\parbox[t]{6.4cm}{
$T_1[2V_L,2V_T],B^{(0)}_1\simeq T_1[3\pi ,v]~~^{(c)}$ \\
  $ T_1[2W_L,2W_T]~~^{(i)} $\\
  $ T_1[4W_T] $\\
  $ T_1[4V_T]~(\neq T_1[4W_T],T_1[4Z_T]) $\\
  $ T_1[4Z_T] $\\
{\footnotesize  $ B_1^{(0)}\simeq T_1[3\pi ,v]~~^{(c,j)} $\\
  $ B_1^{(0)}\simeq T_1[2\pi ,2v]~~^{(c,k)} $\\
  $ B_1^{(0)}\simeq 
                      T_1[2\pi ,2v](\neq T_1[2\pi^\pm ,2v^\pm ])~^{(g)} $ \\
  $ B_1^{(2)}\simeq T_1[\pi^\pm ,2W_T,v^\pm ] $\\
  $ B_1^{(2)}\neq T_1[\pi^\pm ,2W_T,v^\pm ],
     T_1[\pi^0 ,2Z_T,v^0] $\\
  $ B_1^{(2)}\simeq T_1[\pi^0 ,2Z_T,v^0] $ } }
&  
\parbox[t]{2.0cm}{ 
  MD\\
  MD\\
  MI$+$MD\\
  MI$+$MD\\
  MI$+$MD\\
  MD\\
  MI$+$MD\\
  MI$+$MD\\
  MI$+$MD\\
  MI$+$MD\\
  MI$+$MD  }\\
& & &  \\
%\hline\hline 
\end{tabular}
\end{center}
\end{table}

%%%%%% End of Table 5. %%%%%%%%%%%%%%%%%%%%%%%%%%%%%%%%%%%%%%%%%%%%%%%%%

\addtolength{\topmargin}{1.8cm}

\newpage
\addtolength{\topmargin}{-1.8cm}

%%%%%% Beginning of Table 6. %%%%%%%%%%%%%%%%%%%%%%%%%%%%%%%%%%%%%%%%%%%%%%%%%%%

\tabcolsep 1pt
\begin{table}[t]  
\begin{center}

{\bf Table 6.} 
Global classification of sensitivities to probing direct and indirect EWSB\\ 
information from effective operators at the level of $S$-matrix elements (II). 
$^{(a)}$

\vspace{0.5cm}

\small

\begin{tabular}{||c||c|c|c||} 
%\hline\hline
& & &  \\
~Required Precision~
&  Relevant Operators
&  Relevant Amplitudes
&  MI or MD $^{(b)}$ \\
& & & ? \\
\hline\hline
& & & \\
     $O(g^2)$  
&    $~{\cal L}_{\rm MI}~(\equiv 
       {\cal L}_{\rm G}+{\cal L}_{\rm F}+{\cal L}^{(2)}) ~$
& $T_0[q\bar{q};V_LV_L],~ T_0[q\bar{q};V_TV_T]$
& MI \\
& & & \\
\hline
& & & \\
   \parbox[t]{2.5cm}{
        ~~\\
        ~~\\
$O\left(g^2\frac{E^2}{\Lambda^2},~g^3\frac{f_\pi}{E}\right)$ \\ 
        ~~\\          
        ~~ }
&  \parbox[t]{3.0cm}{ 
   ${\cal L}_{2,3,9}$\\
   ${\cal L}_{3,11,12}$\\
   ${\cal L}_{\rm MI}$\\ 
   ${\cal L}_{\rm MI}$\\
   ${\cal L}_{\rm MI}$ }
&  \parbox[t]{4.0cm}{  
   $T_1[q\bar{q};W_LW_L] $\\
   $T_1[q\bar{q};W_LZ_L] $ \\
   $T_0[q\bar{q};V_LV_T] $\\
   $T_1[q\bar{q};V_LV_L] $\\
 {\footnotesize  $B_0^{(1)}$}$\simeq T_0[q\bar{q};V_T,v]$ }
&  \parbox[t]{0.8cm}{ 
  MD\\
  MD\\
  MI\\
  MI\\
  MI }\\
& & & \\
\hline
& & & \\
\parbox[t]{2.5cm}{
~~\\
  $O\left(g^3\frac{Ef_\pi}{\Lambda^2},
   ~g^4\frac{f_\pi^2}{E^2}\right)$\\
~~   }
& \parbox[t]{3.3cm}{ 
${\cal L}_{1,2,3,8,9,11\sim 14}$\\
${\cal L}_{\rm MI}$ \\
${\cal L}_{\rm MI}$  }
& \parbox[t]{4.0cm}{ 
$T_1[q\bar{q};V_LV_T]$\\  
$T_1[q\bar{q};V_LV_T]$\\
{\footnotesize $B_0^{(0)}$}$\simeq T_0[q\bar{q};2v]~^{(c)}$  }
& \parbox[t]{0.8cm}{ MD \\
                     MI \\
                     MI }\\
& & & \\
\hline
& & & \\
\parbox[t]{2.0cm}{ 
~~\\
  $O\left((g^2,g^4)\frac{f_\pi^2}{\Lambda^2}\right)$\\
~~}
& \parbox[t]{3.1cm}{ 
  ${\cal L}^{(2)\prime} $ \\
  ${\cal L}_{1,2,3,8,9,11\sim 14}$\\
  ${\cal L}_{\rm MI}$ } 
& 
\parbox[t]{5.4cm}{$ T_1[q\bar{q};V_LV_L]$\\  
                  $ T_1[q\bar{q};V_TV_T]$,~
 {\footnotesize $B_1^{(0)}$}$\simeq T_1[q\bar{q};\pi ,v]$ \\  
  $ T_1[q\bar{q};V_TV_T]$,~
 {\footnotesize $B_1^{(0)}$}$\simeq T_1[q\bar{q};\pi ,v]$   }
&  \parbox[t]{0.8cm}{ 
  MD\\
  MD\\
  MI  }\\
& & &  \\
%\hline\hline 
\end{tabular}
\end{center}
\end{table}
\begin{table}[t]
\begin{center}

\vspace{0.4cm}
\begin{tabular}{l}
{\footnotesize 
$^{(a)}$ The contributions from $~{\cal L}_{1,2,13}~$ are always associated
with a factor of $~\sin^2\theta_W~$, unless specified otherwise.}\\
{\footnotesize 
~~~$~{\cal L}_{4,5,6,7,10}~$ do not contribute to the 
processes considered in this table. 
Also, for contributions to the $B$-term}\\
{\footnotesize 
~~~~in a given $V_L$-amplitude, 
we list them separately with the $B$-term specified.}\\
{\footnotesize  $^{(b)}$ MI~$=$~model-independent, MD~$=$~model-dependent.
~~}\\
{\footnotesize  $^{(c)}$ Here, $~B_0^{(0)}~$ is dominated by 
$~T_0[q\bar{q};2v]~$ since $~T_0[q\bar{q};\pi,v]~$ contains a 
suppressing factor $\sin^2\theta_W$ as can be}\\
{\footnotesize 
~~~~deduced from $~T_0[q\bar{q};\pi V_T]~$ (cf. Table~3a) 
times the factor $~v^\mu = O\left(\frac{M_W}{E}\right)~$.}\\
\end{tabular}
\end{center}
\end{table}
%%%%%% End of Table 6. %%%%%%%%%%%%%%%%%%%%%%%%%%%%%%%%%%%%%%%%%%%%%%%%%%%

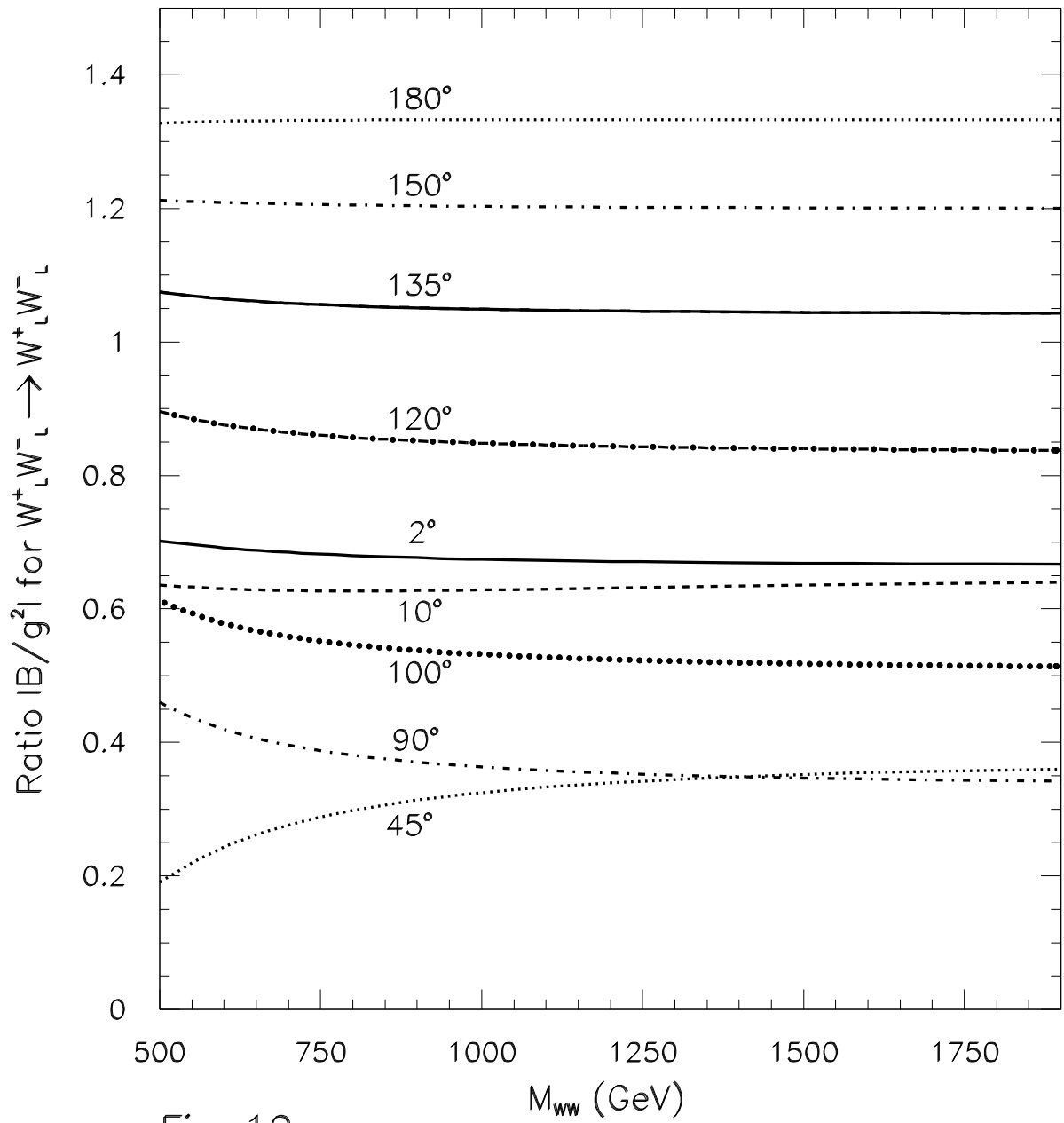
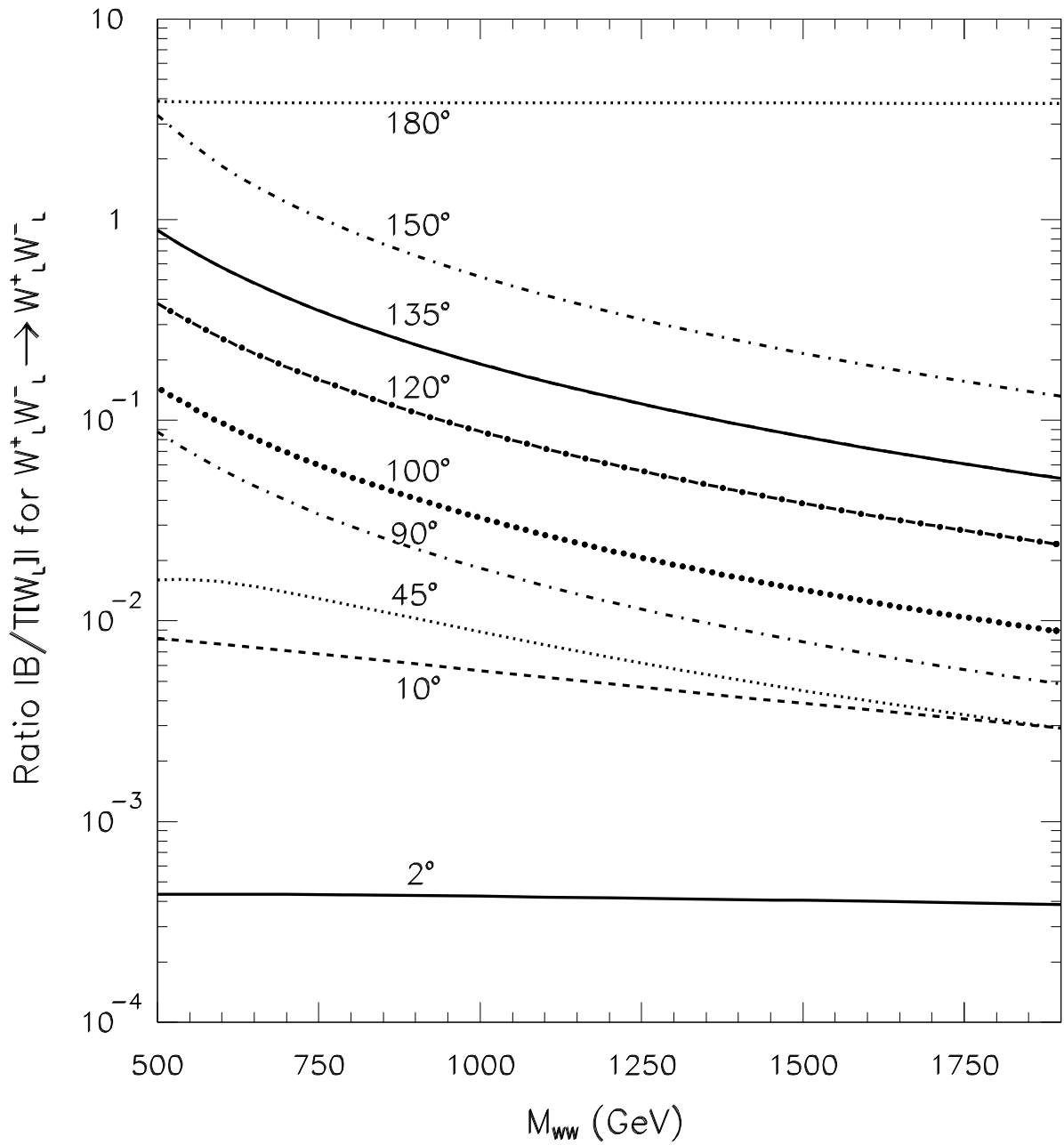
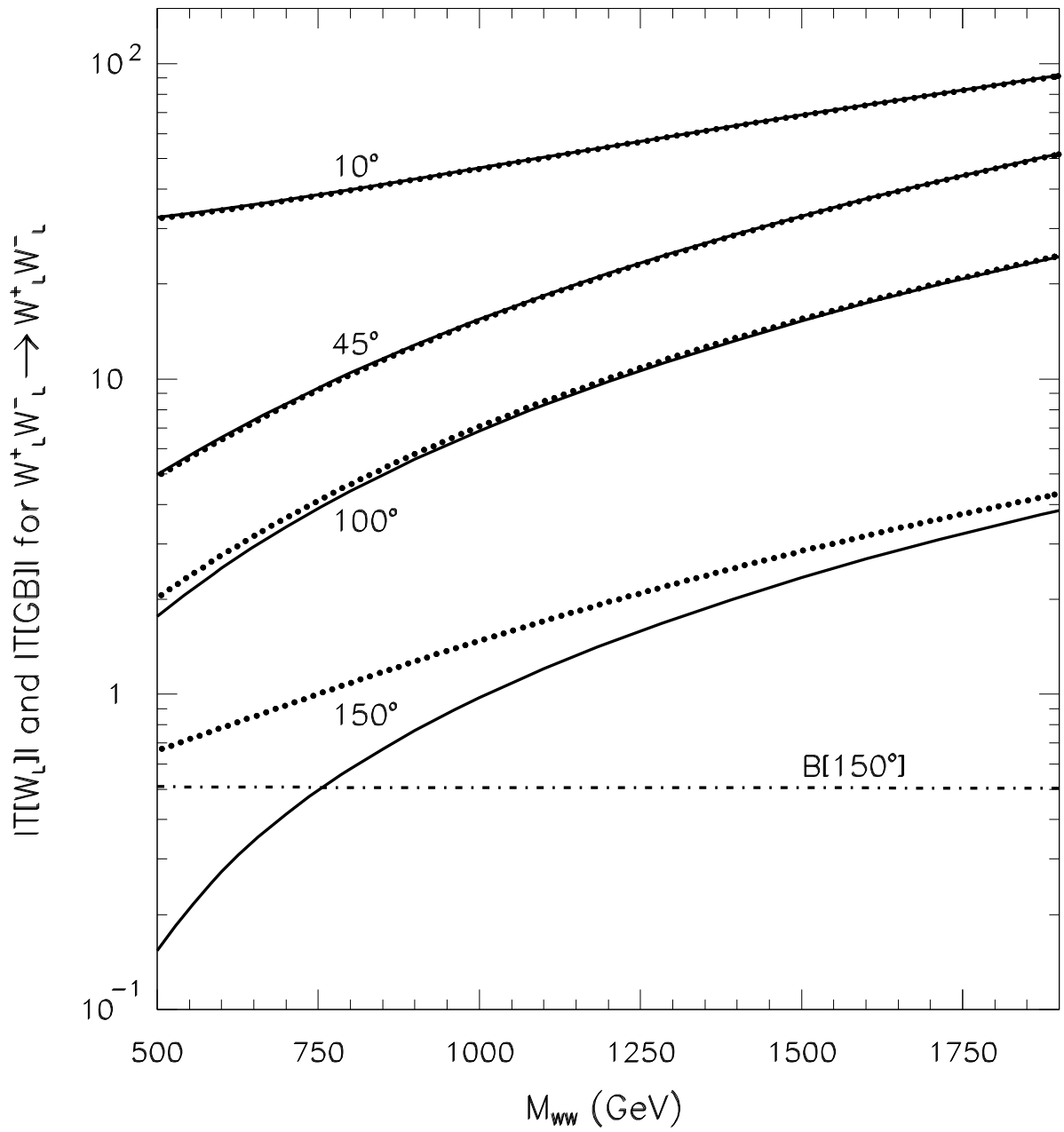
\end{document}